\documentclass[fleqn,usenatbib]{mnras}

\usepackage{newtxtext,newtxmath}
\usepackage[T1]{fontenc}

\usepackage{hyperref}
\usepackage{afterpage}
\usepackage{comment}

\DeclareRobustCommand{\VAN}[3]{#2}
\let\VANthebibliography\thebibliography
\def\thebibliography{\DeclareRobustCommand{\VAN}[3]{##3}\VANthebibliography}

\usepackage{graphicx}	% Including figure files
\usepackage{amsmath}	% Advanced maths commands

\usepackage{float}
\usepackage{xcolor}
\newcommand{\Mbh}{M_\bullet}
\newcommand{\Mdot}{\dot{M}}

\newcommand{\Mdottr}{\dot{M}_{\rm{tr}}}

\newcommand{\chitr}{\chi_{\rm{tr}}}
\newcommand{\chitrnew}{\chi_{\rm{tr}}^{\rm{new}}}
\newcommand{\rhotr}{\rho_{\rm{tr}}}
\newcommand{\sigmatr}{\sigma_{\rm{tr}}}
\newcommand{\sigmagj}{\sigma_{\rm{GJ}}}
\newcommand{\sigmatrq}{\sigma_{\rm{tr},0.25}}
\newcommand{\sigmatrh}{\sigma_{\rm{tr},0.50}}
\newcommand{\sigmatrx}{\sigma_{\rm{tr},x}}
\newcommand{\rg}{\,r_{\rm{g}}}
\newcommand{\rin}{r_{\rm{in}}}
\newcommand{\rmax}{r_{\rm{max}}}

\newcommand{\pmin}{p_{\rm{min}}}

\newcommand{\rhofl}{\rho_{\rm{floor}}}

\newcommand{\rhomin}{\rho_{\rm{min}}}

\newcommand{\sigmamax}{\sigma_{\rm{max}}}
\newcommand{\Gammamax}{\Gamma_{\rm{max}}}
\newcommand{\thetamax}{\theta_{\rm{max}}}
\newcommand{\sqrtg}{\sqrt{-g}}
\newcommand{\rgc}{\,r_{\rm{g}}/c}
\newcommand{\lem}{L_{\rm{EM}}}
\newcommand{\Temrt}{{T_{\rm{EM}}}^r_t}

\newcommand{\bhac}{\texttt{BHAC}}

\newcommand{\ls}[1]{{#1}}

\newcommand{\acc}[1]{{#1}}

\newcommand{\wong}{}
\newcommand{\accc}{}
\newcommand{\lss}{}

\title[]{Baryons in magnetically arrested black hole disks and jets}
\title[]{The baryon  content of magnetically arrested black hole disks and jets}

\author[A. Chow et al.]{Anthony Chow,$^{1}$\thanks{E-mail: kc3058@columbia.edu}
Lorenzo Sironi,$^{1,2}$
Bart Ripperda,$^{3,4,5,6}$ 
Amir Levinson$^{7}$
\\
$^{1}$Department of Astronomy and Columbia Astrophysics Laboratory, Columbia University, New York, NY 10027, USA\\
$^{2}$Center for Computational Astrophysics, Flatiron Institute, 162 Fifth Avenue, New York, NY 10010, USA\\
$^{3}$Canadian Institute for Theoretical Astrophysics, 60 St. George St, Toronto, ON M5S 3H8, Canada\\
$^{4}$Department of Physics, University of Toronto, 60 St. George St, Toronto, ON M5S 1A7, Canada\\
$^{5}$David A. Dunlap Department of Astronomy \& Astrophysics, University of Toronto, 50 St. George St, Toronto, ON M5S 3H4, Canada\\
$^{6}$Perimeter Institute for Theoretical Physics, 31 Caroline St. North, Waterloo, ON N2L 2Y5, Canada\\
$^{7}$School of Physics and Astronomy, Tel Aviv University, Tel Aviv 69978, Israel\\
}

\date{Accepted XXX. Received YYY; in original form ZZZ}

\pubyear{2015}

\begin{document}
\label{firstpage}
\pagerange{\pageref{firstpage}--\pageref{lastpage}}
\maketitle

\begin{abstract}
We study the transport of baryons in magnetically arrested accretion flows and relativistic jets using general relativistic magnetohydrodynamic simulations that incorporate a passive Eulerian tracer. The tracer allows us to reconstruct a proxy for the physical baryon density supplied by the accretion disk while excluding the mass injected numerically to maintain stability in highly magnetized, low-density regions. Applying this method to axisymmetric black hole simulations with varying spin, we show that baryon loading of the jet is intrinsically episodic and regulated by magnetic flux eruption cycles occurring in the inner accretion flow. Each eruption evacuates baryons from the innermost equatorial region, drives reconnection in extended current sheets, and expels moderately magnetized disk material along the funnel wall, establishing a recurrent mass-loading channel. 
% distinct from steady boundary-layer mixing
\acc{In spinning black holes, shear-driven waves along the jet boundary further enhance baryon entrainment, whereas this mechanism is suppressed in the non-spinning case. For parameters representative of the black hole accretion flow in M87, we map the global structure and time evolution of the Goldreich–Julian screening boundary, defined as the surface separating regions where the plasma density is sufficient to supply the charges required to screen electric fields parallel to the magnetic field from regions that are charge starved.} For spinning black holes, we find that the electromagnetic power of the jet is predominantly carried by baryon-poor plasma, with extended time intervals of charge starvation. Our results provide a framework for diagnosing jet composition, charge starvation, and reconnection-driven mass loading in magnetically arrested black hole systems, with direct implications for particle acceleration and non-thermal emission in low-luminosity accretion flows.
%Finally, by partitioning the radial electromagnetic power by magnetization, we assess the degree to which charge starvation is preserved  jets and identify the conditions that enable the outflow to sustain unscreened electric fields capable of powering high-energy emission. 
\end{abstract}


%%%%%%%%%%%%%%%%%%%%%%%%%%%%%%%%%%%%%%%%%%%%%%%%%%

%%%%%%%%%%%%%%%%% BODY OF PAPER %%%%%%%%%%%%%%%%%%
\section{Introduction} \label{sec:introduction}

Relativistic jets are among the most powerful manifestations of black hole activity, playing a fundamental role in both galactic and extragalactic environments (see the reviews by \citealt{Blandford_2019ARA&A, Kara_2025}). They act as efficient channels for feedback, regulating star formation and the thermal state of interstellar and intracluster media \citep[e.g.,][]{Fabian_2012}. Their broadband, non-thermal emission spans the entire electromagnetic spectrum—from radio to X-ray and $\gamma$-ray—and arises from a combination of leptonic and hadronic processes. \acc{In addition to their dynamical and radiative output, relativistic jets have been considered potential sites for accelerating ultra–high–energy cosmic rays (UHECRs) and producing high-energy neutrinos \citep{Kotera_2011,Rodrigues_2021,IceCube_2022_Jul}.} Magnetic reconnection in highly magnetized relativistic outflows \citep{sironi_25} can efficiently accelerate protons to $\gtrsim 10^{19}$eV (e.g., \citealt[][]{Zhang_Sironi_Giannios_2021}), motivating jet-based proton-acceleration models. The $\sim 290\,$TeV neutrino detected in coincidence with a gamma-ray flare in the blazar TXS~0506+056 \citep{IceCube_2018, Padovani_2018} has demonstrated that jets can be powerful neutrino sources, implying the presence of relativistic protons accelerated to energies beyond the PeV scale. 

\acc{In addition to relativistic jets, magnetically powered coronal regions near the black hole have recently been proposed as potential neutrino production sites. In fact, the strongest current candidate for a neutrino emitter, NGC~1068 \citep{IceCube_2022}, shows TeV neutrino emission consistent with originating from the nuclear region, while exhibiting no evidence for a relativistic jet or associated gamma-ray emission \citep{Fischer_2023}.  These observations favor an X-ray–emitting coronal region---rather than a jet-based scenario---as the origin for TeV neutrinos. In systems reaching the magnetically arrested disk (MAD) state—where accretion advects sufficient poloidal magnetic flux to intermittently suppress mass inflow near the black hole horizon \citep{Narayan_2003, Tchekhovskoy_2011}—flux-eruption–driven reconnection layers forming near the black hole are invoked to be promising sites for proton acceleration \citep{Stathopoulos_2024,Hakobyan_2025} and potentially high-energy neutrino production \citep{fiorillo_24,karavola_25,karavola_25b} in non-jetted sources.}
\begin{comment}
Flux-eruption-driven reconnection layers forming near the black hole in MAD states remain promising sites for proton acceleration \citep{Hakobyan_2025} and high-energy neutrino production in non-jetted sources.
\end{comment}

The composition of relativistic jets — and in particular their baryon loading — plays a central role in shaping both their dynamics and their radiative properties. The amount of baryonic matter (protons and heavier ions) entrained in the relativistic outflow determines the plasma density and inertia of the jet. Since the plasma is approximately charge neutral, the electron density is at least comparable to the ion density. Therefore, the ion number density may be compared to the Goldreich–Julian (GJ) density \citep{Goldreich_Julian_1969}, which represents the minimum charge density required to short out the electric fields (parallel to the magnetic field) induced by a rotating magnetosphere, which is a good approximation for the innermost region of the accretion flow \citep{Ripperda_2022, Chael_2023, Wong_Chael_2025}. If the plasma density falls below the GJ value, the available charges are insufficient to supply the currents demanded by the magnetospheric field structure. In this charge-starved regime, vacuum gaps inevitably form — regions where strong unscreened electric fields can accelerate particles to ultra-relativistic energies, enabling pair creation and non-thermal emission \citep{Hirotani_Okamoto_1998, Hirotani_2017, Levinson_2018, Crinquand_2020, Yuan_2025}. On the other hand, if the plasma density is much higher than the GJ density, the magnetosphere remains well-screened, and the presence of baryons (as opposed to a pure electron–positron plasma) increases the inertia of the flow. The degree of baryon loading determines the maximum attainable Lorentz factor of the jet \citep{Toma_Takahara_2012, Sikora_2000, Levinson_2000, Cerruti_2020, Kantzas_2023}.

\acc{Beyond its dynamical role, plasma composition has important implications for the radiative signatures of black hole systems. In sources approaching the MAD state, the jet base and inner magnetosphere become highly magnetised, enabling magnetic pressure to drive violent flux eruptions \citep{Ripperda_2022} and to form reconnecting current sheets. These reconnection layers are promising sites for proton acceleration up to ultra-high energies in low-luminosity sources \citep{Stathopoulos_2024}. The baryon content of these layers  also regulates the efficiency and observational relevance of proton synchrotron emission. As shown by \citet{Hakobyan_2025}, proton synchrotron can still be energetically important even when protons are subdominant in number and mass density; for the plasma conditions expected in M87, protons can tap a significant fraction of the dissipated magnetic energy and dominate the emission in the GeV band during flux eruptions. Thus, a proper assessment of baryon loading within MAD reconnection layers is essential for evaluating their potential as proton accelerators and for understanding the radiative signatures of low-luminosity accreting systems.}

\acc{Plasma composition also influences the radiative properties of relativistic jet boundary layers or sheaths, where dissipation and mixing are expected to be significant. Dissipation associated with jet boundary layers has been observed in the form of limb-brightened radio emission \citep{Janssen_Falcke_2021}. The jet boundary layer has also been proposed as a potential candidate for the X-ray Comptonising corona in luminous sources \citep{Sridhar_2025}. In this scenario, the  optical depth to Compton scattering depends sensitively on whether the plasma at the jet boundary is baryon- or pair-dominated.
%, since the density of scattering electrons—and hence the optical depth—differs substantially between the two cases. 
Constraining the baryon content at jet boundaries is therefore critical for interpreting coronal or jet boundary emission, motivating future global simulations of black hole accretion flows with self-consistent electron thermodynamics and proper radiation physics.}

In summary, accurately constraining the baryon  content of magnetically dominated black hole environments is essential not only for understanding their global electrodynamics — such as gap formation — but also for assessing their mass loading and their ability to act as high-energy particle accelerators and as sources of non-thermal radiation and high-energy neutrinos.

% Leaving the sentence in here seem more reasonable for transition from phyiscal properties to modeling of the system.
The set of equations of general relativistic magnetohydrodynamics (GRMHD), which evolve the plasma mass, momentum and energy densities as well as the magnetic field in curved spacetimes, provides the primary framework for studying black hole accretion and jet dynamics \citep{Gammie_2003, McKinney_2006, Porth_Olivares_Mizuno_2017}. However, GRMHD methods have intrinsic limitations—particularly in low-density, magnetically dominated regions, such as the jet, the black hole magnetosphere, and the reconnection layers created by flux eruptions. To maintain numerical stability, such methods impose upper limits on the plasma magnetization (i.e., on the ratio of magnetic field enthalpy density to plasma enthalpy density) and enforce density floors by continuously injecting mass in regions where the plasma density would otherwise fall below a prescribed minimum. In typical setups, the accretion disk is initialized with a physically motivated magnetization and genuine baryonic matter, while the surrounding atmosphere and jet regions are initialized with an \textit{ad hoc} maximum magnetization and a corresponding \textit{ad hoc} minimum density. As the simulation evolves, the persistent mass injection required to maintain numerical stability erases information about the physical density and plasma content in low-density, magnetically dominated regions. Consequently, such numerical prescriptions, though essential for stability, mask the physical baryon loading and composition in the magnetosphere and the jet.

\wong{\citet{Wong_2021} employed 3D GRMHD simulations of MAD to investigate baryon loading at the jet boundary.} They considered a retrograde disk and identified the disk–jet interface by the sign reversal of the azimuthal four-velocity ($u_\phi$). To track the transport of matter across this interface, they implemented a passive tracer method, in which Lagrangian particles are initialized with probabilities proportional to the local mass density in the initial disk, and they are advected according to the local fluid velocity. This approach allowed them to follow individual tracer particles through space and time, capturing the entrainment of disk material into the jet. By analyzing when the tracer particles’ radial and azimuthal velocities changed sign, \citet{Wong_2021} identified the onset of disk-to-jet mass entrainment and suggested that these episodic mass-loading events arise from Kelvin–Helmholtz (KH) instabilities developing along the jet boundary layer \citep{Chow_2023, Davelaar_2023}.

In this study, we adopt a different approach. Rather than using discrete Lagrangian tracer particles, we employ a passive Eulerian tracer field that records the physical baryon rest-mass density of the accretion flow, independent of the mass added \textit{ad hoc} to maintain the imposed density floors. We initialize the tracer field so that it traces only baryonic material originating from the accretion flow (i.e., it is initialized as zero outside the disk).
%to the same value as the mass density in the disk and to zero elsewhere, so that  
The tracer is passively advected with the fluid, and in regions where numerical density floors inject matter, its value is adjusted so that the local tracer mass density remains conserved. In this way, the tracer behaves as a strictly passive scalar—carrying no dynamical influence—which faithfully tracks the transport of disk-supplied baryons throughout the simulation. This allows us to separate genuine baryonic material supplied by the accretion flow from mass that is numerically injected through density floors---the latter would likely be  dominated by electron-positron pairs. This enables us to follow the transport of disk material into the black hole magnetosphere and jet funnel. We perform axisymmetric (2D) ideal GRMHD simulations of MADs around black holes with three different spins—prograde (with respect to the disk angular momentum), non-spinning, and retrograde—to focus on the dynamics of mass transport and baryonic loading. 
\begin{comment}
Our approach enables us to identify a mass-loading pathway that is distinct from the boundary-layer mixing processes emphasised in \citet{Wong_2021, Wong_2025}: in the MAD regime, magnetic–flux–eruption events expel disk material from the equatorial region at the base of the funnel, producing a moderately baryon-loaded layer that is subsequently advected outward and transported to the jet boundary at larger radii. 
We note that, because our simulations are axisymmetric, non-axisymmetric instabilities—such as Rayleigh–Taylor modes—and fully three-dimensional turbulence may modify our results. A full 3D investigation will therefore be required, building on the strategy we pioneer in this work.
\end{comment}

% Lizhong comments added here.
\acc{We note that all simulations presented in this work are performed in axisymmetry (2D), which imposes well-known limitations. In particular, axisymmetric calculations do not sustain long-lived magnetorotational turbulence, and the turbulent cascade to small scales is suppressed. As a result, some aspects of angular-momentum transport  are not captured in a fully realistic manner. However, in the MAD regime, the global dynamics is dominated by large-scale, coherent magnetic fields rather than by small-scale turbulence. Phenomena central to the MAD state—such as the accumulation of poloidal magnetic flux near the black hole, episodic magnetic-flux eruptions, large-scale reconnection events, and the formation of a strongly magnetized funnel—
%arise from the evolution of these large-scale magnetic structures and 
can  be qualitatively captured in axisymmetric simulations \citep{Ripperda_2020}. We acknowledge, however, that non-axisymmetric modes may drive interchange instabilities in the inner magnetosphere, and thus modify some of the conclusions of this work, as we also discuss in $\S$\ref{sec:EM_partition}.}

%In this sense, 2D MAD simulations provide a controlled and physically transparent framework for isolating the mechanisms that regulate flux eruptions, reconnection , and baryon loading, while deferring the role of fully developed turbulence, as well as long-term evolution, to future three-dimensional studies.

This paper is organized as follows. In $\S$\ref{sec:methods}, we present the method of using a passive Eulerian tracer to reconstruct the physical baryon density field. $\S$\ref{sec:setup} describes the numerical setup of the MAD accretion flow simulations and the GRMHD code used in our study. $\S$\ref{sec:results} contains our main findings: $\S$\ref{sec:mdot_phi} examines how flux-eruption cycles regulate the accretion rate of baryonic matter; $\S$\ref{sec:sigmatr_flux_eruption} analyzes how black-hole spin affects the structure and dynamics of the baryon magnetization (hereafter, the tracer magnetization, which we define as the ratio of magnetic enthalpy density to baryonic rest-mass energy density), in the course of individual eruption events; $\S$\ref{sec:time_polar angle} studies the baryon loading near the equatorial current sheet; $\S$\ref{sec:theta_max} illustrates the geometry of the Goldreich–Julian screening boundary, which governs where and when charge starvation occurs in the jet, for  parameters representative of the black hole accretion flow in M87; and \acc{$\S$\ref{sec:EM_partition} quantifies the degree of baryonic mass loading of the jet, by partitioning the jet electromagnetic power by tracer magnetization.}
%partitions electromagnetic power in the jet region by tracer magnetization and tracks when fixed power percentiles fall below or above the characteristic tracer magnetization evaluated at the Goldreich–Julian density, thereby characterising the jet outflow dynamics.
\begin{comment}
$\S$\ref{sec:EM_partition} investigates how electromagnetic power is partitioned across magnetization regimes to diagnose charge starvation and inertial loading.
\end{comment}
$\S$\ref{sec:discussions} summarizes our conclusions and discusses the implications of our results for jet dynamics and non-thermal emission.

\section{Implementation of the baryon tracer density} \label{sec:methods}
\begin{comment}
\ls{Amir: Perhaps we should use "tracer plasma",  although "baryon" should be fine (understanding that it is neutral).  You may want to add a sentence at the beginning of sec 2 clarifying that the tracer density reflects the composition of the disk at radii < 100 rg or so (at much larger radii, the disk is cold and contains also dust particles and conceivably molecules).}
\end{comment}

The primary aim of this work is to investigate the baryon loading of the highly magnetized jet and magnetosphere of the black hole. We employ the GRMHD code \bhac~ 
 \citep{Porth_Olivares_Mizuno_2017, Olivares_2019}, which regulates the maximum plasma magnetization through the injection of density floors. The code imposes a maximum value of the cold magnetization,
\begin{align}
    \sigmamax=\frac{b^2}{4\pi c^2 \rhofl},
\end{align}
\acc{where $b$ is the co-moving magnetic field. Mass injection is activated when the cold magnetization exceeds some prescribed $\sigmamax=100$. By injecting floor density, $\rhofl$, the code enforces $\sigma < \sigmamax$ and maintains numerical stability in magnetically dominated regions.}
\begin{comment}
where $b$ is the co-moving magnetic field, and activates mass injection whenever the rest-mass density $\rho$ falls below the threshold $\rhofl$. In such cases, the density is reset to $\rho'=\rhofl$, thereby preventing the magnetization from exceeding $\sigmamax$ and maintaining numerical stability in magnetically dominated regions.
\end{comment}
This artificial mass injection, however, limits our ability to recover the physical baryon density in highly magnetized regions. To address this, we make use of a passive scalar tracer.

\begin{comment}
It is important to note that the tracer density reflects the composition of the inner disk (typically at radii $\lesssim 100r_g$), where the flow is hot and fully ionized; at much larger radii the disk becomes cooler and may contain dust and molecular material not captured by this simplified tracer prescription.
\end{comment}

A passive tracer $\chitr$ is a quantity that is advected with the fluid flow but does not influence the system’s dynamics \citep{Porth_2014}. For the sake of comparison, the evolution of the mass density $\rho$ is governed by the mass continuity equation
\begin{equation} \label{equ:mass_continutiy}
\nabla_\mu(\rho u^\mu)=0,
\end{equation}
where $u^{\mu}$ is the fluid 4-velocity and Greek indices $\mu=0,1,2,3$ denote spacetime components, with $\nabla_\mu$ the covariant derivative associated with the spacetime metric. The tracer density $\rhotr=\rho\chitr$ evolves according to the tracer continuity equation
\begin{equation} \label{equ:tracer_continutiy}
\nabla_\mu(\rhotr u^\mu)=0,
\end{equation}
i.e., it is advected with the same four-velocity $u^\mu$ as the MHD fluid. The evolution of energy and momentum of the MHD fluid depends only on Equation~\eqref{equ:mass_continutiy} and not on Equation~\eqref{equ:tracer_continutiy}. This implicitly requires that the global dynamics is nearly independent of the value chosen for $\sigmamax$. In Appendix~\ref{sec:sigma_max_sensitivity}, we verify that our results do not depend on the maximum magnetization $\sigmamax$.
%assumption is appropriate provided that the density-injection prescription does not significantly modify the global flow dynamics, which we verify by varying  (see Appendix~\ref{sec:sigma_max_sensitivity}).
The tracer field is inevitably subject to numerical diffusion; we assess the degree of numerical convergence with respect to grid resolution in Appendix~\ref{sec:convergence}.

For the purposes of this study, we interpret $\rhotr$ as a proxy for the baryon mass density unaffected by the injection of density floors. To preserve the value of the tracer density when floor mass is injected, the updated tracer fraction (i.e., the tracer fraction $\chitrnew$ after floor injection) must satisfy
\begin{align}
    \rhofl \chitrnew=\rho \chitr.
\end{align}
This essentially states that the tracer mass density $\rho_{\rm tr}$ is unaffected by the addition of density floors.

At the start of the simulation, we initialize the passive tracer fraction to tag matter originating in the disk by setting $\chitr = 1$ inside the accretion disk and $\chitr = 0$ in the ambient medium. Interpreting the tracer as a proxy for baryonic mass, this choice allows us to track the transport and redistribution of baryons originating from the disk into the whole domain, including magnetically dominated regions.
\begin{comment}
At the start of the simulation, the accretion disk is assumed to be baryon-rich (ion-dominated), while the surrounding atmosphere contains no baryonic material. The passive tracer fraction is therefore initialized as $\chitr = 1$ within the accretion disk and $\chitr = 0$ in the ambient medium, enabling us to track how disk-originating materials, representing baryons, are transported into and distributed throughout the magnetically dominated regions.
\end{comment}

\section{Numerical Setup} \label{sec:setup}
 This work employs axisymmetric ideal GRMHD simulations performed with $\bhac$ in a logarithmic Kerr-Schild coordinate system \citep{McKinney_2004}, with $t$ representing time and $r,\theta,\phi$ the radial, polar, and toroidal coordinates (for a logarithmic Kerr-Schild metric, the radial coordinate is $\log r$). We investigate MAD configurations with three representative values of the dimensionless spin parameter, $a \in [-0.9375, 0, 0.9375]$, corresponding respectively to near-extremal retrograde Kerr, Schwarzschild, and near-extremal prograde Kerr black holes.

With the gravitational radius defined as $\rg=GM_\bullet/c^2$, where $M_\bullet$ is the black hole mass, we measure lengths and times in units of $\rg$ and $\rgc$, respectively. We also adopt the Lorentz-Heaviside units, where a factor of $1/\sqrt{4\pi}$ is absorbed into the electromagnetic fields. Each simulation is initialized with a hydrodynamic equilibrium torus \citep{Fishbone_1976} threaded by a single weak poloidal magnetic field loop, defined through the vector potential
\begin{align}
A_\phi \propto \max(q, 0),
\end{align}
where
\begin{align}
q = \frac{\rho}{\rho_*} \left( \frac{r}{r_{\text{in}}} \right)^3 \sin^3 \theta \cdot \exp \left( - \frac{r}{400\rg} \right) - 0.2,
\end{align}
with an inner radius $\rin = 20\rg$ and a density maximum $\rho_*$ located at $\rmax$. To enable a consistent comparison across simulations with different spins, we adopt $\rmax=41\rg, 41.4\rg~\text{and}~42\rg$ for $a=0.9375, 0, -0.9375$, respectively. These choices of $\rmax$ ensure that the disks have comparable sizes for the three different spins.
%by maintaining a nearly constant aspect ratio.

The initial magnetic field strength is normalized such that the plasma beta $2p/b^2 = 100$, where $p$ denotes the plasma pressure. We set an atmospheric rest-mass density $\rho_{\rm atm} = \rhomin (r/\rg)^{-3/2}$ and pressure $p_{\rm atm} = \pmin (r/\rg)^{-5/2}$, where $\rhomin = 10^{-4}$ and $\pmin = 10^{-6}/3$, as in \citet{Ripperda_2020}. In the course of the evolution, we apply floors on rest-mass density, pressure, and Lorentz factor $\Gamma < \Gammamax=20$, and we maintain a cold magnetization $\sigma=b^2/\rho c^2 < \sigmamax=100$ and plasma beta $\beta=2p/b^2>\beta_{\rm min}=1/(10\,
\sigmamax)^{(\hat{\gamma}-1)}$, where $\hat{\gamma}$ is the adiabatic index. We employ the equation of state of a relativistic ideal gas with $\hat{\gamma} = 4/3$. 

The magnetorotational instability \citep{velikhov_1959, chandrasekhar_1960, Balbus_Hawley_1991} is triggered by perturbing the equilibrium pressure as $p = p_{\rm eq} (1 + X_{\rm p})$, where $p_{\rm eq}$ is the analytic equilibrium pressure of the Fishbone–Moncrief torus, and  $X_{\rm p}$ is a uniformly distributed random variable in the range $[-0.02, 0.02]$.

The base grid resolution is $384 \times 192$ cells in $(\log r, \theta)$, and the simulations employ five levels of adaptive mesh refinement (AMR), yielding an effective resolution of $6144$ radial and $3072$ polar cells. We performed a resolution-convergence study to verify the robustness of our results; details are provided in Appendix \ref{sec:convergence}.

\section{Results} \label{sec:results}
\subsection{Temporal evolution of baryon accretion rates} \label{sec:mdot_phi}
In MADs, magnetic flux eruptions intermittently generate strongly magnetized, low-density regions near the event horizon. These magnetically dominated regions trigger the injection of numerical density floors, causing the total accretion rate to deviate from the physical baryon accretion rate. In this section, we examine how black-hole spin affects the near-horizon accretion dynamics.

\begin{figure} 
	\centering
	\includegraphics[width=0.48\textwidth]{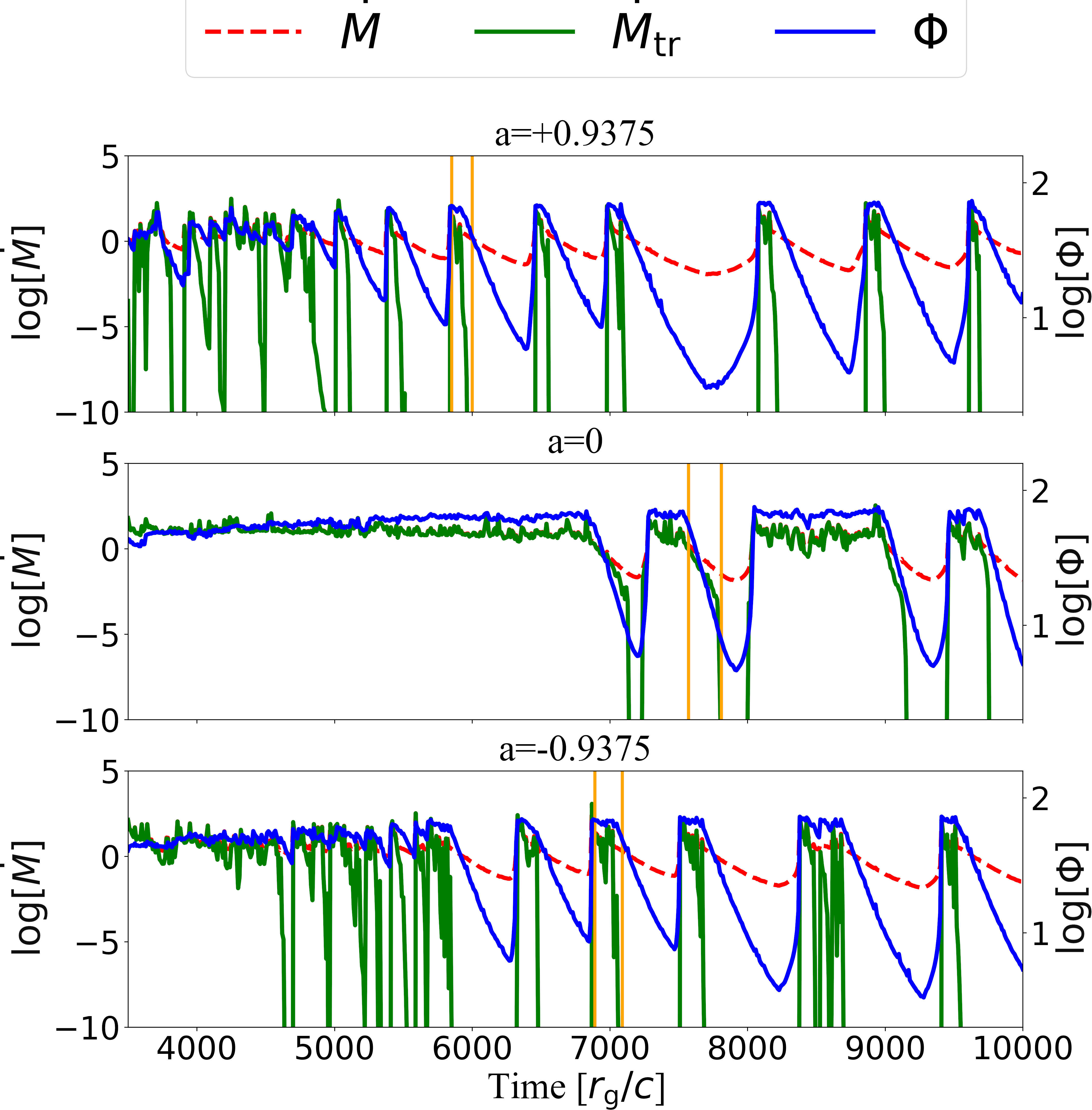}
	\caption{Time evolution of the mass accretion rate $\dot{M}$ (red dashed line), the tracer mass accretion rate $\Mdottr$ (green line) and the magnetic flux $\Phi$ threading the horizon (blue line)  for spin parameter $a=0.9375$ (top panel), $a=0$ (middle panel) and $a=-0.9375$ (bottom panel). $\Mdot$, $\Mdottr$ and $\Phi$ are all in code units.}
    \label{fig:mdot_phi}
\end{figure}

Fig. \ref{fig:mdot_phi} shows the time evolution of the mass accretion rates and of the magnetic flux evaluated on the event horizon. For a Kerr black hole, the outer horizon lies at
\begin{align}
    r_+=\rg \left ( 1 + \sqrt{1-a^2} \right ),
\end{align}
giving $r_+/\rg= 1.35, 2, 1.35$ for spins $a=0.9375,0,-0.9375$, respectively. Let $u^r$ and $B^r$ denote the radial components of the four-velocity and magnetic field, respectively. Then the mass accretion rate (red dashed lines in Fig.~\ref{fig:mdot_phi}) is defined as
\begin{align}
    \dot{M}=-2\pi \int_0^\pi \rho u^r\sqrtg~d\theta,   
\end{align}
while the tracer mass accretion rate (green solid lines) is
\begin{align}
    \dot{M}_{\rm{tr}}=-2\pi \int_0^\pi \rhotr u^r\sqrtg~ d\theta.
\end{align}
The magnetic flux threading the horizon (blue solid lines) is 
\begin{align}
    \Phi=\pi \int_0^\pi |B^r|\sqrtg ~d\theta.
\end{align}

\acc{Following an initial transient phase during which the magnetorotational instability grows and accretion is established, we begin our analysis once the system has reached a quasi-steady state, as indicated by the fact that the mass accretion rate attains a nearly-constant value prior to the onset of magnetic flux eruptions.} 
In prograde accretion (top panel), the time evolution of the magnetic flux from $t\approx 3000\rgc$ to $t\approx 5000\rgc$ is characterized by small, episodic flux eruptions. \acc{After $t\approx 5000\rgc$, the eruptions become stronger, produce deeper flux-suppression episodes, and persist for longer durations, although this late-time evolution may be influenced by the assumption of axisymmetry and should be confirmed with fully three-dimensional simulations.} A similar behavior is observed in retrograde accretion (bottom panel), with flux eruptions of comparable durations and magnitudes as in the prograde case. 

In contrast, for the non-spinning case (middle panel), no small-scale flux eruptions are present at early times ($t<6700\rgc$), and large-scale eruptions appear only at later times, after $t\approx 6700\rgc$. This delayed onset arises because, in the absence of black-hole spin, frame dragging does not contribute to twisting field lines near the horizon (which would amplify the toroidal field, increasing the overall field strength), and the magnetospheric configuration evolves more slowly toward a highly pressured state. \acc{Continued accretion leads to a gradual accumulation of poloidal flux in the funnel region, eventually
%until a delayed reconnection episode is triggered—likely, once the inflow channel becomes sufficiently thin to activate the tearing instability—
producing a large-scale flux eruption.}
\begin{comment}
Continued accretion leads to a gradual accumulation of poloidal flux in the funnel region, until a delayed reconnection episode is eventually triggered, producing a large-scale flux eruption.
\end{comment}
\begin{comment}
The intense eruptions observed in both the prograde and retrograde cases are also seen in 3D simulations \cite{Ripperda_2022, Salas_2025}, although the statistics in 3D remains limited at high enough resolution to study magnetic reconnection. These large-scale eruptions are driven by the progressive magnetization of the inner region, which allows magnetic energy to accumulate until it is released in bursts. In three dimensions, flux eruptions drive magnetic reconnection that relieves the accumulated magnetic stress, while Rayleigh–Taylor and interchange instabilities subsequently reopen accretion channels and restart the cycle toward the MAD state. In contrast, two-dimensional simulations lack a third spatial dimension for such streams, allowing magnetic energy to build up more rapidly. Consequently, eruptions in 2D occur more frequently and with greater intensity, leading to faster overall magnetization of the disk compared to the 3D case. These differences highlight the critical role of dimensionality and plasma instabilities in regulating the dynamics and energetics of magnetized accretion flows.
\end{comment} 

Flux eruptions are accompanied by the formation of magnetospheric reconnection layers \citep[e.g.,][]{Ripperda_2020}.
 As accretion drags large-scale poloidal magnetic fields into the black hole, the magnetic pressure at the funnel base accumulates until it becomes comparable to the inflow’s ram pressure. At this point, oppositely directed field lines in the inner disk reconnect violently, producing a flux eruption. This rapid reconfiguration of the magnetic topology drives plasma outward along the reconnected field lines, evacuating the innermost region. As plasma is expelled during the eruption, the  density in the vicinity of the black hole drops sharply, giving rise to a magnetically dominated cavity, i.e., a magnetosphere.

During flux eruptions, the drop in density in the inner region is reflected by the decline in the mass accretion rate $\Mdot$ (red dashed line) and the tracer mass accretion rate $\Mdottr$ (green solid line) for all spin cases. Because the magnetization near the black hole often reaches the maximum $\sigmamax$, numerical floor mass is injected. This explains the divergence between $\Mdot$  and $\Mdottr$ during flux eruptions, with $\Mdottr$ dropping much faster. In fact, Fig. \ref{fig:mdot_phi} demonstrates that most of the mass accreted during flux eruptions comes from numerical floors.

%, the decrease in $\Mdot$ is less pronounced than in $\Mdottr$. 
%The much steeper decline of $\Mdottr$ during flux eruptions indicates that the magnetically dominated cavity has a very low baryon density.

Near the end of a flux eruption, ram pressure begins to dominate over magnetic pressure, allowing baryonic matter to accrete onto the black hole again (in fully three-dimensional systems, this picture may be modified by magnetic interchange instabilities, see \citealt{Ripperda_2022}). \acc{Consequently, the divergence between the total mass accretion rate $\Mdot$ and the tracer accretion rate $\Mdottr$ is reduced during phases of high accretion in between consecutive flux eruptions, as the inflow advects baryons back from the disk and the inner accretion channel remains well below the maximum magnetization $\sigmamax$.}

\begin{comment}
Consequently, the divergence between the total mass accretion rate $\Mdot$ and the tracer accretion rate $\Mdottr$ is reduced during the phases of high mass accretion rate in between consecutive flux eruptions, and injection of numerical floor density is not triggered in the near-horizon accretion funnel. 
\end{comment}
\begin{figure*}
	\centering
	\includegraphics[width=0.9\textwidth]{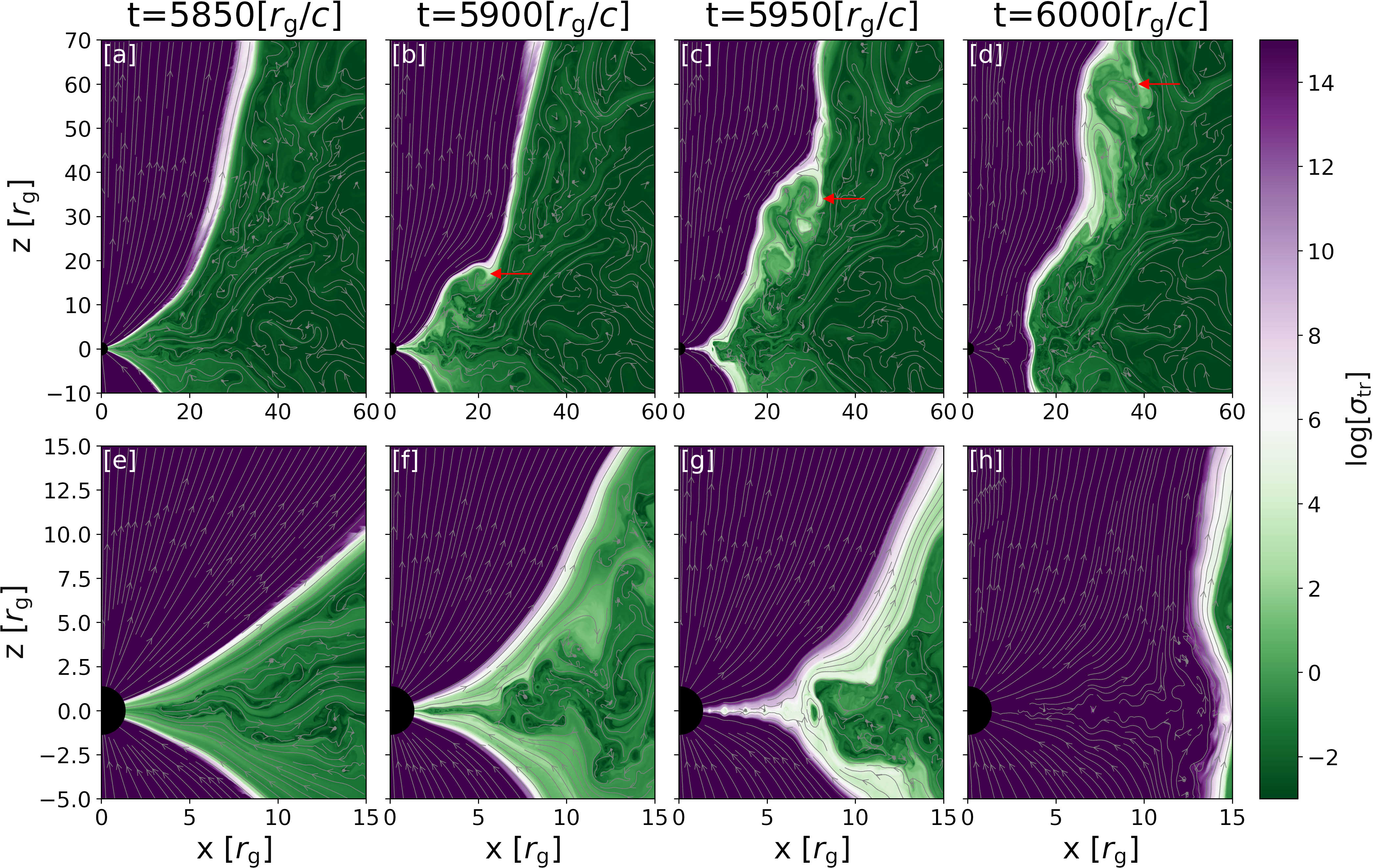}
	\caption{$x$–$z$ slices of the tracer magnetization, $\sigmatr$, at simulation times $t = 5850\rgc$, $5900\rgc$, $5950\rgc$, \text{and } $6000\rgc$, for a black hole with spin parameter $a = 0.9375$. The upper panels display the large-scale structure, while the lower panels provide a zoomed-in view of the near-horizon region, highlighting the evolution of magnetized outflows and instabilities. Red arrows in panels [b]–[d] point to the location of a growing vortex.}
    \label{fig:2D_grid_spin+ve}
\end{figure*}

\begin{figure*}
	\centering
	\includegraphics[width=0.9\textwidth]{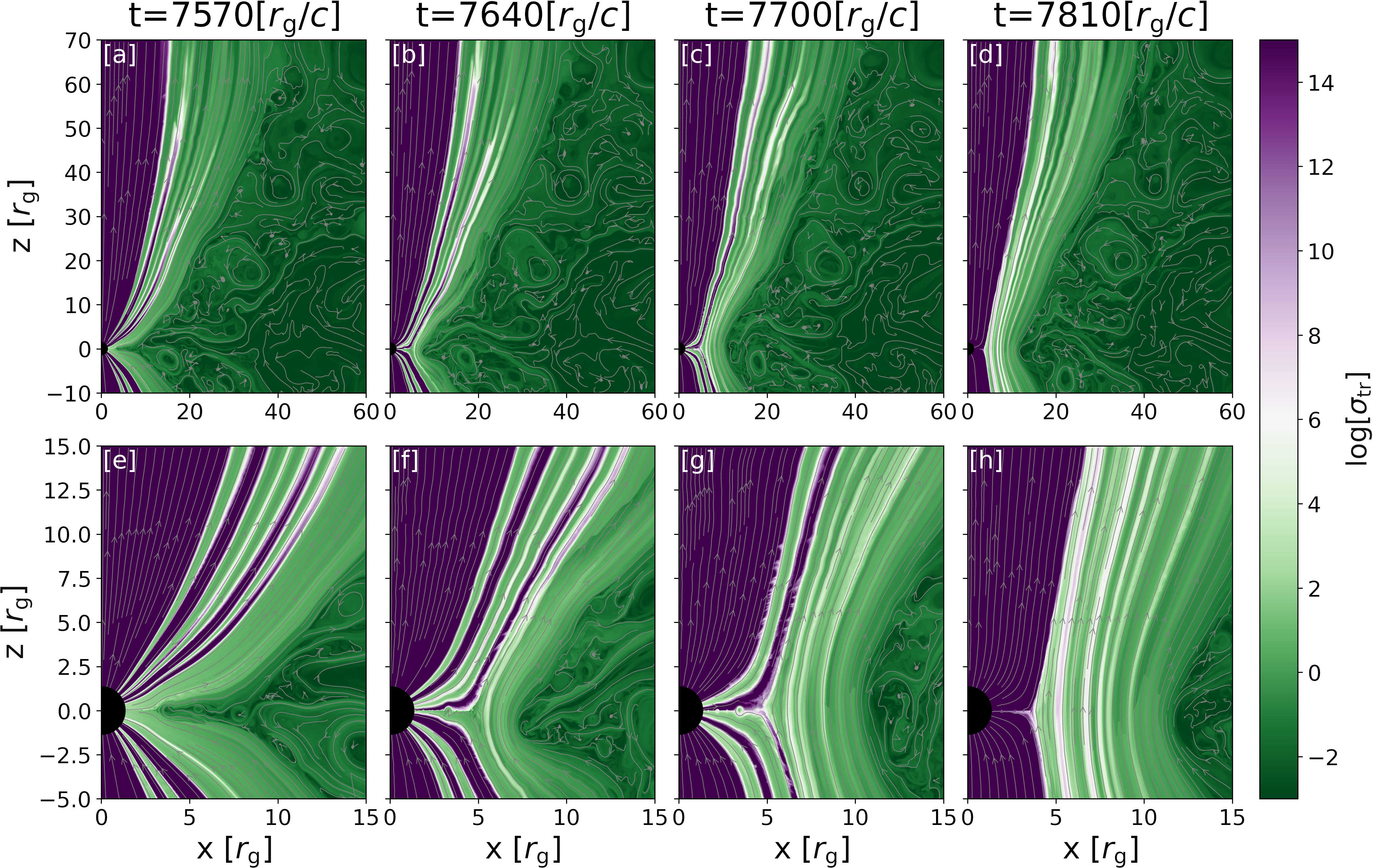}
	\caption{Same as Fig. \ref{fig:2D_grid_spin+ve}, but for a black hole with spin parameter $a=0$, shown at simulation times $t = 7570\rgc$, $7640\rgc$, $7700\rgc$, \text{and } $7810\rgc$.}
    \label{fig:2D_grid_spin0}
\end{figure*}

\begin{figure*}
	\centering
	\includegraphics[width=0.9\textwidth]{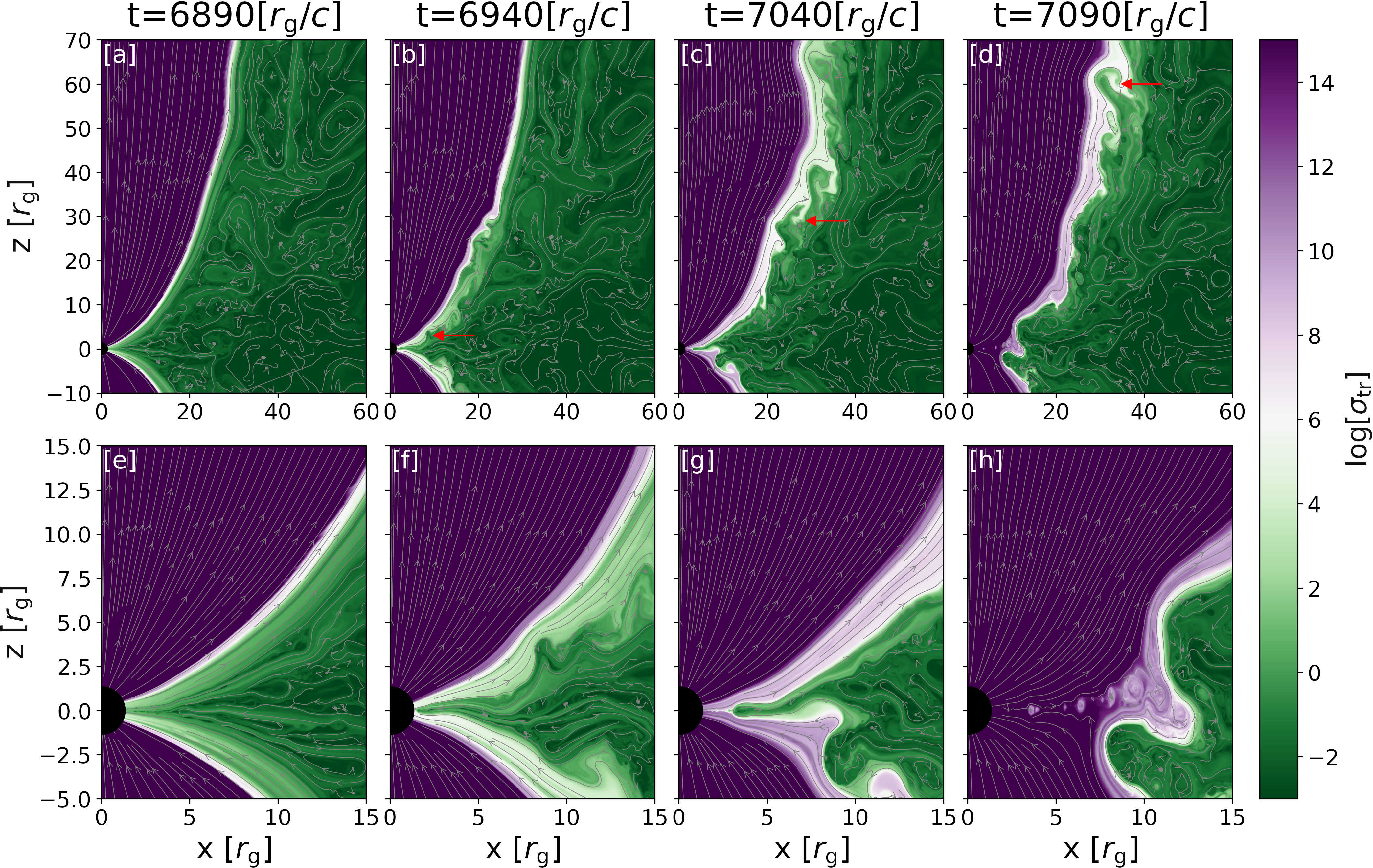}
	\caption{Same as Fig. \ref{fig:2D_grid_spin+ve}, but for a black hole with spin parameter $a=-0.9375$, shown at simulation times $t = 6890\rgc$, $6940\rgc$, $7040\rgc$, \text{and } $7090\rgc$.}
    \label{fig:2D_grid_spin-ve}
\end{figure*}

\subsection{Baryon dynamics during a single flux-eruption cycle} \label{sec:sigmatr_flux_eruption}
In this section, we focus on a  single eruption cycle and we characterize the  evolution of the baryon content at the jet boundary and in the inner equatorial region. We highlight the similarities that emerge among the two rotating black-hole models and the contrasting behaviour of the non-rotating case. Using our passive-tracer modeling of baryons originating in the disk, we identify the physical processes that govern baryon entrainment within the jet funnel.

For a cold plasma, the tracer magnetization is defined as
\begin{align}
    \sigmatr=\frac{b^2}{\rhotr c^2}=\frac{b^2}{\rho\chitr c^2},
\end{align}
which has the advantage of being insensitive to the magnetization ceiling $\sigmamax$. Figs.~\ref{fig:2D_grid_spin+ve}-\ref{fig:2D_grid_spin-ve} present the spatial structure of $\sigmatr$ obtained from our simulations with spin parameter of $a=0.9375,0,-0.9375$ respectively (the corresponding plots for $\chitr$ are displayed in Appendix \ref{sec:chi_plots}). The top row panels display the profile of  $\sigmatr$ over the spatial domain $x=[0,60]\rg$ and $z=[-10,70]\rg$ while the bottom row provides a zoomed-in view of the inner region, $x=[0,15]\rg$ and $z=[-5,15]\rg$. We focus on the upper hemisphere, as both hemispheres are statistically equivalent. Each column in the figure corresponds to a different simulation time, with $t=[5850,5900,5950,6000]\rgc$ for the prograde case, $t=[7570,7640,7700,7810]\rgc$ for Schwarzschild and $t=[6890,6940,7040,7090]\rgc$ for the retrograde case. These time snapshots are selected to illustrate the temporal evolution of $\sigmatr$ from the early phase of a magnetic flux eruption (the first orange vertical line in Fig. \ref{fig:mdot_phi}) to the stage when the tracer accretion rate approaches zero (the second orange vertical line in Fig. \ref{fig:mdot_phi}). Figs.~\ref{fig:2D_grid_spin+ve}-\ref{fig:2D_grid_spin-ve} are also overlaid with magnetic field streamlines (grey lines).

For both the prograde (Fig.~\ref{fig:2D_grid_spin+ve}) and the retrograde case (Fig.~\ref{fig:2D_grid_spin-ve}), panel~[a] shows that the jet boundary prior to the flux eruption is laminar, and the tracer magnetization $\sigmatr$ increases sharply across a very thin layer, from $\sigmatr\sim 10$ (green) up to $10^6$ (white). At this time, the  system is accreting in a non-erupting state.

During the initial phase of a flux eruption, reconnection between oppositely oriented magnetic field lines near the equatorial plane gives rise to a baryon-rich, low-$\sigmatr$ plasmoid chain (panel~[f]). At the same time, the magnetic field lines in the jet boundary are perturbed at small radii, and shear-driven vortices
\begin{comment}
the magnetic field is perturbed at small radii, and Kelvin–Helmholtz-like vortices
\end{comment}
(red arrow in panel~[b]) develop at the interface between the highly magnetized, low-density white layer and the weakly magnetized, high-density green region. %\citep{Wong_2021, Chow_2023}. 
The representative small vortex seen in panel~[b] advects outward to larger distances, drawing in ambient baryons as it propagates. As it grows into a larger vortex (red arrow in panel~[c]), it leads to further entrainment of low-$\sigmatr$ baryonic matter into the jet boundary layer.

\ls{Later in the eruptive phase (panels~[c] and [g]), magnetic field lines that were initially loaded with baryon-rich disk material are now threading the black hole, and their baryonic matter starts flowing out along the jet boundary. On such field lines, baryons can no longer be replenished from the disk, leading to a reduction in the local tracer density and a corresponding increase in $\sigmatr$ (see the evolution from green to white at the jet boundary between panels [f] and [g]).}
%reconnection redirects baryon-loaded field lines from accretion to ejection. Once their footpoints thread the black hole, 

%of the equatorial current sheet (panel~[g]).}

Towards the end of the flux eruption (panels~[d] and [h]), the inner region becomes baryon-depleted (panel~[h]). In this phase, the equatorial current sheet and its plasmoids are filled primarily with the numerically-injected floor density (in realistic systems, we expect that the layer will be dominated by a pair plasma). Meanwhile, the vortex originating in panel~[c] entrains additional baryons as it propagates outward, growing into an even larger vortex (red arrow in panel~[d]) that effectively thickens the jet boundary layer.

For the non-rotating black hole case, Fig.~\ref{fig:2D_grid_spin0} corresponds to the second major flux-eruption event as shown by the vertical orange lines in the middle panel of Fig.~\ref{fig:mdot_phi}. Compared to prograde and retrograde models, the equatorial reconnection layer is shorter, $\sim 5 \rg$ (panel~[h]). Across panels~[a]–[d], the disk–funnel interface remains largely laminar throughout the whole duration of the flux-eruption event and exhibits a clear sequence of alternating high- and low-$\sigmatr$ stripes. We find that this stratification correlates with the sign of the radial plasma velocity (low-$\sigmatr$ for material coming from the disk towards the black hole, high-$\sigmatr$ in the opposite case); however, we do not further investigate its physical origin here, since it is likely that three-dimensional effects would smooth or erase such a stratification.

These results show that baryon entrainment within the funnel boundary is governed by two distinct physical processes. First, during the initial baryon-loaded stage of a flux-eruption event, baryons are expelled from the funnel base, thus injecting low-$\sigmatr$ (i.e., baryon-rich) material directly into the boundary layer, which then propagates to larger radii; this process is present for all three spins, including the non-rotating case. Second, in the rotating black-hole models, shear-driven waves along the jet boundary lead to the formation of vortices as described by \citet{Wong_2021}, which propagate outward and entrain additional ambient baryons, thereby thickening the mixing layer. No such vortices are observed in the non-spinning model.

\begin{figure}
	\centering
	\includegraphics[width=0.48\textwidth]{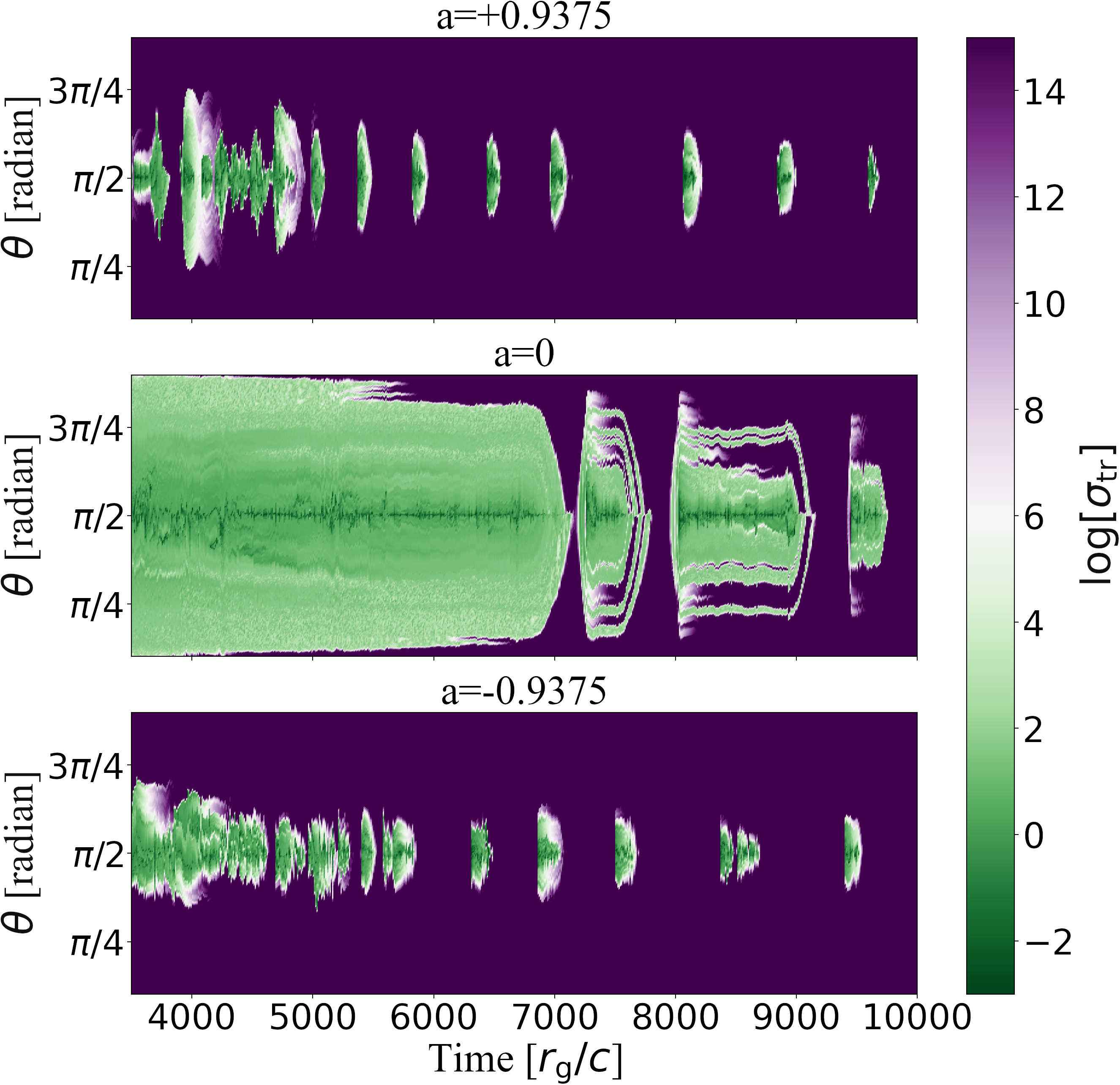}
	\caption{Time evolution of the logarithm of $\sigmatr$ at different polar angles $\theta$ for a fixed radial coordinate $r = 3\rg$ measured from the center of the black hole. The three panels correspond to different black hole spins: (top) $a = 0.9375$, (middle) $a = 0$, and (bottom) $a = -0.9375$.}
    \label{fig:theta_vs_time}
\end{figure}

\subsection{Baryon loading of the equatorial current sheet} \label{sec:time_polar angle}
The equatorial current sheet that forms during MAD flux eruptions can be a source of bright high-energy emission \citep{Hakobyan_2023}. It remains unclear whether reconnection in this layer is fed purely by pair plasma from photon-photon collisions or whether protons originating from the disk also participate. \citet{Hakobyan_2025} performed local particle-in-cell simulations of relativistic reconnection in a pair-proton plasma and demonstrated that even when protons are subdominant in mass and number, they can acquire a sizeable fraction of the dissipated magnetic energy, so their radiative signature (through proton synchrotron emission) can be significant. A reliable quantification of the baryonic content in this region is therefore essential to predict its high-energy signatures. In this subsection, we focus on the innermost magnetospheric region around the current sheet, and use the passive tracer to track its baryon content both in time and polar angle, thereby identifying the composition of the material flowing into the reconnection layer.

Fig. \ref{fig:theta_vs_time} shows the temporal evolution of the tracer magnetization $\sigmatr$ across polar angle $\theta$ at fixed radius $r=3\rg$ for three black hole spins: $a=0.9375$ (top), $a=0$ (middle) and $a=-0.9375$ (bottom). Throughout the evolution of the  prograde and retrograde cases,  the regions near the polar axis ($\theta<\pi/4$ and $\theta>3\pi/4$) remain persistently high-$\sigmatr$, forming the base of the baryon-poor jet core.
Closer to the equator, the early evolution ($4000\rgc\lesssim t\lesssim 5000\rgc$) is characterized by a broad low-$\sigmatr$ (green) region \acc{extended over a wide range of $\theta$ around the equatorial plane ($\theta=\pi/2$),} produced by steady accretion and intermittently perturbed by small flux-eruption events (white-purple). These early eruptions reduce—but do not fully remove—the baryon content near the mid-plane. At later times, both rotating models develop a sequence of isolated low-$\sigmatr$ (green) patches separated by extended high-$\sigmatr$ (purple) time intervals.
This alternating pattern mirrors the sequence of
brief episodes of significant baryon loading during high accretion states followed by baryon depletion during large-scale eruptions. 
 In  the course of each eruption, the baryon content at a given polar angle above or below the equator (e.g., take $\theta=3\pi/8$) is progressively depleted, as marked by the progression from dark green to white and finally purple in Fig.~\ref{fig:theta_vs_time}. This evolution in $\sigmatr$ indicates that the plasma flowing into the equatorial current sheet evolves from a baryon-rich to a baryon-poor state in the course of each flux eruption. This can inform hadronic synchrotron emission models, as in the case of the accretion flow in M87 discussed by \citet{Hakobyan_2025}.

 The build-up of magnetic pressure generated by reconnection exhausts during a flux eruption prevents the inflow of disk material and produces extended high-$\sigmatr$ purple intervals. \acc{When the magnetic barrier weakens and accretion resumes, baryon-rich plasma appears as the next low-$\sigmatr$ green patch. The relatively short duration of the low-$\sigmatr$ phases is likely a consequence of the axisymmetric nature of the simulations, which suppresses non-axisymmetric magnetic Rayleigh–Taylor–like instabilities that may facilitate sustained accretion in three-dimensional MADs \citep[e.g.][]{Ripperda_2022}. }
\begin{comment}
When the magnetic barrier weakens and accretion resumes, baryon-rich plasma appears as the next low-$\sigmatr$ green patch near the mid-plane. Throughout this evolution, the high polar angle region ($\theta<\pi/4$ or $\theta>3\pi/4$) remains persistently high-$\sigmatr$, forming the base of the baryon-poor jet core.
\end{comment}

The behavior of the non-spinning model sharply  contrasts  with the spinning cases. In most of its evolution, the baryon-poor funnel remains narrowly collimated, 
%and the correspondingly small polar opening angle
which
permits low-$\sigmatr$ (green) material to extend up to substantially higher latitudes than in the rotating cases. 
%As a result, the green region appears more vertically extended throughout the early evolution. 
Flux-eruption activity develops more gradually than in the rotating cases, with significant eruptions emerging only after $t\approx7000 \rgc$, consistent with the slower accumulation of magnetic pressure  expected in the absence of frame dragging, as we have discussed above. The resulting high-$\sigmatr$ (purple) intervals are correspondingly shorter than in the spinning models, indicating that the eruptions themselves are weaker and less capable of sustaining prolonged baryon depletion in the inner magnetosphere.

\subsection{Structure and evolution of the Goldreich–Julian screening boundary} \label{sec:theta_max}
The Goldreich--Julian (GJ) density provides a quantitative criterion for assessing whether the local plasma supply is sufficient to screen parallel electric fields and maintain force-free conditions. Using our passive-tracer--based estimate of the baryon density, $\rho_{\rm tr}$, we identify regions where the plasma content falls below the GJ threshold, thereby mapping the global structure and temporal evolution of the GJ screening boundary.
%This boundary marks regions where numerical GRMHD solutions carry larger physical uncertainty. 
In charge-starved regions where the plasma density falls below the GJ value, gap formation and self-consistent pair creation---processes not captured within the standard GRMHD framework---would be required to restore force-free conditions.
%In charge-starved regions where the density is below the GJ value, gap formation and self-consistent pair creation---processes that cannot be captured within the GRMHD framework---would be needed to re-establish the force-free dynamics.
%In GRMHD simulations, the plasma content in these regions is set by ad hoc mass injection. In reality, we expect much lower densities, as suggested by the density tracer, where 

\accc{We restrict this analysis to the rotating black-hole models. For a non-spinning black hole, the GJ density associated with black-hole rotation formally vanishes, $n_{\rm GJ}=0$, corresponding to $\sigmagj\propto 1/n_{\rm GJ} \rightarrow \infty$. 
Although $\sigmatr$ remains a meaningful measure of baryon loading in the $a=0$ simulation, comparison with a finite GJ threshold cannot be used to diagnose charge starvation in this case.}
 
\begin{figure}
	\centering
	\includegraphics[width=0.48\textwidth]{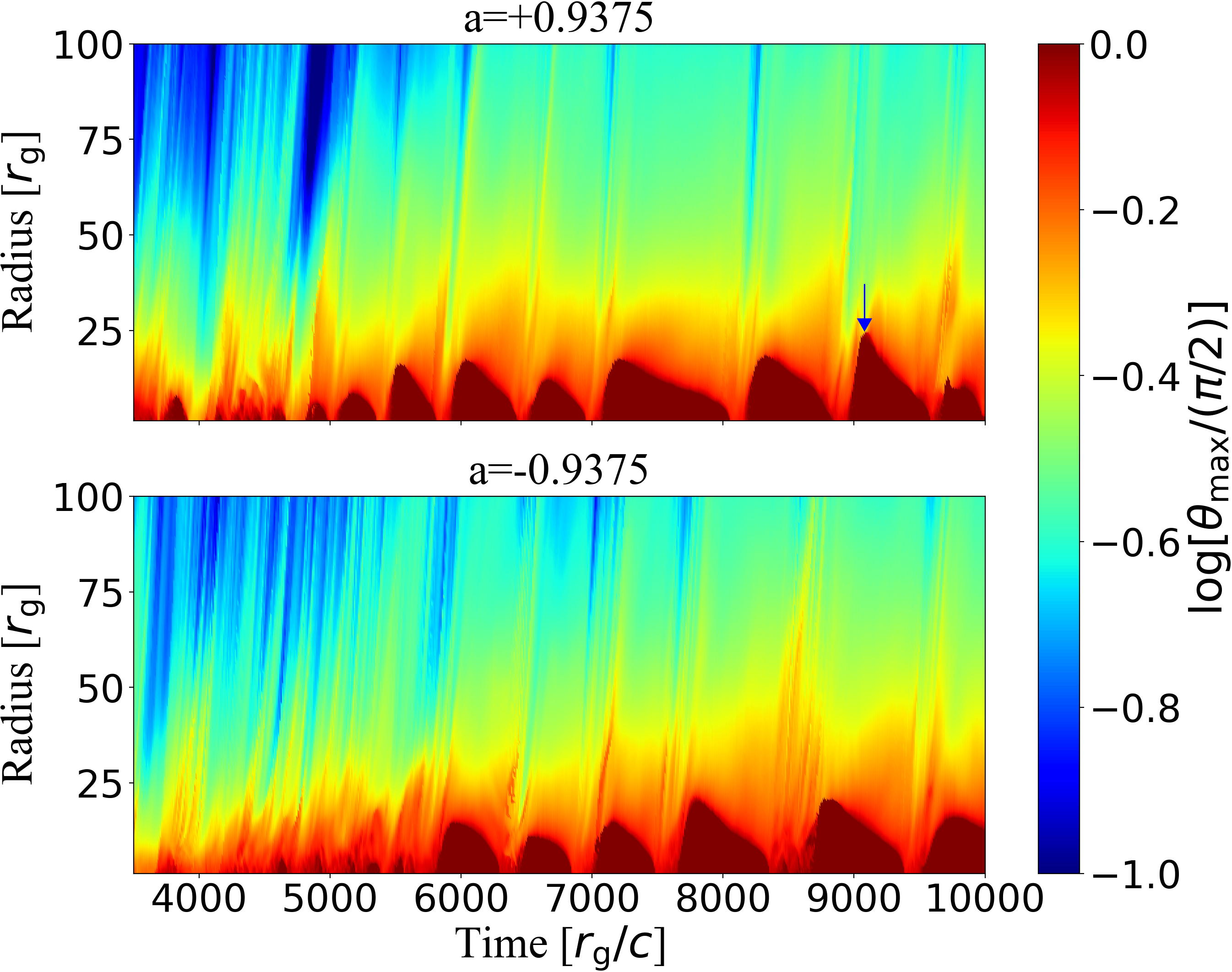}
    \caption{\accc{Time–radius diagrams of the maximum polar angle (in radians) in the upper hemisphere $(\theta \leq \pi/2)$ at which the tracer magnetization consistently satisfies $\sigmatr > \sigma_{\rm GJ}= 10^{10}$. The horizontal axis shows time, and the vertical axis corresponds to radial distance from the black hole, spanning $r \in [0,100]\rg$. The two panels correspond to the rotating black-hole models: (top) $a = 0.9375$ and (bottom) $a = -0.9375$. The blue arrow in the top panel points to the peak of the dark red patch, which corresponds to the maximum extent of the equatorial charge-starved region for $a=0.9375$.}}
    \label{fig:theta_max_vs_time_1}
\end{figure}

To quantify the structure of the GJ boundary \accc{in the rotating models}, we measure at each radius the maximum polar angle in the upper hemisphere, $\theta_{\max}\leq \pi/2$, such that all cells with $\theta < \theta_{\max}$ satisfy $\sigma_{\rm tr} > \sigma_{\rm GJ}=10^{10}$. \accc{This fiducial threshold is appropriate, to order of magnitude, for the conditions in M87*; its dependence on magnetic-field strength and black-hole mass, as well as its estimate for Sgr A*, is discussed in Appendix~\ref{appendix:GJ_magnetization}}. The critical angle $\theta_{\max}$ therefore traces the outer edge of the charge-starved region in the jet spine and provides a quantitative measure of how its opening angle evolves in response to magnetic-flux eruptions and reconnection events.
%To quantify the structure of the GJ boundary, we measure at each radius the maximum polar angle in the upper hemisphere, $\theta_{\max}\leq \pi/2$, such that all cells with $\theta < \theta_{\max}$ satisfy $\sigma_{\rm tr} >\sigma_{\rm GJ}= 10^{10}$; this threshold corresponds to the GJ density for a magnetic field of $\sim 100\,\mathrm{G}$ at the horizon, as appropriate for M87 (see Appendix~\ref{appendix:GJ_magnetization}). The critical angle $\thetamax$ then traces the outer edge of the charge-starved region in the jet spine and provides a quantitative measure of how its opening angle evolves in response to magnetic-flux eruptions and reconnection events.
Fig.~\ref{fig:theta_max_vs_time_1} presents the time evolution of $\thetamax$ as a function of radius and time. We consider only the upper hemisphere and show results for \accc{the two rotating models, $a=0.9375$ and $a=-0.9375$}. The colour scale shows $\log(\thetamax/(\pi/2))$, with darker (redder) regions indicating larger values of $\thetamax$. The lower hemisphere is statistically symmetric and exhibits similar behaviour.

%We first consider the rotating models (top and bottom panels). 
At large radii ($r>25\rg$), $\theta_{\max}$ decreases with distance, indicating that the charge-starved jet spine becomes progressively more collimated while the jet boundary gets increasingly baryon-loaded as the jet propagates outward. Superimposed on this global trend are slanted striations produced by disturbances launched near the black hole during flux eruptions. These disturbances propagate outward along the jet and modulate the angular extent of the GJ boundary.

%We first discuss the rotating cases (top and bottom panels). At large radii ($r>25\rg$), $\thetamax$ decreases with distance, indicating that the charge-starved jet spine becomes progressively more collimated and the jet boundary more baryon loaded, as the jet propagates outward. Superimposed on this global trend are slanted striations, produced by disturbances launched near the black hole during magnetic-flux eruptions; these propagate outward along the jet and modulate the angular extent of the GJ boundary.

At small radii ($r<25\rg$), intermittent dark-red patches mark episodes in which strong flux eruptions completely evacuate baryons from the inner magnetosphere. The resulting magnetic pressure creates an equatorial cavity, causing $\theta_{\max}$ to approach $\pi/2$. This indicates that the charge-starved region extends down to the equator (and, by symmetry, across all polar angles). The peak time of each dark-red patch coincides with a minimum in the tracer accretion rate $\Mdottr$ and with the maximum radial extent of the equatorial reconnection layer (or, more generally, of the charge-starved magnetosphere). For example, the blue arrow in the top panel of Fig.~\ref{fig:theta_max_vs_time_1} (prograde case) marks a charge-starved cavity extending to $22.8\rg$. As accretion resumes, baryons gradually refill the cavity and $\theta_{\max}$ decreases. This cycle of baryon evacuation, disk displacement, and subsequent refilling reflects the episodic nature of magnetic-flux eruptions and their impact on the extent of the charge-starved region.

\wong{We note that the comparison between the tracer magnetization $\sigmatr$ and the Goldreich–Julian threshold $\sigmagj$ only measures the ability of disk-entrained plasma to supply the charges required to screen the magnetosphere. Regions with $\sigmatr > \sigmagj$ therefore indicate that disk entrainment alone may be insufficient to provide this charge supply. However, this does not necessarily imply that the region is charge starved, since additional charges may be supplied through pair-production processes (e.g., Breit–Wheeler pairs produced from upscattered disk photons \citep{Wong_Ryan_2021}), which are not captured by the tracer analysis.}

\subsection{Jet power partitioned by baryonic magnetization} \label{sec:EM_partition}
To quantify the degree of baryon loading in the Poynting-dominated jet, we examine how the radial electromagnetic (EM) power is distributed as a function of $\sigma_{\rm tr}$. By identifying which magnetization regimes dominate the outward EM energy flux, and how these regimes evolve during flux-eruption episodes, we assess whether\accc{, for conditions typical of M87*,} the jet remains predominantly charge-starved or becomes sufficiently screened and potentially inertia-dominated. In this subsection we focus on the rotating models and exclude the non-spinning case ($a=0$), which does not produce a sustained powerful jet.

%To investigate the degree of baryon loading to the Poynting-dominated jet, we analyze how the radial component of the electromagnetic (EM) power is partitioned as a function of $\sigmatr$. By identifying which magnetization regimes dominate the outward EM energy flux, and how these regimes evolve during flux-eruption episodes, we assess whether the jet remains predominantly charge-starved or instead becomes sufficiently screened and potentially even inertia-dominated. In this subsection, we focus only on the rotating models and exclude the non-spinning case ($a=0$) because it produces no sustained powerful jet.

To quantify how the radial component of the EM power is distributed as a function of the tracer magnetization, let $\partial \Temrt / \partial \sigmatr$ denote the amount of radial EM power carried by plasma of a given magnetization $\sigmatr$. We quantify the cumulative electromagnetic power carried by regions with $\sigmatr<\sigma_{\rm tr,x}$ at radius $r$ and time $t$  by computing
\begin{align} \label{eq:partial_EM_power}
    L(\sigmatrx) = 2\pi \int_0^{\sigmatrx}\int_0^{\pi/2} \frac{\partial \Temrt}{\partial \sigmatr} \sqrtg ~d\theta~d\sigmatr,
\end{align}
where we call $\sigmatrx$ the tracer magnetization value such that a fraction $x$ of the total radial EM power is carried by plasma with $\sigmatr<\sigma_{\rm tr,x}$. Equivalently, if we call $L_{\rm EM}$  the total EM power, we  define $\sigmatrx$ such that $L(\sigmatrx) /\lem=x$. In this subsection, we track the time and radial dependence of $\sigmatrq$ and $\sigmatrh$, and compare them with the GJ magnetization $\sigma_{\rm GJ}=10^{10}$ appropriate for M87* \lss{(we remark that the GJ magnetization scales linearly with magnetic field strength and black hole mass, see Appendix~\ref{appendix:GJ_magnetization})}. 
 %where $\sigmatrx$ is the magnetization threshold at which x\% of the total radial EM luminosity, $L_{\rm EM} \equiv L(\infty)$, is contained (i.e. $L(\sigmatrx) /\lem=x$). We consider the cases of in this study.
\begin{comment}
Because the Poynting flux in certain ranges of $\sigma_{\rm tr}$ can be undersampled at individual time snapshots, we apply a moving average to 
$\partial T^{r}_{\rm EM}/\partial \sigma_{\rm tr} \cdot \sqrt{-g}$
using a time window of width~200 to smooth the distribution.
\end{comment}

\begin{figure}
	\centering
	\includegraphics[width=0.48\textwidth]{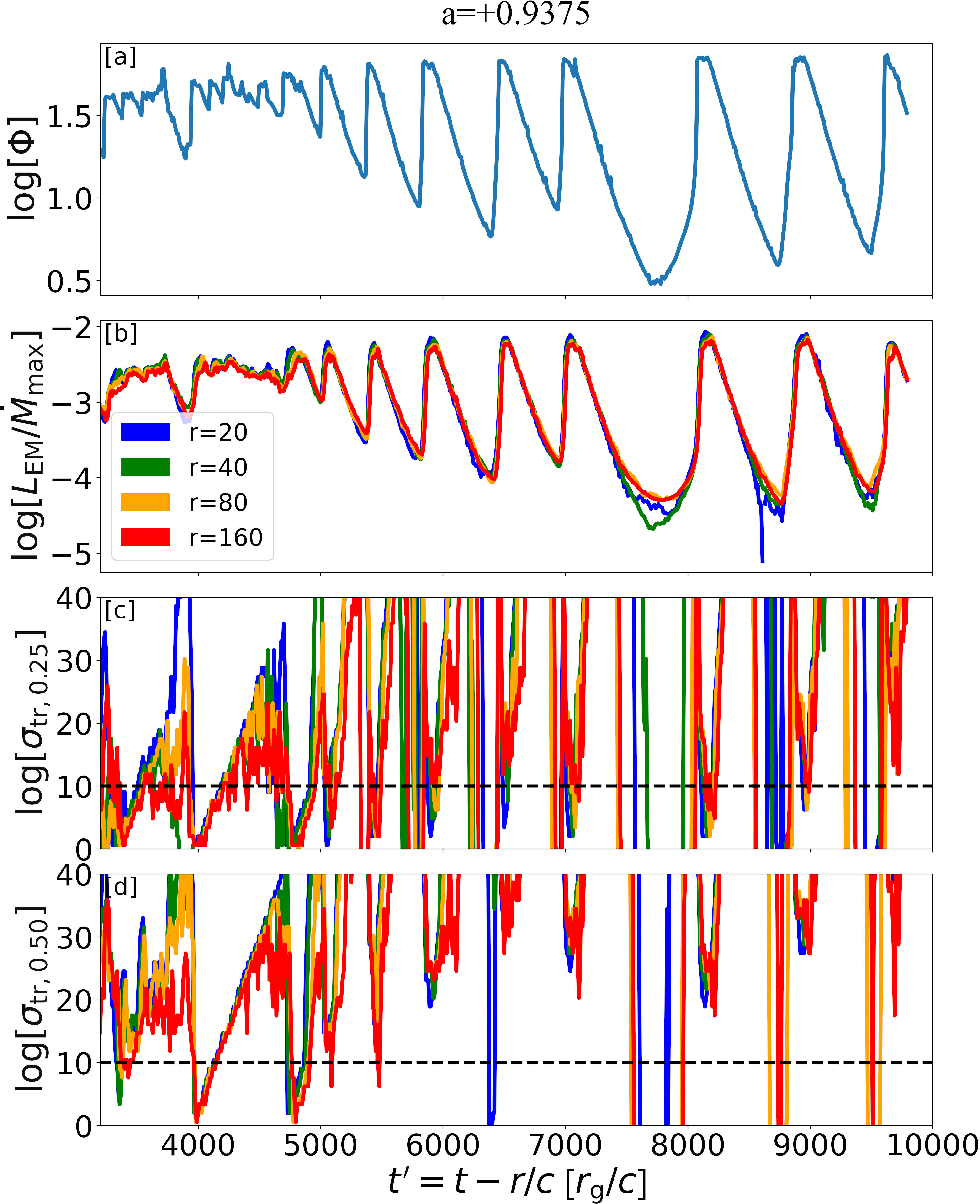}
    \caption{Electromagnetic jet power diagnostics as a function of time for spin $a=0.9375$. 
    \textbf{[a]} Magnetic flux threading the black hole horizon (same as in Fig.~\ref{fig:mdot_phi}).
    \textbf{[b]} Total radial electromagnetic (EM) power normalized by ${\Mdot}_{\rm{max}}c^2$, measured at the radii indicated in the legend. ${\Mdot}_{\rm{max}}=295.82$ is defined as the maximum accretion rate at the event horizon for $t\in[0,5000]\rgc$.
    \textbf{[c]} Tracer magnetization value $\sigma_{\rm tr,0.25}$, defined as the threshold at which $25\%$ of the EM energy flux is carried by material with $\sigmatr<\sigma_{\rm tr,0.25}$.
    %and $75\%$ lies above, thereby identifying the characteristic magnetization range that carries the dominant portion of the outflow power.
    \textbf{[d]} Tracer magnetization value $\sigma_{\rm tr,0.5}$, corresponding to the median of the EM power distribution. The color scheme for panels (c) and (d) is the same as for panel (b). The black dashed horizontal lines in panels (c) and (d) mark the characteristic GJ magnetization $\sigmagj = 10^{10}$ appropriate for M87*. 
    Time is shifted in a radius-dependent way to account for light-travel time effects (for panel (a), we take $r=1.348\,r_{\rm g}$).}   \label{fig:sigmatr_EMpower_25_50_spin+ve}
\end{figure}

Fig.~\ref{fig:sigmatr_EMpower_25_50_spin+ve} summarizes the evolution of the magnetic flux and EM power partition in the prograde case. Panel~[a] shows the magnetic flux $\Phi$ threading the black hole horizon (see also Fig.~\ref{fig:mdot_phi}), while panel~[b] displays the total EM energy flux $\lem$ measured at $r = 20\rg, 40\rg, 80\rg,$ and $160\rg$ (see legend). The two quantities are well correlated: during flux accumulation, magnetic flux is advected onto the horizon, strengthening the Blandford--Znajek (BZ) outflow; during flux-eruption episodes, the horizon-threading flux decreases, reducing the jet power. Panels~[c] and [d] show the time evolution of $\sigmatrq$ and $\sigmatrh$, respectively, with black dashed horizontal lines marking the characteristic GJ magnetization $\sigmagj = 10^{10}$ appropriate for M87* (see $\S$\ref{sec:theta_max} and Appendix~\ref{appendix:GJ_magnetization}). The fraction of time for which $\sigmatrq$ and $\sigmatrh$ fall below $\sigmagj$ at $r=80\rg$ is listed in Table~\ref{tab:sigma_GJ_fraction}. In the prograde model, $\sigmatrh < \sigmagj$ for only 16\% of the time, while $\sigmatrq < \sigmagj$ occurs for 43\% of the time. 
\lss{For the GJ magnetization appropriate for Sgr A*, $\sigmagj = 2\times 10^6-9\times 10^6$ (see Appendix~\ref{appendix:GJ_magnetization}), these fractions would be even smaller.}
Overall, we find that most of the jet electromagnetic power is carried by baryon-poor plasma with density below the GJ threshold.

\begin{figure}
	\centering
	\includegraphics[width=0.48\textwidth]{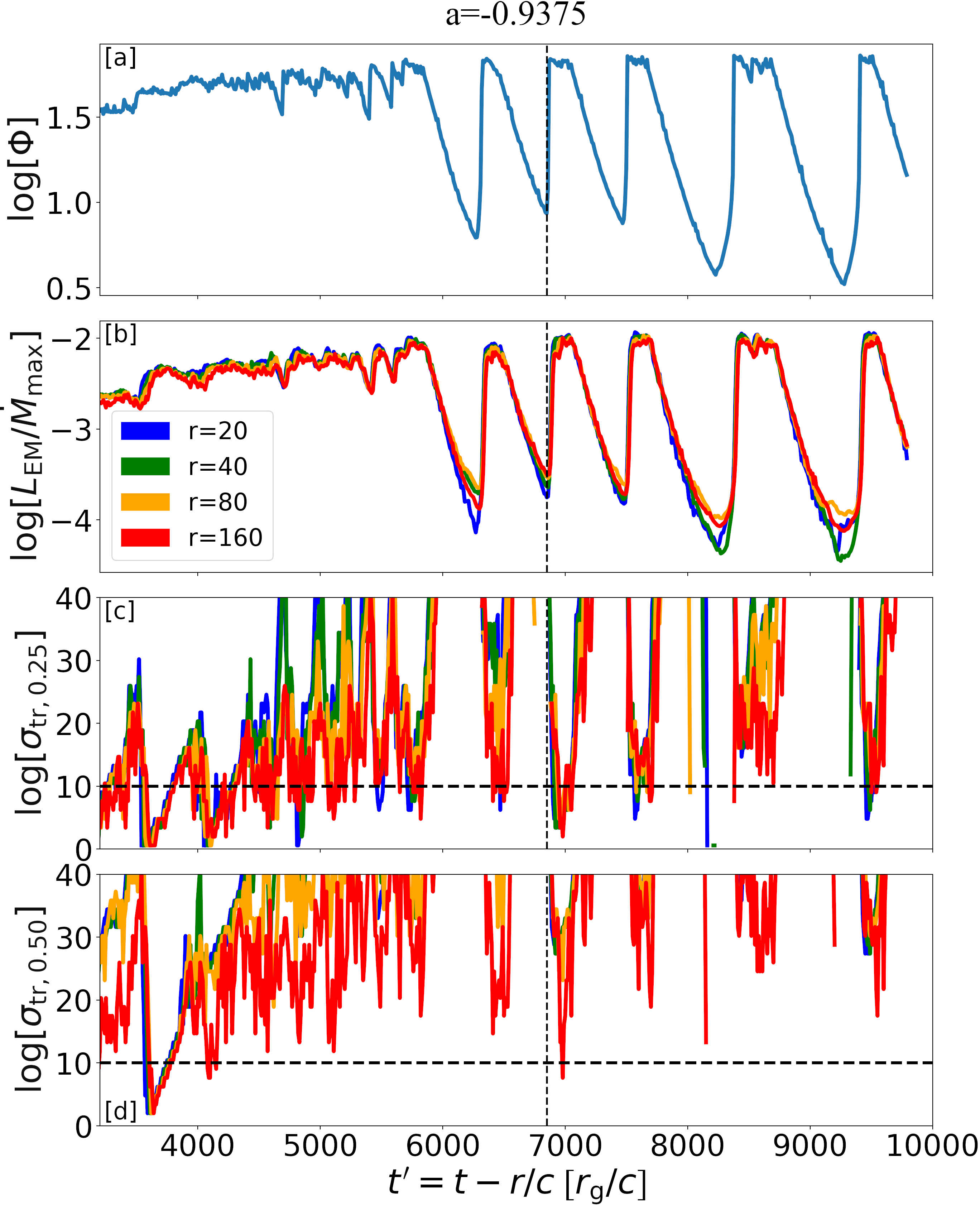}
    \caption{Same as Fig. \ref{fig:sigmatr_EMpower_25_50_spin+ve} but for spin parameter $a=-0.9375$. Here, $\dot{M}_{\rm max}=196.69$.}
    \label{fig:sigmatr_EMpower_25_50_spin-ve}
\end{figure}

Fig.~\ref{fig:sigmatr_EMpower_25_50_spin-ve} shows the corresponding results for the retrograde case, in the same format as Fig.~\ref{fig:sigmatr_EMpower_25_50_spin+ve}. In this simulation, sub-GJ excursions are even rarer: $\sigmatrq < \sigmagj=10^{10}$ (as for M87*) occurs only 15\% of the time, and $\sigmatrh < \sigmagj$ only 2\% of the time. This indicates that the retrograde jet remains predominantly charge-starved, with only brief and infrequent episodes of efficient screening.
%Fig.~\ref{fig:sigmatr_EMpower_25_50_spin-ve} presents the corresponding results for the retrograde case, in the same format as Fig.~\ref{fig:sigmatr_EMpower_25_50_spin+ve}. In the retrograde simulation, sub-GJ excursions are even rarer: only 15\% of the time satisfies $\sigmatrq<\sigmagj$, and merely 2\% satisfies $\sigmatrh<\sigmagj$. These fractions indicate that the retrograde jet remains predominantly charge-starved, with only brief and infrequent intervals when screening becomes significant. 
\begin{comment}
At relatively small radii, the difference between prograde and retrograde cases can be explained as follows: because the retrograde innermost stable circular orbit lies at a larger radius than the prograde case, the inner accretion flow enters the plunging region farther from the black hole, where the gas density is lower, leaving less baryonic material available to load the jet. 
\end{comment}
\ls{For the retrograde case in Fig.~\ref{fig:sigmatr_EMpower_25_50_spin-ve}, a vertical dashed black line at $t' = t - r/c = 6850\rgc$ marks the onset of the increase in magnetic flux (panel~[a]) and electromagnetic power (panel~[b]) following a flux eruption. The corresponding vertical lines in panels~[c] and~[d] show that the decrease in $\sigmatrq$ and $\sigmatrh$ occurs simultaneously with the rise in $\Phi$ and $\lem$. This temporal alignment is not unique to this event: similar correlations hold during other eruption events in the same simulation, as well as in the prograde case shown in Fig.~\ref{fig:sigmatr_EMpower_25_50_spin+ve}. Together, these correlations further support the conclusion that flux-eruption events play a key role in regulating the jet mass loading.}

%For the retrograde case in Fig.~\ref{fig:sigmatr_EMpower_25_50_spin-ve}, a vertical dashed black line at $t'=t-r/c=6850$ is overplotted in each panel to mark the onset of the increase in magnetic flux (panel~[a]) and electromagnetic power (panel~[b]) following a flux eruption. The corresponding lines in panels~[c] and~[d] show that the decrease in $\sigmatrq$ and $\sigmatrh$ coincides with the rise in $\Phi$ and $\lem$. This temporal alignment is not unique to this event: similar correlations are observed for other eruption events in the same simulation, as well as in the prograde case shown in Fig.~\ref{fig:sigmatr_EMpower_25_50_spin+ve}. Taken together, these correlations futher support the interpretation that flux-eruption events play a key role in regulating the mass loading of the jet.}


\begin{table}
\centering
\caption{Fraction of time (within $t\in[3000,10000]\rgc$) for which $\sigmatrq$ and 
$\sigmatrh$ at radius $r=80\rg$ fall below the GJ magnetization 
$\sigma_{\rm GJ}=10^{10}$ \lss{appropriate for M87*}.}
\begin{tabular}{lcc}
\hline
Quantity & Prograde (Fig. \ref{fig:sigmatr_EMpower_25_50_spin+ve}) & Retrograde (Fig. \ref{fig:sigmatr_EMpower_25_50_spin-ve}) \\
\hline
$\sigmatrq < \sigma_{\rm GJ}$ & 43\% & 15\% \\
$\sigmatrh < \sigma_{\rm GJ}$ & 16\% & 2\% \\
\hline
\end{tabular}
\label{tab:sigma_GJ_fraction}
\end{table}

\section{Summary \& discussion}\label{sec:discussions}
In this work, we introduce a passive Eulerian tracer that enables reconstruction of the physical baryon density in GRMHD simulations of magnetically arrested accretion flows. By decoupling the evolution of disk-supplied baryons from the mass injected \textit{ad hoc} in highly magnetized, low-density regions, this method reveals the true baryon content of black hole magnetospheres and jets. Applying this framework to axisymmetric simulations of prograde, Schwarzschild, and retrograde black holes, we identify the mechanisms that regulate when, where, and how baryons are entrained from the disk into the jet, providing a quantitative approach to measuring baryon loading in magnetically dominated regions of global GRMHD simulations. 

%In this work, we introduce a passive Eulerian tracer that enables the reconstruction of the physical baryon density in GRMHD simulations of magnetically arrested accretion flows. By decoupling the evolution of disk-supplied baryons from the mass injected \textit{ad hoc} in highly magnetized, low density regions, this method reveals the physical baryon content of the black hole magnetosphere and jet. Applying this framework to prograde, Schwarzschild, and retrograde black holes, we identify the mechanisms that regulate when, where, and how baryons are entrained from the disk into the jet funnel, thereby providing a strategy for quantifying the baryon loading of magnetically-dominated regions in global GRMHD simulations.

Our simulations show that magnetic-flux–eruption cycles are the primary regulators of baryon loading in the MAD regime. Each eruption expels disk material from the inner equatorial region, producing moderately magnetized, baryon-rich outflows that are advected along the jet boundary. For spinning black holes, shear-driven vortices \citep{Wong_2021} develop near the funnel wall, further entraining disk material and thickening the jet boundary layer; this shear-driven mixing is absent in the non-rotating case.

%Our simulations show that magnetic-flux–eruption cycles are the primary regulators of baryon loading in the MAD regime. Each eruption expels disk material from the inner equatorial region, creating moderately magnetized, baryon-rich outflows that are advected along the jet boundary. In cases with nonzero black hole spin, shear-driven vortices \citep{Wong_2021} develop near the funnel wall, further entraining disk material and thicken the jet boundary layer; such shear-driven mixing is absent for non-rotating black holes.

We also use the passive tracer to quantify the baryon content of the plasma feeding the equatorial current sheet that forms during flux eruptions. We find that the upstream plasma transitions from baryon-rich to baryon-poor over the course of each eruption. This evolution has implications for hadronic emission models powered by pair–proton reconnection, such as those proposed for the accretion flow in M87 \citep{Hakobyan_2025}. In this sense, our results provide the global GRMHD context that complements local studies of particle acceleration and nonthermal emission in pair–proton reconnection.

%We also use the passive tracer to quantify the baryon content of the plasma feeding the equatorial current sheet that forms during flux eruptions. We find that the upstream plasma transitions from baryon-rich to baryon-poor over the course of each eruption. This evolution has implications for hadronic emission models powered by pair–proton reconnection, such as those proposed for the accretion flow in M87 \citep{Hakobyan_2025}. In this sense, our results provide the global GRMHD context and temporal evolution for local studies of particle acceleration and nonthermal emission in pair-proton reconnection.
%, suggesting that large-scale magnetic-flux eruptions may control both jet mass loading and the conditions under which reconnection-powered high-energy radiation becomes observationally relevant.

%We also use our passive tracer to quantify the baryon content of the plasma feeding the equatorial current sheet that forms during flux eruptions. We find that the material upstream of the reconnection layer transitions from baryon-rich to baryon poor in the course of each flux eruption. This can inform hadronic emission models powered by pair-proton reconnection, as in the case of the accretion flow in M87 discussed by \citep{Hakobyan_2025}.

Using the tracer-based baryon density, we map the GJ screening boundary in space and time, identifying where the jet transitions between charge-starved, force-free, and inertia-dominated regimes, \lss{for conditions representative of M87*}. Large flux-eruption events temporarily evacuate the inner magnetosphere, extending the charge-starved region—normally confined to the jet funnel—across the full range of polar angles, before accretion gradually refills the cavity. By partitioning the radial electromagnetic jet power by tracer magnetization, we determine whether the Poynting flux is predominantly carried by baryon-rich or baryon-poor plasma. \accc{For both prograde and retrograde models, we find that most of the jet electromagnetic power is carried by baryon-poor plasma with density often below the GJ threshold appropriate for M87* \lss{(see  Appendix~\ref{appendix:GJ_magnetization} for the dependence of GJ magnetization on system parameters)}.}
%For both prograde and retrograde models, we find that most of the jet electromagnetic power is carried by baryon-poor plasma with density often below the GJ threshold.
%In prograde jets, extended intervals occur in which material with $\sigma_{\rm tr} < \sigma_{\rm GJ}$ carries a significant fraction of the Poynting flux; in reality, this would partially screen the electric fields and limit gap formation. In contrast, retrograde jets remain predominantly in the high-$\sigma_{\rm tr}$ regime, with only brief and infrequent sub-GJ episodes.

%With the tracer-derived baryon density, we map the GJ screening boundary throughout the domain, identifying where the jet transitions between charge-starved, force-free and inertia-dominated regimes. Large flux-eruption events temporarily evacuate the inner magnetosphere, extending the charge-starved region (which usually is confined to the jet funnel) to the whole range of polar angles, before accretion refills the cavity.

%By partitioning the radial EM power by tracer magnetization, we quantify whether most of the EM power is carried by baryon-rich or baryon-poor material. Prograde jets exhibit substantial intervals where $\sigmatr<\sigmagj$ material carries a non-negligible fraction of the Poynting flux, which in reality would result in partial screening of the electric fields and limiting gap formation. Retrograde jets maintain predominantly high-$\sigmatr$, with only brief and infrequent sub-GJ episodes.

Overall, these results show that baryon loading in MAD systems is inherently time-dependent, regulated by eruption-driven ejections from the inner magnetosphere and by turbulent mixing along the jet boundary. The passive tracer introduced here provides a transparent and robust diagnostic of baryon content, jet composition, and charge starvation. It also establishes a framework for future three-dimensional studies, potentially including radiative cooling. Ultimately, this approach will help constrain the jet composition and baryon loading, as well as the radiative signatures of the equatorial current sheet and the jet sheath.

%Overall, these results show that baryon loading in MAD systems is inherently time-dependent, regulated by eruption-driven ejections from the inner magnetosphere and by turbulent mixing along the jet boundary. The passive tracer introduced here provides a transparent and robust diagnostic of baryon content, jet composition, and charge starvation. It also establishes a framework for future three-dimensional studies, potentially including radiative cooling. Ultimately, this approach will help constrain the jet composition and baryon loading, as well as the radiative signatures of the equatorial current sheet and of the jet sheath.

%Overall, these results show that baryon loading in MAD systems is inherently time-dependent, governed by eruption-driven ejections from the innermost equatorial current sheet and turbulent mixing along the jet boundary. The passive tracer introduced in this work provides a transparent and robust means of diagnosing baryon content, jet composition, and charge starvation, and it offers a foundation for future 3D investigations, potentially including cooling losses. Ultimately, this strategy will help determine the jet composition and terminal Lorentz factor, as well as radiative signatures from the equatorial current sheet and the jet sheath.

\wong{A useful comparison for our results is provided by \citet{Wong_2021},} who studied baryon entrainment in 3D GRMHD simulations using passive Lagrangian tracer particles. They considered a retrograde disk and identified the disk–jet interface through a sign reversal of the azimuthal four-velocity. By tracking particles crossing this interface, they measured a time-averaged entrainment rate of $\sim 0.01\dot{M}$, where $\dot{M}$ is the accretion rate at the horizon, over the interval \wong{$t \in [6200,7000]\rgc$}. Our approach differs, in that we use a passive Eulerian tracer field to reconstruct the physical baryon density. Consistent with \citet{Wong_2021}, we find that baryon entrainment at large radii is dominated by shear-driven mixing associated with vortices developing along the jet boundary. However, one of our main results—that most of the jet electromagnetic power is carried by plasma with density below the GJ threshold \accc{appropriate for M87*}—appears at odds with the large entrainment rate reported by \citet{Wong_2021}.

%A useful point of comparison for our results is the study of baryon entrainment by \citet{Wong_2021, Wong_2025}, who investigated jet mass loading in 3D GRMHD simulations using passive Lagrangian tracer particles. They considered a retrograde disk and identified the disk–jet interface by the sign reversal of the azimuthal four-velocity.By tracking the particles crossing this interface, they reported a time-averaged entrainment rate of $\sim 0.01\dot{M}$, where $\dot{M}$ is the accretion rate measured at the horizon, over the time interval $t \in [1200,2000]\rgc$. We use an alternative strategy: rather than Lagrangian particles, our approach employs a passive Eulerian tracer field that reconstructs the physical baryon density. In agreement with \citet{Wong_2021, Wong_2025}, we find that baryon entrainment at large radii is dominated by shear-driven mixing associated with growing vortices along the jet boundary. On the other hand, one of our main results---that most of the jet electromagnetic power is carried by plasma whose density is below the GJ threshold---appears in contradiction with the large entrainment rate measured by \citet{Wong_2021}. 

\wong{This discrepancy may arise for several reasons. First, the entrainment rate reported by \citet{Wong_2021} is averaged over the interval $t \in [6200,7000]\rgc$. } 
\begin{comment}
In our simulations, this period corresponds to an early evolutionary stage, before the system reaches a quasi-steady state and prior to the onset of large-scale flux eruptions.
\end{comment}
\wong{The time-averaged entrainment rate may therefore not capture the two phases governed by eruptions versus accretion.} In addition, our results indicate that baryon loading in MAD systems is inherently time-dependent and regulated by magnetic-flux–eruption cycles; a time-averaged estimate may therefore miss the intrinsically episodic nature of the entrainment process. Consistent with this interpretation, Fig.~10 of \citet{Wong_2021} shows substantial temporal variability in the entrainment rate, which in our framework can be naturally interpreted as a consequence of episodic flux eruptions.

\wong{Second, a potential source of discrepancy between our results and those of \citet{Wong_2021} is the dimensionality of the simulations. While their study is performed in three dimensions, our analysis is restricted to axisymmetric (2D) GRMHD simulations. This difference may have important implications for the efficiency of baryon entrainment. In particular, fully three-dimensional flows can sustain non-axisymmetric instabilities, turbulence, and vortex stretching that are inherently suppressed in axisymmetry. These effects are expected to enhance mixing across the jet–disk interface and promote the transport of baryons into the funnel. Consequently, entrainment in 3D may be systematically stronger than in 2D, implying that our results could represent a lower bound on the true entrainment rate.}

\wong{Third, the entrainment rate measured by \citet{Wong_2021} does not distinguish how deeply tracer particles penetrate into the jet, i.e. whether they reach the region carrying most of the electromagnetic power.} Our results could therefore be reconciled with \citet{Wong_2021} if most of the baryons entering the jet surface (identified there by the sign reversal of azimuthal four-velocity) remain confined near the jet boundary, while the jet spine—carrying most of the electromagnetic flux—remains baryon-poor. We discuss this possibility further in Appendix~\ref{sec:wong}, where we systematically compare our simulations with those of \citet{Wong_2021}.

Our passive tracer identifies charge-starved regions within the global black-hole accretion–jet system, \lss{where the tracer magnetization exceeds the GJ magnetization (see Appendix~\ref{appendix:GJ_magnetization}). For parameters appropriate for M87*, charge-starved regions include} the polar funnel and the innermost equatorial reconnection layer during flux-eruption episodes. In such regions, insufficient charge supply leads to a breakdown of ideal GRMHD and the emergence of unscreened electric fields capable of accelerating particles to ultra-relativistic energies. Charge starvation at the jet base is a persistent feature at all times, including during phases of high accretion; the physics of spark gaps and subsequent pair creation in the polar region has been studied extensively \citep[e.g.,][]{Hirotani_2017, Levinson_2018, Crinquand_2020, Yuan_2025}. In contrast, charge starvation in the equatorial current sheet is intrinsically episodic and occurs primarily during the late stages of magnetic flux eruptions. During these episodes, pair creation can supply sufficient plasma to screen the magnetosphere and self-regulate the magnetization of the equatorial current sheet \citep{Hakobyan_2023}.

While the passive-tracer framework developed in this work provides a physically transparent assessment of baryon loading in MAD accretion flows, an important limitation is that our study is conducted in axisymmetry. This suppresses inherently three-dimensional instabilities and the associated turbulence that may play a central role in the physics of accretion and mass entrainment. In fully three-dimensional MAD flows, Rayleigh–Taylor-like modes and magnetic interchange instabilities at the disk–magnetosphere interface \citep{Tchekhovskoy_2011, McKinney_2012, Ripperda_2022} enable sustained, non-axisymmetric accretion streams even during eruptive phases, whereas in 2D models magnetic-flux eruptions can temporarily halt accretion. Similarly, mixing and baryon entrainment at the jet–disk interface may be enhanced by three-dimensional effects \citep{Wong_2021, Davelaar_2023}.

%While the passive-tracer framework developed in this work provides a physically transparent view of baryon loading in MAD accretion flows, an important limitation is that our study is conducted in axisymmetry. This suppresses the development of inherently three-dimensional instabilities and their resulting turbulence that may play a central role in accretion and mass entrainment. In fully 3D MAD flows, ayleigh–Taylor-like modes and magnetic interchange instabilities at the disk–magnetosphere interface \citep{Tchekhovskoy_2011, McKinney_2012, Ripperda_2022} enable sustained, non-axisymmetric accretion streams even during eruptive phases (in contrast to 2D models, where magnetic flux eruptions completely halt accretion). 
%As a result, large-scale magnetic flux eruptions become less frequent and typically clear only localized azimuthal sectors rather than the entire magnetosphere \citep{Ripperda_2022}. 
%Fully 3D waves—seen prominently in recent simulations —are therefore expected to modify both the spatial distribution and temporal variability of the baryonic component. These instabilities could enhance mixing, seed additional turbulent entrainment channels, and potentially increase the degree of baryon loading as compared to our 2D results.

Our results show that baryon entrainment and jet mass loading are regulated by the duty cycle and strength of magnetic-flux eruptions—driven by reconnection in the equatorial current sheet—so it is natural to ask how sensitive these processes are to departures from ideal MHD. In ideal GRMHD, reconnection is mediated by numerical resistivity, whereas in resistive GRMHD the reconnection rate and current-sheet structure are controlled by a prescribed conductivity \citep{Komissarov_2007, Ripperda_2020, Grehan_2025}. Studies with uniform resistivity indicate that, at sufficiently high resolution, resistive and ideal GRMHD produce similar reconnection behaviour and global dynamics \citep[e.g.][]{Ripperda_2020, Grehan_2025}. However, in the reconnection layer of MAD disks—and in many relativistic astrophysical plasmas—the Coulomb mean free path is so large that resistive MHD is formally inapplicable, requiring a kinetic description. In this collisionless regime, particle-in-cell simulations show that reconnection proceeds at rates typically an order of magnitude faster than the asymptotic MHD value \citep{sironi_25}. Recent work has proposed  a kinetically motivated subgrid prescription for non-uniform resistivity \citep{Moran_2025} that reproduces kinetic reconnection rates within MHD simulations. Such a prescription leads to faster reconnection and shorter flux-eruption cycles \citep{Bransgrove_2021, Galishnikova_2023, Ripperda_2026}. Systematic comparisons between ideal GRMHD and models incorporating kinetically motivated, non-uniform resistivity will therefore be essential to assess how robust the mass-loading pathways identified here remain once collisionless plasma effects—particularly in reconnection dynamics—are taken into account.

\section*{Acknowledgements}
This work was supported by a grant from the Simons Foundation (MP-SCMPS-00001470).
This research was supported in part by grant NSF PHY-2309135 to the Kavli Institute for Theoretical Physics (KITP).
L.S. acknowledges support from DoE Early Career Award DE-SC0023015, NASA ATP 80NSSC24K1238, NASA ATP 80NSSC24K1826,  NSF AST-2307202, and the Multimessenger Plasma Physics Center (MPPC), grant NSF PHY-2206609.
B.R. is supported by grant 23JWGO2A01 from the Canadian Space Agency, and by grants 613413 and RGPIN-2023-04844 from the  Natural Sciences and Engineering Research Council of Canada (NSERC).

The computational resources and services used in this work were partially provided by facilities supported by the VSC (Flemish Supercomputer Center), funded by the Research Foundation Flanders (FWO) and the Flemish Government – department EWI, by Compute Ontario and the Digital Research Alliance of Canada (alliancecan.ca) compute allocation rrg-ripperda, and the CCA at the Flatiron Institute supported by the Simons Foundation.

%%%%%%%%%%%%%%%%%%%%%%%%%%%%%%%%%%%%%%%%%%%%%%%%%%
\section*{Data Availability}
The data underlying this article were generated from numerical simulations performed with \bhac. Owing to the large size of the simulation outputs, these data are not publicly archived.

%\newpage


\bibliographystyle{mnras}
\bibliography{IonMixing.bib} % if your bibtex file is called example.bib

@ARTICLE{Wong_Ryan_2021,
       author = {{Wong}, George N. and {Ryan}, Benjamin R. and {Gammie}, Charles F.},
        title = "{Pair Drizzle around Sub-Eddington Supermassive Black Holes}",
      journal = {\apj},
         year = 2021,
        month = feb,
       volume = {907},
       number = {2},
          eid = {73},
        pages = {73},
          doi = {10.3847/1538-4357/abd0f9},
archivePrefix = {arXiv},
       eprint = {2012.04658},
 primaryClass = {astro-ph.HE},
       adsurl = {https://ui.adsabs.harvard.edu/abs/2021ApJ...907...73W}
}

@ARTICLE{sironi_25,
       author = {{Sironi}, Lorenzo and {Uzdensky}, Dmitri A. and {Giannios}, Dimitrios},
        title = "{Relativistic Magnetic Reconnection in Astrophysical Plasmas: A Powerful Mechanism of Nonthermal Emission}",
      journal = {\araa},
         year = 2025,
        month = aug,
       volume = {63},
       number = {1},
        pages = {127-178},
          doi = {10.1146/annurev-astro-020325-115713},
archivePrefix = {arXiv},
       eprint = {2506.02101},
 primaryClass = {astro-ph.HE},
       adsurl = {https://ui.adsabs.harvard.edu/abs/2025ARA&A..63..127S}
}

@ARTICLE{fiorillo_24,
       author = {{Fiorillo}, Damiano F.~G. and {Petropoulou}, Maria and {Comisso}, Luca and {Peretti}, Enrico and {Sironi}, Lorenzo},
        title = "{TeV Neutrinos and Hard X-Rays from Relativistic Reconnection in the Corona of NGC 1068}",
      journal = {\apjl},
         year = 2024,
        month = jan,
       volume = {961},
       number = {1},
          eid = {L14},
        pages = {L14},
          doi = {10.3847/2041-8213/ad192b},
archivePrefix = {arXiv},
       eprint = {2310.18254},
 primaryClass = {astro-ph.HE},
       adsurl = {https://ui.adsabs.harvard.edu/abs/2024ApJ...961L..14F}
}

@ARTICLE{karavola_25,
       author = {{Karavola}, D. and {Petropoulou}, M. and {Fiorillo}, D.~F.~G. and {Comisso}, L. and {Sironi}, L.},
        title = "{Neutrino and pair creation in reconnection-powered coronae of accreting black holes}",
      journal = {\jcap},
         year = 2025,
        month = apr,
       volume = {2025},
       number = {4},
          eid = {075},
        pages = {075},
          doi = {10.1088/1475-7516/2025/04/075},
archivePrefix = {arXiv},
       eprint = {2410.12638},
 primaryClass = {astro-ph.HE},
       adsurl = {https://ui.adsabs.harvard.edu/abs/2025JCAP...04..075K}
}

@ARTICLE{karavola_25b,
       author = {{Karavola}, D. and {Petropoulou}, M. and {Fiorillo}, D.~F.~G. and {Georgakakis}, A. and {Comisso}, L. and {Sironi}, L.},
        title = "{Diffuse neutrino flux from relativistic reconnection in AGN coronae}",
      journal = {arXiv e-prints},
         year = 2026,
        month = jan,
          eid = {arXiv:2601.01533},
        pages = {arXiv:2601.01533},
archivePrefix = {arXiv},
       eprint = {2601.01533},
 primaryClass = {astro-ph.HE},
       adsurl = {https://ui.adsabs.harvard.edu/abs/2026arXiv260101533K}
}

@article{Galishnikova_2023,
  title = {Collisionless Accretion onto Black Holes: Dynamics and Flares},
  author = {Galishnikova, Alisa and Philippov, Alexander and Quataert, Eliot and Bacchini, Fabio and Parfrey, Kyle and Ripperda, Bart},
  journal = {Phys. Rev. Lett.},
  volume = {130},
  issue = {11},
  pages = {115201},
  numpages = {6},
  year = {2023},
  month = {Mar},
  publisher = {American Physical Society},
  doi = {10.1103/PhysRevLett.130.115201},
  url = {https://link.aps.org/doi/10.1103/PhysRevLett.130.115201}
}

@article{Bransgrove_2021,
  title = {Magnetic Hair and Reconnection in Black Hole Magnetospheres},
  author = {Bransgrove, Ashley and Ripperda, Bart and Philippov, Alexander},
  journal = {Phys. Rev. Lett.},
  volume = {127},
  issue = {5},
  pages = {055101},
  numpages = {6},
  year = {2021},
  month = {Jul},
  publisher = {American Physical Society},
  doi = {10.1103/PhysRevLett.127.055101},
  url = {https://link.aps.org/doi/10.1103/PhysRevLett.127.055101}
}

@ARTICLE{Ripperda_2026,
       author = {{Ripperda}, B. and {Grehan}, M.~P. and {Moran}, A. and {Selvi}, S. and {Sironi}, L. and {Philippov}, A. and {Bransgrove}, A. and {Porth}, O.},
        title = "{Magnetic reconnection with a 0.1 rate: Effective resistivity in general relativistic magnetohydrodynamics}",
      journal = {arXiv e-prints},
         year = 2026,
        month = jan,
          eid = {arXiv:2601.02460},
        pages = {arXiv:2601.02460},
          doi = {10.48550/arXiv.2601.02460},
archivePrefix = {arXiv},
       eprint = {2601.02460},
 primaryClass = {astro-ph.HE},
       adsurl = {https://ui.adsabs.harvard.edu/abs/2026arXiv260102460R}
}

@article{Kotera_2011,
    author = "Kotera, Kumiko and Olinto, Angela V.",
    title = "{The Astrophysics of Ultrahigh Energy Cosmic Rays}",
    eprint = "1101.4256",
    archivePrefix = "arXiv",
    primaryClass = "astro-ph.HE",
    doi = "10.1146/annurev-astro-081710-102620",
    journal = "Ann. Rev. Astron. Astrophys.",
    volume = "49",
    pages = "119--153",
    year = "2011"
}

@article{IceCube_2022_Jul,
  title = {Search for neutrino emission from cores of active galactic nuclei},
  author = {Abbasi, R. and Ackermann, M. and Adams, J. and Aguilar, J. A. and Ahlers, M. and Ahrens, M. and Alameddine, J. M. and Alispach, C. and Alves, A. A. and Amin, N. M. and Andeen, K. and Anderson, T. and Anton, G. and Arg\"uelles, C. and Ashida, Y. and Axani, S. and Bai, X. and Balagopal V., A. and Barbano, A. and Barwick, S. W. and Bastian, B. and Basu, V. and Baur, S. and Bay, R. and Beatty, J. J. and Becker, K.-H. and Becker Tjus, J. and Bellenghi, C. and BenZvi, S. and Berley, D. and Bernardini, E. and Besson, D. Z. and Binder, G. and Bindig, D. and Blaufuss, E. and Blot, S. and Boddenberg, M. and Bontempo, F. and Borowka, J. and B\"oser, S. and Botner, O. and B\"ottcher, J. and Bourbeau, E. and Bradascio, F. and Braun, J. and Brinson, B. and Bron, S. and Brostean-Kaiser, J. and Browne, S. and Burgman, A. and Burley, R. T. and Busse, R. S. and Campana, M. A. and Carnie-Bronca, E. G. and Chen, C. and Chen, Z. and Chirkin, D. and Choi, K. and Clark, B. A. and Clark, K. and Classen, L. and Coleman, A. and Collin, G. H. and Conrad, J. M. and Coppin, P. and Correa, P. and Cowen, D. F. and Cross, R. and Dappen, C. and Dave, P. and De Clercq, C. and DeLaunay, J. J. and Delgado L\'opez, D. and Dembinski, H. and Deoskar, K. and Desai, A. and Desiati, P. and de Vries, K. D. and de Wasseige, G. and de With, M. and DeYoung, T. and Diaz, A. and D\'{\i}az-V\'elez, J. C. and Dittmer, M. and Dujmovic, H. and Dunkman, M. and DuVernois, M. A. and Dvorak, E. and Ehrhardt, T. and Eller, P. and Engel, R. and Erpenbeck, H. and Evans, J. and Evenson, P. A. and Fan, K. L. and Fazely, A. R. and Fedynitch, A. and Feigl, N. and Fiedlschuster, S. and Fienberg, A. T. and Filimonov, K. and Finley, C. and Fischer, L. and Fox, D. and Franckowiak, A. and Friedman, E. and Fritz, A. and F\"urst, P. and Gaisser, T. K. and Gallagher, J. and Ganster, E. and Garcia, A. and Garrappa, S. and Gerhardt, L. and Ghadimi, A. and Glaser, C. and Glauch, T. and Gl\"usenkamp, T. and Gonzalez, J. G. and Goswami, S. and Grant, D. and Gr\'egoire, T. and Griswold, S. and G\"unther, C. and Gutjahr, P. and Haack, C. and Hallgren, A. and Halliday, R. and Halve, L. and Halzen, F. and Ha Minh, M. and Hanson, K. and Hardin, J. and Harnisch, A. A. and Haungs, A. and Hebecker, D. and Helbing, K. and Henningsen, F. and Hettinger, E. C. and Hickford, S. and Hignight, J. and Hill, C. and Hill, G. C. and Hoffman, K. D. and Hoffmann, R. and Hokanson-Fasig, B. and Hoshina, K. and Huang, F. and Huber, M. and Huber, T. and Hultqvist, K. and H\"unnefeld, M. and Hussain, R. and Hymon, K. and In, S. and Iovine, N. and Ishihara, A. and Jansson, M. and Japaridze, G. S. and Jeong, M. and Jin, M. and Jones, B. J. P. and Kang, D. and Kang, W. and Kang, X. and Kappes, A. and Kappesser, D. and Kardum, L. and Karg, T. and Karl, M. and Karle, A. and Katz, U. and Kauer, M. and Kellermann, M. and Kelley, J. L. and Kheirandish, A. and Kin, K. and Kintscher, T. and Kiryluk, J. and Klein, S. R. and Koirala, R. and Kolanoski, H. and Kontrimas, T. and K\"opke, L. and Kopper, C. and Kopper, S. and Koskinen, D. J. and Koundal, P. and Kovacevich, M. and Kowalski, M. and Kozynets, T. and Kun, E. and Kurahashi, N. and Lad, N. and Lagunas Gualda, C. and Lanfranchi, J. L. and Larson, M. J. and Lauber, F. and Lazar, J. P. and Lee, J. W. and Leonard, K. and Leszczy\ifmmode \acute{n}\else \'{n}\fi{}ska, A. and Li, Y. and Lincetto, M. and Liu, Q. R. and Liubarska, M. and Lohfink, E. and Lozano Mariscal, C. J. and Lu, L. and Lucarelli, F. and Ludwig, A. and Luszczak, W. and Lyu, Y. and Ma, W. Y. and Madsen, J. and Mahn, K. B. M. and Makino, Y. and Mancina, S. and Mari\ifmmode \mbox{\c{s}}\else \c{s}\fi{}, I. C. and Martinez-Soler, I. and Maruyama, R. and Mase, K. and McElroy, T. and McNally, F. and Mead, J. V. and Meagher, K. and Mechbal, S. and Medina, A. and Meier, M. and Meighen-Berger, S. and Micallef, J. and Mockler, D. and Montaruli, T. and Moore, R. W. and Morse, R. and Moulai, M. and Naab, R. and Nagai, R. and Naumann, U. and Necker, J. and Nguyễn, L. V. and Niederhausen, H. and Nisa, M. U. and Nowicki, S. C. and Obertacke Pollmann, A. and Oehler, M. and Oeyen, B. and Olivas, A. and O'Sullivan, E. and Pandya, H. and Pankova, D. V. and Park, N. and Parker, G. K. and Paudel, E. N. and Paul, L. and P\'erez de los Heros, C. and Peters, L. and Peterson, J. and Philippen, S. and Pieper, S. and Pittermann, M. and Pizzuto, A. and Plum, M. and Popovych, Y. and Porcelli, A. and Prado Rodriguez, M. and Price, P. B. and Pries, B. and Przybylski, G. T. and Raab, C. and Raissi, A. and Rameez, M. and Rawlins, K. and Rea, I. C. and Rehman, A. and Reichherzer, P. and Reimann, R. and Renzi, G. and Resconi, E. and Reusch, S. and Rhode, W. and Richman, M. and Riedel, B. and Roberts, E. J. and Robertson, S. and Roellinghoff, G. and Rongen, M. and Rott, C. and Ruhe, T. and Ryckbosch, D. and Rysewyk Cantu, D. and Safa, I. and Saffer, J. and Sanchez Herrera, S. E. and Sandrock, A. and Sandroos, J. and Santander, M. and Sarkar, S. and Sarkar, S. and Satalecka, K. and Schaufel, M. and Schieler, H. and Schindler, S. and Schmidt, T. and Schneider, A. and Schneider, J. and Schr\"oder, F. G. and Schumacher, L. and Schwefer, G. and Sclafani, S. and Seckel, D. and Seunarine, S. and Sharma, A. and Shefali, S. and Silva, M. and Skrzypek, B. and Smithers, B. and Snihur, R. and Soedingrekso, J. and Soldin, D. and Spannfellner, C. and Spiczak, G. M. and Spiering, C. and Stachurska, J. and Stamatikos, M. and Stanev, T. and Stein, R. and Stettner, J. and Steuer, A. and Stezelberger, T. and St\"urwald, T. and Stuttard, T. and Sullivan, G. W. and Taboada, I. and Ter-Antonyan, S. and Tilav, S. and Tischbein, F. and Tollefson, K. and T\"onnis, C. and Toscano, S. and Tosi, D. and Trettin, A. and Tselengidou, M. and Tung, C. F. and Turcati, A. and Turcotte, R. and Turley, C. F. and Twagirayezu, J. P. and Ty, B. and Unland Elorrieta, M. A. and Valtonen-Mattila, N. and Vandenbroucke, J. and van Eijndhoven, N. and Vannerom, D. and van Santen, J. and Verpoest, S. and Walck, C. and Watson, T. B. and Weaver, C. and Weigel, P. and Weindl, A. and Weiss, M. J. and Weldert, J. and Wendt, C. and Werthebach, J. and Weyrauch, M. and Whitehorn, N. and Wiebusch, C. H. and Williams, D. R. and Wolf, M. and Woschnagg, K. and Wrede, G. and Wulff, J. and Xu, X. W. and Yanez, J. P. and Yoshida, S. and Yu, S. and Yuan, T. and Zhang, Z. and Zhelnin, P.},
  collaboration = {IceCube Collaboration},
  journal = {Phys. Rev. D},
  volume = {106},
  issue = {2},
  pages = {022005},
  numpages = {14},
  year = {2022},
  month = {Jul},
  publisher = {American Physical Society},
  doi = {10.1103/PhysRevD.106.022005},
  url = {https://link.aps.org/doi/10.1103/PhysRevD.106.022005}
}

@article{Rodrigues_2021,
  title = {Active Galactic Nuclei Jets as the Origin of Ultrahigh-Energy Cosmic Rays and Perspectives for the Detection of Astrophysical Source Neutrinos at EeV Energies},
  author = {Rodrigues, Xavier and Heinze, Jonas and Palladino, Andrea and van Vliet, Arjen and Winter, Walter},
  journal = {Phys. Rev. Lett.},
  volume = {126},
  issue = {19},
  pages = {191101},
  numpages = {6},
  year = {2021},
  month = {May},
  publisher = {American Physical Society},
  doi = {10.1103/PhysRevLett.126.191101},
  url = {https://link.aps.org/doi/10.1103/PhysRevLett.126.191101}
}

@article{Grehan_2025,
   title={Comparison of magnetic diffusion and reconnection in ideal and resistive relativistic magnetohydrodynamics, ideal magnetodynamics, and resistive force-free electrodynamics},
   volume={112},
   ISSN={2470-0029},
   url={http://dx.doi.org/10.1103/8xf2-x2nq},
   DOI={10.1103/8xf2-x2nq},
   number={6},
   journal={Physical Review D},
   publisher={American Physical Society (APS)},
   author={Grehan, Michael P. and Ghosal, Tanisha and Beattie, James R. and Ripperda, Bart and Porth, Oliver and Bacchini, Fabio},
   year={2025},
   month=sep }

@ARTICLE{Ripperda_2020,
       author = {{Ripperda}, Bart and {Bacchini}, Fabio and {Philippov}, Alexander A.},
        title = "{Magnetic Reconnection and Hot Spot Formation in Black Hole Accretion Disks}",
      journal = {\apj},
         year = 2020,
        month = sep,
       volume = {900},
       number = {2},
          eid = {100},
        pages = {100},
          doi = {10.3847/1538-4357/ababab},
archivePrefix = {arXiv},
       eprint = {2003.04330},
 primaryClass = {astro-ph.HE},
       adsurl = {https://ui.adsabs.harvard.edu/abs/2020ApJ...900..100R}
}

@ARTICLE{Hakobyan_2023,
       author = {{Hakobyan}, H. and {Ripperda}, B. and {Philippov}, A.~A.},
        title = "{Radiative Reconnection-powered TeV Flares from the Black Hole Magnetosphere in M87}",
      journal = {\apjl},
         year = 2023,
        month = feb,
       volume = {943},
       number = {2},
          eid = {L29},
        pages = {L29},
          doi = {10.3847/2041-8213/acb264},
archivePrefix = {arXiv},
       eprint = {2209.02105},
 primaryClass = {astro-ph.HE},
       adsurl = {https://ui.adsabs.harvard.edu/abs/2023ApJ...943L..29H}
}

@ARTICLE{Olivares_2019,
       author = {{Olivares}, Hector and {Porth}, Oliver and {Davelaar}, Jordy and {Most}, Elias R. and {Fromm}, Christian M. and {Mizuno}, Yosuke and {Younsi}, Ziri and {Rezzolla}, Luciano},
        title = "{Constrained transport and adaptive mesh refinement in the Black Hole Accretion Code}",
      journal = {\aap},
         year = 2019,
        month = sep,
       volume = {629},
          eid = {A61},
        pages = {A61},
          doi = {10.1051/0004-6361/201935559},
archivePrefix = {arXiv},
       eprint = {1906.10795},
 primaryClass = {astro-ph.HE},
       adsurl = {https://ui.adsabs.harvard.edu/abs/2019A&A...629A..61O}
}

@ARTICLE{Chael_2023,
       author = {{Chael}, Andrew and {Lupsasca}, Alexandru and {Wong}, George N. and {Quataert}, Eliot},
        title = "{Black Hole Polarimetry I. A Signature of Electromagnetic Energy Extraction}",
      journal = {\apj},
         year = 2023,
        month = nov,
       volume = {958},
       number = {1},
          eid = {65},
        pages = {65},
          doi = {10.3847/1538-4357/acf92d},
archivePrefix = {arXiv},
       eprint = {2307.06372},
 primaryClass = {astro-ph.HE},
       adsurl = {https://ui.adsabs.harvard.edu/abs/2023ApJ...958...65C}
}

@ARTICLE{Wong_Chael_2025,
       author = {{Wong}, George N. and {Chael}, Andrew and {Lupsasca}, Alexandru and {Quataert}, Eliot},
        title = "{Black Hole Polarimetry II: The Connection Between Spin and Polarization}",
      journal = {arXiv e-prints},
         year = 2025,
        month = sep,
          eid = {arXiv:2509.22639},
        pages = {arXiv:2509.22639},
          doi = {10.48550/arXiv.2509.22639},
archivePrefix = {arXiv},
       eprint = {2509.22639},
 primaryClass = {astro-ph.HE},
       adsurl = {https://ui.adsabs.harvard.edu/abs/2025arXiv250922639W}
}

@article{McKinney_2012,
  author       = {McKinney, Jonathan C. and Tchekhovskoy, Alexander and Blandford, Roger D.},
  title        = {General relativistic magnetohydrodynamic simulations of magnetically choked accretion flows around black holes},
  journal      = {Monthly Notices of the Royal Astronomical Society},
  volume       = {423},
  number       = {4},
  pages        = {3083--3117},
  year         = {2012},
  doi          = {10.1111/j.1365-2966.2012.21074.x}
}

@ARTICLE{Moran_2025,
       author = {{Moran}, Abigail and {Sironi}, Lorenzo and {Levis}, Aviad and {Ripperda}, Bart and {Most}, Elias R. and {Selvi}, Sebastiaan},
        title = "{Effective Resistivity in Relativistic Reconnection: A Prescription Based on Fully Kinetic Simulations}",
      journal = {\apjl},
         year = 2025,
        month = jan,
       volume = {978},
       number = {2},
          eid = {L45},
        pages = {L45},
          doi = {10.3847/2041-8213/ada158},
archivePrefix = {arXiv},
       eprint = {2501.04800},
 primaryClass = {astro-ph.HE},
       adsurl = {https://ui.adsabs.harvard.edu/abs/2025ApJ...978L..45M}
}

@ARTICLE{Davelaar_2023,
       author = {{Davelaar}, J. and {Ripperda}, B. and {Sironi}, L. and {Philippov}, A.~A. and {Olivares}, H. and {Porth}, O. and {Berg}, B. van den and {Bronzwaer}, T. and {Chatterjee}, K. and {Liska}, M.},
        title = "{Synchrotron Polarization Signatures of Surface Waves in Supermassive Black Hole Jets}",
      journal = {\apjl},
         year = 2023,
        month = dec,
       volume = {959},
       number = {1},
          eid = {L3},
        pages = {L3},
          doi = {10.3847/2041-8213/ad0b79},
archivePrefix = {arXiv},
       eprint = {2309.07963},
 primaryClass = {astro-ph.HE},
       adsurl = {https://ui.adsabs.harvard.edu/abs/2023ApJ...959L...3D}
}

@article{Kara_2025,
  author = {Kara, Erin and García, Javier},
  title = {Supermassive Black Holes in X-Rays: From Standard Accretion to Extreme Transients},
  journal = {Annual Review of Astronomy and Astrophysics},
  year = {2025},
  volume = {63},
  number = {1},
  pages = {379--430},
  doi = {10.1146/annurev-astro-071221-052844},
}

@article{IceCube_2022,
  author       = {{IceCube Collaboration} and Abbasi, R. and et al.},
  title        = {Evidence for neutrino emission from the nearby active galaxy NGC 1068},
  journal      = {Science},
  year         = {2022},
  volume       = {378},
  number       = {6619},
  pages        = {538--543},
  doi          = {10.1126/science.abg3395},
  url          = {https://ui.adsabs.harvard.edu/abs/2022Sci...378..538I/abstract}
}

@ARTICLE{Fischer_2023,
       author = {{Fischer}, Travis C. and {Johnson}, Megan C. and {Secrest}, Nathan J. and {Crenshaw}, D. Michael and {Kraemer}, Steven B.},
        title = "{No Small-scale Radio Jets Here: Multiepoch Observations of Radio Continuum Structures in NGC 1068 with the VLBA}",
      journal = {\apj},
         year = 2023,
        month = aug,
       volume = {953},
       number = {1},
          eid = {87},
        pages = {87},
          doi = {10.3847/1538-4357/ace1f0},
archivePrefix = {arXiv},
       eprint = {2306.15047},
 primaryClass = {astro-ph.GA},
       adsurl = {https://ui.adsabs.harvard.edu/abs/2023ApJ...953...87F}
}

@article{Blandford_2019ARA&A,
  author       = {Blandford, Roger and Meier, David and Readhead, Anthony},
  title        = {Relativistic Jets from Active Galactic Nuclei},
  journal      = {Annual Review of Astronomy and Astrophysics},
  volume       = {57},
  pages        = {467--509},
  year         = {2019},
  doi          = {10.1146/annurev-astro-081817-051948},
  url          = {https://ui.adsabs.harvard.edu/abs/2019ARA&A..57..467B}
}

@article{Komissarov_2007,
  author  = {Komissarov, S. S. and Barkov, M. V. and Vlahakis, N. and K\"onigl, A.},
  title   = {Magnetic acceleration of relativistic jets},
  journal = {Monthly Notices of the Royal Astronomical Society},
  volume  = {380},
  number  = {1},
  pages   = {51--70},
  year    = {2007},
  doi     = {10.1111/j.1365-2966.2007.12050.x}
}

@article{Yuan_2025,
       author = {{Yuan}, Yajie and {Chen}, Alexander Y. and {Luepker}, Martin},
        title = "{Physics of Pair-producing Gaps in Black Hole Magnetospheres: Two-dimensional General Relativistic Particle-in-cell Simulations}",
      journal = {\apj},
         year = 2025,
        month = jun,
       volume = {985},
       number = {2},
          eid = {159},
        pages = {159},
          doi = {10.3847/1538-4357/adce79},
archivePrefix = {arXiv},
       eprint = {2503.08487},
 primaryClass = {astro-ph.HE},
       adsurl = {https://ui.adsabs.harvard.edu/abs/2025ApJ...985..159Y}
}

@article{Crinquand_2020,
  title        = {Multidimensional Simulations of Ergospheric Pair Discharges around Black Holes},
  author       = {Crinquand, B. and Cerutti, B. and Parfrey, K. and Philippov, A. and Dubus, G.},
  journal      = {Physical Review Letters},
  volume       = {124},
  number       = {14},
  pages        = {145101},
  year         = {2020},
  doi          = {10.1103/PhysRevLett.124.145101}
}

@ARTICLE{Chow_2023,
       author = {{Chow}, Anthony and {Davelaar}, Jordy and {Rowan}, Michael E. and {Sironi}, Lorenzo},
        title = "{The Kelvin-Helmholtz Instability at the Boundary of Relativistic Magnetized Jets}",
      journal = {\apjl},
         year = 2023,
        month = jul,
       volume = {951},
       number = {2},
          eid = {L23},
        pages = {L23},
          doi = {10.3847/2041-8213/acdfcf},
archivePrefix = {arXiv},
       eprint = {2209.13699},
 primaryClass = {astro-ph.HE},
       adsurl = {https://ui.adsabs.harvard.edu/abs/2023ApJ...951L..23C}
}

@article{Narayan_2003,
  author       = {Narayan, Ramesh and Igumenshchev, Igor V. and Abramowicz, Marek A.},
  title        = {Magnetically Arrested Disk: an Energetically Efficient Accretion Flow},
  journal      = {Publications of the Astronomical Society of Japan},
  year         = {2003},
  volume       = {55},
  number       = {6},
  pages        = {L69--L72},
  doi          = {10.1093/pasj/55.6.L69},
  url          = {https://doi.org/10.1093/pasj/55.6.L69}
}

@article{Padovani_2018,
  author       = {Padovani, P. and Giommi, P. and Resconi, E. and Glauch, T. and Arsioli, B. and Sahakyan, N. and Huber, M.},
  title        = {A Multi-Messenger View of Blazars: Implications for IceCube-170922A},
  journal      = {Monthly Notices of the Royal Astronomical Society},
  year         = {2018},
  volume       = {480},
  number       = {1},
  pages        = {192--203},
  doi          = {10.1093/mnras/sty1852},
  url          = {https://doi.org/10.1093/mnras/sty1852}
}

@article{Zhang_Sironi_Giannios_2021,
       author = {{Zhang}, Hao and {Sironi}, Lorenzo and {Giannios}, Dimitrios},
        title = "{Fast Particle Acceleration in Three-dimensional Relativistic Reconnection}",
      journal = {\apj},
         year = 2021,
        month = dec,
       volume = {922},
       number = {2},
          eid = {261},
        pages = {261},
          doi = {10.3847/1538-4357/ac2e08},
archivePrefix = {arXiv},
       eprint = {2105.00009},
 primaryClass = {astro-ph.HE},
       adsurl = {https://ui.adsabs.harvard.edu/abs/2021ApJ...922..261Z}
}

@article{Fabian_2012,
  author       = {Fabian, A. C.},
  title        = {Observational Evidence of Active Galactic Nuclei Feedback},
  journal      = {Annual Review of Astronomy and Astrophysics},
  year         = {2012},
  volume       = {50},
  pages        = {455--489},
  doi          = {10.1146/annurev-astro-081811-125521}
}

@ARTICLE{Hakobyan_2025,
       author = {{Hakobyan}, Hayk and {Levinson}, Amir and {Sironi}, Lorenzo and {Philippov}, Alexander and {Ripperda}, Bart},
        title = "{Reconnection-driven Flares in M87*: Proton─Synchrotron-powered GeV Emission}",
      journal = {\apjl},
         year = 2025,
        month = dec,
       volume = {995},
       number = {2},
          eid = {L73},
        pages = {L73},
          doi = {10.3847/2041-8213/ae286a},
archivePrefix = {arXiv},
       eprint = {2507.14002},
 primaryClass = {astro-ph.HE},
       adsurl = {https://ui.adsabs.harvard.edu/abs/2025ApJ...995L..73H}
}

@article{Stathopoulos_2024,
  author       = {S. I. Stathopoulos and M. Petropoulou and L. Sironi and D. Giannios},
  title        = {The role of magnetospheric current sheets in pair enrichment and ultra-high energy proton acceleration in M87*},
  journal      = {Journal of Cosmology and Astroparticle Physics},
  year         = {2024},
  number       = {12},
  pages        = {009},
  doi          = {10.1088/1475-7516/2024/12/009},
}

@article{Sridhar_2025,
  author       = {Sridhar, Navin and Ripperda, Bart and Sironi, Lorenzo and Davelaar, Jordy and Beloborodov, Andrei M.},
  title        = {Bulk Motions in the Black Hole Jet Sheath as a Candidate for the Comptonizing Corona},
  journal      = {The Astrophysical Journal},
  year         = {2025},
  volume       = {979},
  number       = {2},
  pages        = {199},
  doi          = {10.3847/1538-4357/ada385},
}

@article{Salas_2025,
  title        = {Two‐temperature treatments in magnetically arrested disk GRMHD simulations more accurately predict light curves of Sagittarius A*},
  author       = {Salas, L. D. S. and Liska, Matthew and Markoff, Sera and Chatterjee, Koushik and Musoke, Gibwa and Porth, Oliver and Ripperda, Bart and Yoon, Doosoo and Mulaudzi, Wanga},
  journal      = {Monthly Notices of the Royal Astronomical Society},
  volume       = {538},
  number       = {2},
  pages        = {698--710},
  year         = {2025},
  doi          = {10.1093/mnras/staf240},
  eprint       = {2411.09556},
  archivePrefix= {arXiv},
  primaryClass = {astro-ph.HE}
}

@article{Ripperda_2022,
  title   = {Black Hole Flares: Ejection of Accreted Magnetic Flux through 3D Plasmoid-mediated Reconnection},
  author  = {Ripperda, Bart and Liska, Matthew and Chatterjee, Koushik and Musoke, Gibwa and Philippov, Alexander A. and Markoff, Sera B. and Tchekhovskoy, Alexander and Younsi, Ziri},
  journal = {The Astrophysical Journal Letters},
  volume  = {924},
  pages   = {L32},
  year    = {2022},
  doi     = {10.3847/2041-8213/ac46a1}
}

@article{Porth_2014,
  author       = {Porth, O. and Xia, C. and Hendrix, T. and Moschou, S. P. and Keppens, R.},
  title        = {{MPI-AMRVAC for Solar and Astrophysics}},
  journal      = {The Astrophysical Journal Supplement Series},
  year         = {2014},
  volume       = {214},
  number       = {1},
  pages        = {4},
  doi          = {10.1088/0067-0049/214/1/4},
  eprint       = {arXiv:1407.2052 [astro-ph.IM]},
  note         = {Open-source adaptive mesh refinement framework}
}

@article{Gammie_2003,
  author       = {Gammie, Charles F. and McKinney, Jonathan C. and Tóth, Gábor},
  title        = {HARM: A Numerical Scheme for General Relativistic Magnetohydrodynamics},
  journal      = {The Astrophysical Journal},
  year         = {2003},
  volume       = {589},
  pages        = {444--457},
  doi          = {10.1086/374594},
  url          = {https://doi.org/10.1086/374594}
}

@article{McKinney_2006,
  author       = {McKinney, Jonathan C.},
  title        = {General relativistic magnetohydrodynamic simulations of the jet formation and large-scale propagation from black hole accretion systems},
  journal      = {Monthly Notices of the Royal Astronomical Society},
  year         = {2006},
  volume       = {368},
  number       = {4},
  pages        = {1561--1582},
  doi          = {10.1111/j.1365-2966.2006.10256.x},
  url          = {https://doi.org/10.1111/j.1365-2966.2006.10256.x}
}

@ARTICLE{Kantzas_2023,
       author = {{Kantzas}, D. and {Markoff}, S. and {Lucchini}, M. and {Ceccobello}, C. and {Chatterjee}, K.},
        title = "{Exploring the role of composition and mass loading on the properties of hadronic jets}",
      journal = {\mnras},
         year = 2023,
        month = apr,
       volume = {520},
       number = {4},
        pages = {6017-6039},
          doi = {10.1093/mnras/stad521},
archivePrefix = {arXiv},
       eprint = {2301.06382},
 primaryClass = {astro-ph.HE},
       adsurl = {https://ui.adsabs.harvard.edu/abs/2023MNRAS.520.6017K}
}

@article{Cerruti_2020,
  author = {Cerruti, M.},
  year = {2020},
  title = {Leptonic and Hadronic Radiative Processes in Supermassive Black Hole Jets},
  journal = {Galaxies},
  volume = {8},
  number = {4},
  pages = {72},
  doi = {10.3390/galaxies8040072},
  url = {https://www.mdpi.com/2075-4434/8/4/72}
}

@article{Sikora_2000,
doi = {10.1086/308756},
url = {https://doi.org/10.1086/308756},
year = {2000},
month = {may},
publisher = {},
volume = {534},
number = {1},
pages = {109},
author = {Sikora, M. and Madejski, G.},
title = {On Pair Content and Variability of Subparsec Jets in
Quasars},
journal = {The Astrophysical Journal}
}

@ARTICLE{Hirotani_Okamoto_1998,
       author = {{Hirotani}, Kouichi and {Okamoto}, Isao},
        title = "{Pair Plasma Production in a Force-free Magnetosphere around a Supermassive Black Hole}",
      journal = {\apj},
         year = 1998,
        month = apr,
       volume = {497},
       number = {2},
        pages = {563-572},
          doi = {10.1086/305479},
       adsurl = {https://ui.adsabs.harvard.edu/abs/1998ApJ...497..563H}
}

@ARTICLE{Hirotani_2017,
       author = {{Hirotani}, Kouichi and {Pu}, Hung-Yi and {Lin}, Lupin Chun-Che and {Kong}, Albert K.~H. and {Matsushita}, Satoki and {Asada}, Keiichi and {Chang}, Hsiang-Kuang and {Tam}, Pak-Hin T.},
        title = "{Lepton Acceleration in the Vicinity of the Event Horizon: Very High Energy Emissions from Supermassive Black Holes}",
      journal = {\apj},
         year = 2017,
        month = aug,
       volume = {845},
       number = {1},
          eid = {77},
        pages = {77},
          doi = {10.3847/1538-4357/aa7895},
archivePrefix = {arXiv},
       eprint = {1706.03766},
 primaryClass = {astro-ph.HE},
       adsurl = {https://ui.adsabs.harvard.edu/abs/2017ApJ...845...77H}
}

@ARTICLE{Levinson_2000,
       author = {{Levinson}, Amir},
        title = "{Particle Acceleration and Curvature TeV Emission by Rotating, Supermassive Black Holes}",
      journal = {\prl},
         year = 2000,
        month = jul,
       volume = {85},
       number = {5},
        pages = {912-915},
          doi = {10.1103/PhysRevLett.85.912},
       adsurl = {https://ui.adsabs.harvard.edu/abs/2000PhRvL..85..912L}
}

@ARTICLE{Toma_Takahara_2012,
       author = {{Toma}, K. and {Takahara}, F.},
        title = "{Baryon Loading of Active Galactic Nucleus Jets Mediated by Neutrons}",
      journal = {\apj},
         year = 2012,
        month = aug,
       volume = {754},
       number = {2},
          eid = {148},
        pages = {148},
          doi = {10.1088/0004-637X/754/2/148},
archivePrefix = {arXiv},
       eprint = {1205.6868},
 primaryClass = {astro-ph.HE},
       adsurl = {https://ui.adsabs.harvard.edu/abs/2012ApJ...754..148T}
}

@article{Goldreich_Julian_1969,
  author       = {Goldreich, P. and Julian, W. H.},
  title        = {Pulsar Electrodynamics},
  journal      = {The Astrophysical Journal},
  year         = {1969},
  volume       = {157},
  pages        = {869--880},
  doi          = {10.1086/150119},
  url          = {https://doi.org/10.1086/150119}
}

@ARTICLE{Porth_Olivares_Mizuno_2017,
       author = {{Porth}, Oliver and {Olivares}, Hector and {Mizuno}, Yosuke and {Younsi}, Ziri and {Rezzolla}, Luciano and {Moscibrodzka}, Monika and {Falcke}, Heino and {Kramer}, Michael},
        title = "{The black hole accretion code}",
      journal = {Computational Astrophysics and Cosmology},
         year = 2017,
        month = may,
       volume = {4},
       number = {1},
          eid = {1},
        pages = {1},
          doi = {10.1186/s40668-017-0020-2},
archivePrefix = {arXiv},
       eprint = {1611.09720},
 primaryClass = {gr-qc},
       adsurl = {https://ui.adsabs.harvard.edu/abs/2017ComAC...4....1P}
}

@article{McKinney_2004,
   title={A Measurement of the Electromagnetic Luminosity of a Kerr Black Hole},
   volume={611},
   ISSN={1538-4357},
   url={http://dx.doi.org/10.1086/422244},
   DOI={10.1086/422244},
   number={2},
   journal={The Astrophysical Journal},
   publisher={American Astronomical Society},
   author={McKinney, Jonathan C. and Gammie, Charles F.},
   year={2004},
   month=aug, pages={977–995} }

@ARTICLE{Fishbone_1976,
       author = {{Fishbone}, L.~G. and {Moncrief}, V.},
        title = "{Relativistic fluid disks in orbit around Kerr black holes.}",
      journal = {\apj},
         year = 1976,
        month = aug,
       volume = {207},
        pages = {962-976},
          doi = {10.1086/154565},
       adsurl = {https://ui.adsabs.harvard.edu/abs/1976ApJ...207..962F}
}

@ARTICLE{velikhov_1959,
       author = {{Velikhov}, Evgeny Pavlovich},
        title = "{Stability of an Ideally Conducting Liquid Flowing between Cylinders Rotating in a Magnetic Field}",
      journal = {Soviet Journal of Experimental and Theoretical Physics},
         year = 1959,
        month = nov,
       volume = {9},
       number = {5},
        pages = {995-998},
       adsurl = {https://ui.adsabs.harvard.edu/abs/1959JETP....9..995V}
}

@article{chandrasekhar_1960,
  author    = {S. Chandrasekhar},
  title     = {The Stability of Non-Dissipative Couette Flow in Hydromagnetics},
  journal   = {Proceedings of the National Academy of Sciences of the United States of America},
  year      = {1960},
  volume    = {46},
  number    = {2},
  pages     = {253--257},
  doi       = {10.1073/pnas.46.2.253},
  url       = {https://doi.org/10.1073/pnas.46.2.253}
}

@ARTICLE{Balbus_Hawley_1991,
       author = {{Balbus}, Steven A. and {Hawley}, John F.},
        title = "{A Powerful Local Shear Instability in Weakly Magnetized Disks. I. Linear Analysis}",
      journal = {\apj},
         year = 1991,
        month = jul,
       volume = {376},
        pages = {214},
          doi = {10.1086/170270},
       adsurl = {https://ui.adsabs.harvard.edu/abs/1991ApJ...376..214B}
}

@article{Tchekhovskoy_2011,
   title={Efficient generation of jets from magnetically arrested accretion on a rapidly spinning black hole},
   volume={418},
   ISSN={1745-3933},
   url={http://dx.doi.org/10.1111/j.1745-3933.2011.01147.x},
   DOI={10.1111/j.1745-3933.2011.01147.x},
   number={1},
   journal={Monthly Notices of the Royal Astronomical Society: Letters},
   publisher={Oxford University Press (OUP)},
   author={Tchekhovskoy, Alexander and Narayan, Ramesh and McKinney, Jonathan C.},
   year={2011},
   month=nov, pages={L79–L83} }

@article{IceCube_2018,
  title        = {Multimessenger observations of a flaring blazar coincident with high-energy neutrino IceCube-170922A},
  author       = {{IceCube Collaboration} and {Fermi-LAT collaboration} and {MAGIC collaboration} and AGILE and ASAS-SN and HAWC and H.E.S.S. and INTEGRAL and {Kanata, Kiso, and Subaru observing teams} and Kapteyn and {Liverpool telescope} and Swift/NuSTAR and VERITAS and {VLA/17B-403 team}},
  journal      = {Science},
  volume       = {361},
  number       = {6398},
  pages        = {eaat1378},
  year         = {2018},
  month        = jul,
  doi          = {10.1126/science.aat1378},
  note         = {Article eaat1378}
}

@ARTICLE{Wong_2021,
       author = {{Wong}, George N. and {Du}, Yufeng and {Prather}, Ben S. and {Gammie}, Charles F.},
        title = "{The Jet-disk Boundary Layer in Black Hole Accretion}",
      journal = {\apj},
         year = 2021,
        month = jun,
       volume = {914},
       number = {1},
          eid = {55},
        pages = {55},
          doi = {10.3847/1538-4357/abf8b8},
archivePrefix = {arXiv},
       eprint = {2104.07035},
 primaryClass = {astro-ph.HE},
       adsurl = {https://ui.adsabs.harvard.edu/abs/2021ApJ...914...55W}
}

@article{Wong_2025,
       author = {{Wong}, George N. and {Medeiros}, Lia and {Stone}, James M.},
        title = "{Mass Transport, Turbulent Mixing, and Inflow in Black Hole Accretion}",
      journal = {\apj},
         year = 2025,
        month = dec,
       volume = {995},
       number = {1},
          eid = {119},
        pages = {119},
          doi = {10.3847/1538-4357/ae14fd},
archivePrefix = {arXiv},
       eprint = {2509.14202},
 primaryClass = {astro-ph.HE},
       adsurl = {https://ui.adsabs.harvard.edu/abs/2025ApJ...995..119W}
}

@ARTICLE{Levinson_2018,
       author = {{Levinson}, Amir and {Cerutti}, Beno{\^\i}t},
        title = "{Particle-in-cell simulations of pair discharges in a starved magnetosphere of a Kerr black hole}",
      journal = {\aap},
         year = 2018,
        month = aug,
       volume = {616},
          eid = {A184},
        pages = {A184},
          doi = {10.1051/0004-6361/201832915},
archivePrefix = {arXiv},
       eprint = {1803.04427},
 primaryClass = {astro-ph.HE},
       adsurl = {https://ui.adsabs.harvard.edu/abs/2018A&A...616A.184L}
}

@ARTICLE{Janssen_Falcke_2021,
       author = {{Janssen}, Michael and {Falcke}, Heino and {Kadler}, Matthias and {Ros}, Eduardo and {Wielgus}, Maciek and {Akiyama}, Kazunori and {Balokovi{\'c}}, Mislav and {Blackburn}, Lindy and {Bouman}, Katherine L. and {Chael}, Andrew and {Chan}, Chi-kwan and {Chatterjee}, Koushik and {Davelaar}, Jordy and {Edwards}, Philip G. and {Fromm}, Christian M. and {G{\'o}mez}, Jos{\'e} L. and {Goddi}, Ciriaco and {Issaoun}, Sara and {Johnson}, Michael D. and {Kim}, Junhan and {Koay}, Jun Yi and {Krichbaum}, Thomas P. and {Liu}, Jun and {Liuzzo}, Elisabetta and {Markoff}, Sera and {Markowitz}, Alex and {Marrone}, Daniel P. and {Mizuno}, Yosuke and {M{\"u}ller}, Cornelia and {Ni}, Chunchong and {Pesce}, Dominic W. and {Ramakrishnan}, Venkatessh and {Roelofs}, Freek and {Rygl}, Kazi L.~J. and {van Bemmel}, Ilse and {Event Horizon Telescope Collaboration} and {Alberdi}, Antxon and {Alef}, Walter and {Algaba}, Juan Carlos and {Anantua}, Richard and {Asada}, Keiichi and {Azulay}, Rebecca and {Baczko}, Anne-Kathrin and {Ball}, David and {Ball}, David and {Barrett}, John and {Benson}, Bradford A. and {Bintley}, Dan and {Bintley}, Dan and {Blundell}, Raymond and {Boland}, Wilfred and {Boland}, Wilfred and {Bower}, Geoffrey C. and {Boyce}, Hope and {Bremer}, Michael and {Brinkerink}, Christiaan D. and {Brissenden}, Roger and {Britzen}, Silke and {Broderick}, Avery E. and {Broguiere}, Dominique and {Bronzwaer}, Thomas and {Byun}, Do-Young and {Carlstrom}, John E. and {Chatterjee}, Shami and {Chen}, Ming-Tang and {Chen}, Yongjun and {Chesler}, Paul M. and {Cho}, Ilje and {Christian}, Pierre and {Conway}, John E. and {Cordes}, James M. and {Crawford}, Thomas M. and {Crew}, Geoffrey B. and {Cruz-Osorio}, Alejandro and {Cui}, Yuzhu and {Cui}, Yuzhu and {De Laurentis}, Mariafelicia and {Deane}, Roger and {Dempsey}, Jessica and {Desvignes}, Gregory and {Dexter}, Jason and {Doeleman}, Sheperd S. and {Eatough}, Ralph P. and {Farah}, Joseph and {Farah}, Joseph and {Fish}, Vincent L. and {Fomalont}, Ed and {Ford}, H. Alyson and {Fraga-Encinas}, Raquel and {Friberg}, Per and {Friberg}, Per and {Fuentes}, Antonio and {Galison}, Peter and {Gammie}, Charles F. and {Garc{\'\i}a}, Roberto and {Gelles}, Zachary and {Gentaz}, Olivier and {Georgiev}, Boris and {Georgiev}, Boris and {Gold}, Roman and {Gold}, Roman and {G{\'o}mez-Ruiz}, Arturo I. and {Gu}, Minfeng and {Gurwell}, Mark and {Hada}, Kazuhiro and {Haggard}, Daryl and {Hecht}, Michael H. and {Hesper}, Ronald and {Himwich}, Elizabeth and {Ho}, Luis C. and {Ho}, Paul and {Honma}, Mareki and {Huang}, Chih-Wei L. and {Huang}, Lei and {Hughes}, David H. and {Ikeda}, Shiro and {Inoue}, Makoto and {Inoue}, Makoto and {James}, David J. and {Jannuzi}, Buell T. and {Jeter}, Britton and {Jiang}, Wu and {Jimenez-Rosales}, Alejandra and {Jorstad}, Svetlana and {Jung}, Taehyun and {Karami}, Mansour and {Karuppusamy}, Ramesh and {Kawashima}, Tomohisa and {Keating}, Garrett K. and {Kettenis}, Mark and {Kim}, Dong-Jin and {Kim}, Jae-Young and {Kim}, Jongsoo and {Kino}, Motoki and {Kofuji}, Yutaro and {Koyama}, Shoko and {Kramer}, Michael and {Kramer}, Carsten and {Kuo}, Cheng-Yu and {Lauer}, Tod R. and {Lee}, Sang-Sung and {Levis}, Aviad and {Li}, Yan-Rong and {Li}, Zhiyuan and {Lindqvist}, Michael and {Lico}, Rocco and {Lindahl}, Greg and {Liu}, Kuo and {Lo}, Wen-Ping and {Lobanov}, Andrei P. and {Loinard}, Laurent and {Lonsdale}, Colin and {Lu}, Ru-Sen and {MacDonald}, Nicholas R. and {Mao}, Jirong and {Marchili}, Nicola and {Marscher}, Alan P. and {Mart{\'\i}-Vidal}, Iv{\'a}n and {Matsushita}, Satoki and {Matthews}, Lynn D. and {Medeiros}, Lia and {Menten}, Karl M. and {Mizuno}, Izumi and {Moran}, James M. and {Moriyama}, Kotaro and {Moscibrodzka}, Monika and {Moscibrodzka}, Monika and {Musoke}, Gibwa and {Mej{\'\i}as}, Alejandro Mus and {Nagai}, Hiroshi and {Nagar}, Neil M. and {Nakamura}, Masanori and {Narayan}, Ramesh and {Narayanan}, Gopal and {Natarajan}, Iniyan and {Nathanail}, Antonios and {Neilsen}, Joey and {Neri}, Roberto and {Noutsos}, Aristeidis and {Nowak}, Michael A. and {Okino}, Hiroki and {Olivares}, H{\'e}ctor and {Ortiz-Le{\'o}n}, Gisela N. and {Oyama}, Tomoaki and {{\"O}zel}, Feryal and {Palumbo}, Daniel C.~M. and {Park}, Jongho and {Patel}, Nimesh and {Pen}, Ue-Li and {Pi{\'e}tu}, Vincent and {Plambeck}, Richard and {PopStefanija}, Aleksandar and {Porth}, Oliver and {P{\"o}tzl}, Felix M. and {Prather}, Ben and {Preciado-L{\'o}pez}, Jorge A. and {Psaltis}, Dimitrios and {Pu}, Hung-Yi and {Pu}, Hung-Yi and {Rao}, Ramprasad},
        title = "{Event Horizon Telescope observations of the jet launching and collimation in Centaurus A}",
      journal = {Nature Astronomy},
         year = 2021,
        month = jul,
       volume = {5},
        pages = {1017-1028},
          doi = {10.1038/s41550-021-01417-w},
archivePrefix = {arXiv},
       eprint = {2111.03356},
 primaryClass = {astro-ph.GA},
       adsurl = {https://ui.adsabs.harvard.edu/abs/2021NatAs...5.1017J}
}

%\begin{thebibliography}{99}
%\end{thebibliography}

%%%%%%%%%%%%%%%%%%%%%%%%%%%%%%%%%%%%%%%%%%%%%%%%%%

%%%%%%%%%%%%%%%%% APPENDICES %%%%%%%%%%%%%%%%%%%%%

\appendix

\section{Dependence on the Magnetization Ceiling $\sigmamax$} \label{sec:sigma_max_sensitivity}
As discussed in $\S$\ref{sec:methods}, GRMHD simulations impose a numerical ceiling $\sigmamax$ on the magnetization parameter to maintain stability in highly magnetized, low-density regions. Dynamical quantities may therefore, in principle, depend on the adopted magnetization ceiling.

\begin{figure}
	\centering
	\includegraphics[width=0.48\textwidth]{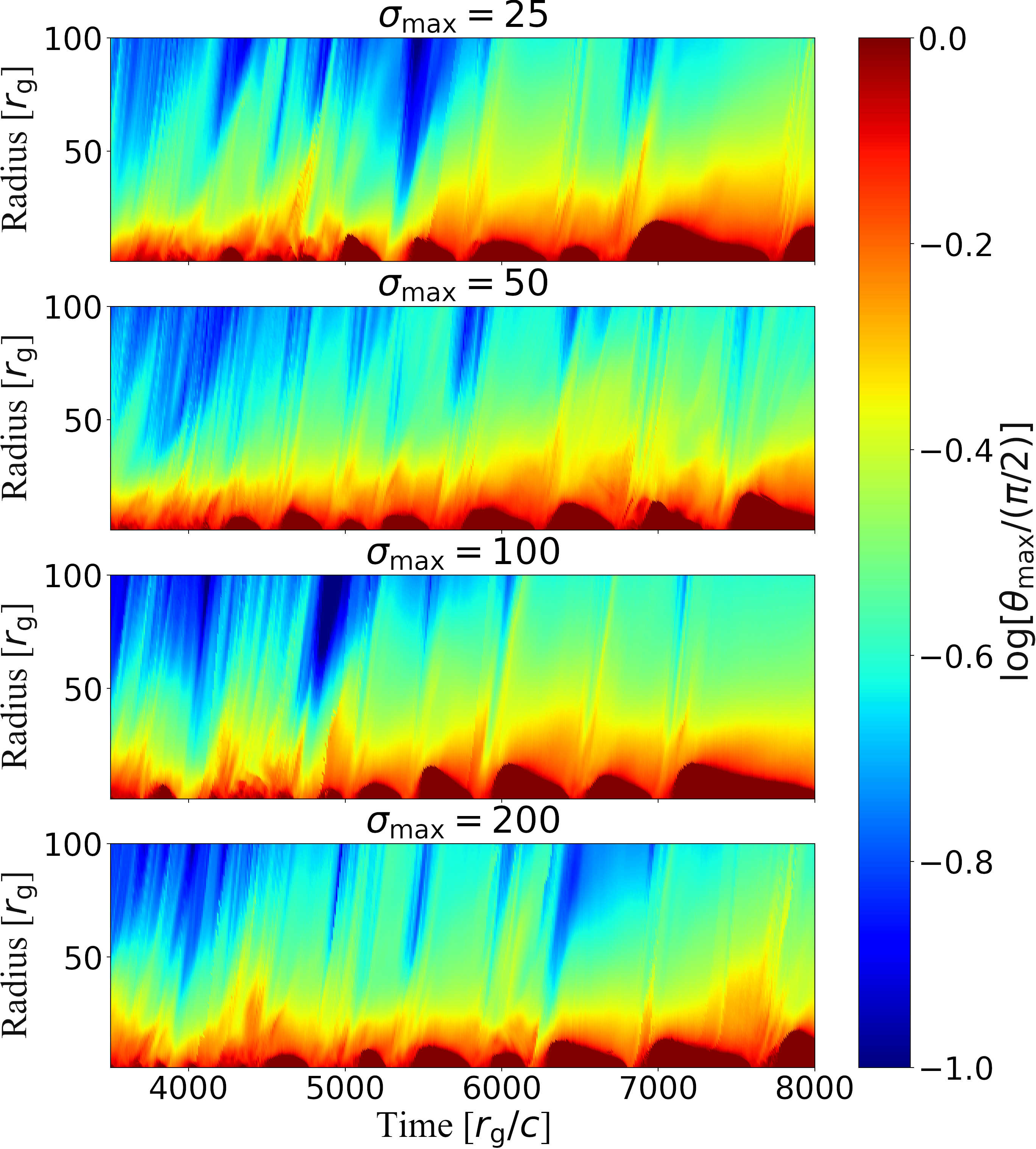}
    \caption{Time–radius diagrams of the maximum polar angle (in radians) in the upper hemisphere $(\theta \leq \pi/2)$ at which the tracer magnetization satisfies $\sigmatr > \sigma_{\rm GJ}=10^{10}$. The horizontal axis shows time, and the 
    vertical axis corresponds to radial distance from the black hole, spanning $r \in [0,100]\rg$. Each panel shows results for a different magnetization ceiling $\sigmamax$ $=25,50,100,200$.}    \label{fig:resol_theta_max_vs_time_sigmax}
\end{figure}

\begin{figure}
	\centering
	\includegraphics[width=0.48\textwidth]{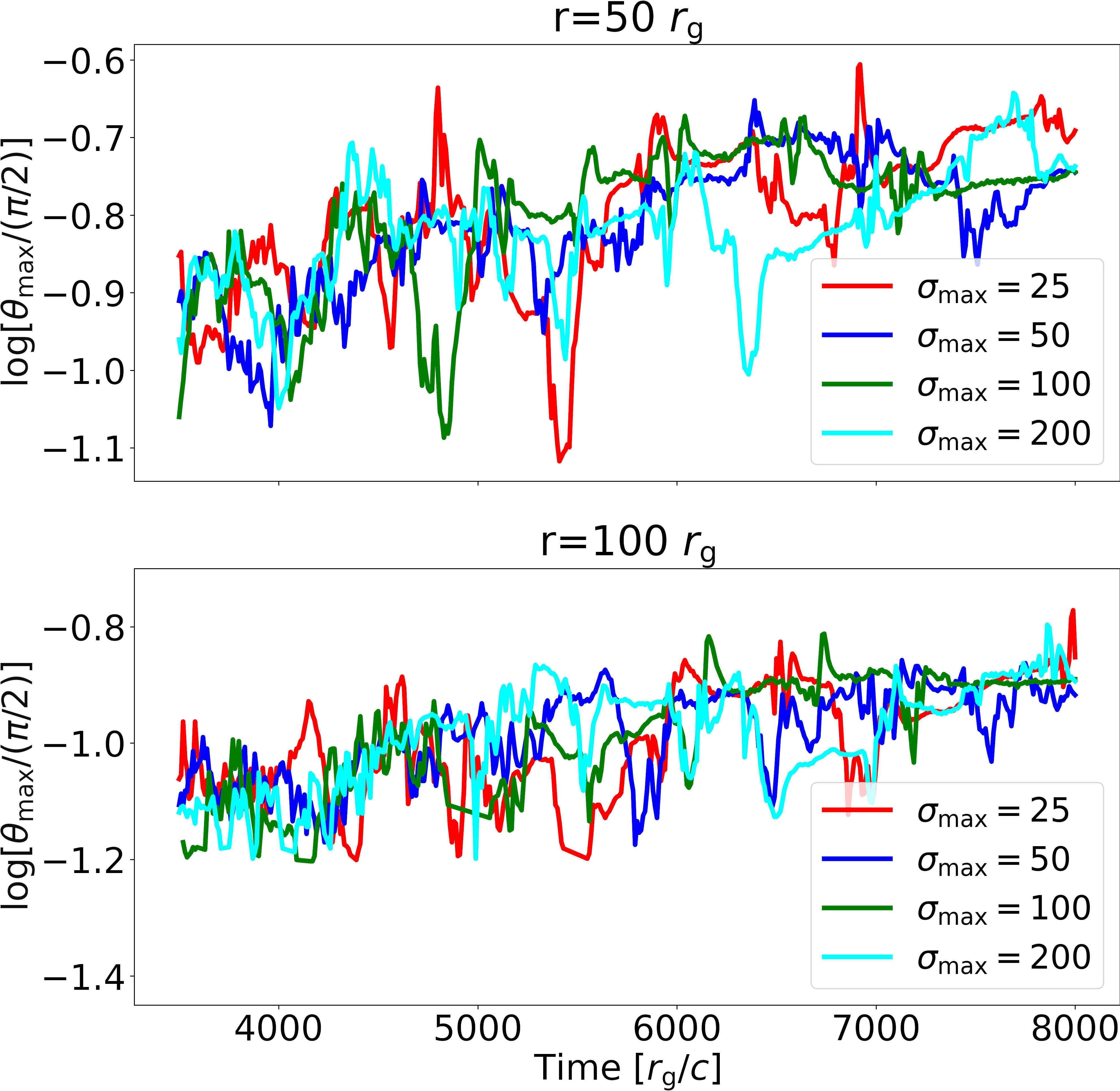}
    \caption{Time evolution of the maximum polar angle $\thetamax$ at a fixed radius of $r=50\rg$ (top panel) and $r=100\rg$ (bottom panel), from the four simulations shown in Fig. \ref{fig:resol_theta_max_vs_time_sigmax}. The curves show $\log[\thetamax/(\pi/2)]$ for different magnetization ceilings $\sigmamax=25,50,100,200$ as indicated in the legend. The horizontal axis shows time in units of $\rg/c.$}
    \label{fig:resol_converge_sigmax}
\end{figure}

To test the robustness of our results, we perform a parameter study varying $\sigmamax = 25, 50, 100, 200$ in the prograde case. Because mass injection occurs in the most magnetized and least dense regions, any dependence on $\sigmamax$ should be most evident in the structure of the baryon-depleted jet core.
Fig.~\ref{fig:resol_theta_max_vs_time_sigmax} shows time–radius diagrams of the maximum polar angle $\thetamax$ bounding the $\sigmatr > \sigma_{\rm GJ}=10^{10}$ region for each value of $\sigmamax$. Despite the factor-of-eight variation in $\sigmamax$, all panels display qualitatively similar temporal behaviour, including comparable sequences of flux-eruption events (dark red patches).
To quantify these differences further, Fig.~\ref{fig:resol_converge_sigmax} shows the evolution of $\thetamax$ at fixed radii of $r=50\rg$ and $r=100\rg$. The curves for all four values of $\sigmamax$ remain within the same range and show no systematic trend with the magnetization ceiling. We therefore conclude that the system dynamics and tracer-based diagnostics presented here are robust to the adopted value of $\sigmamax$.


\section{Dependence on numerical resolution} \label{sec:convergence}
\begin{figure}
	\centering
	\includegraphics[width=0.48\textwidth]{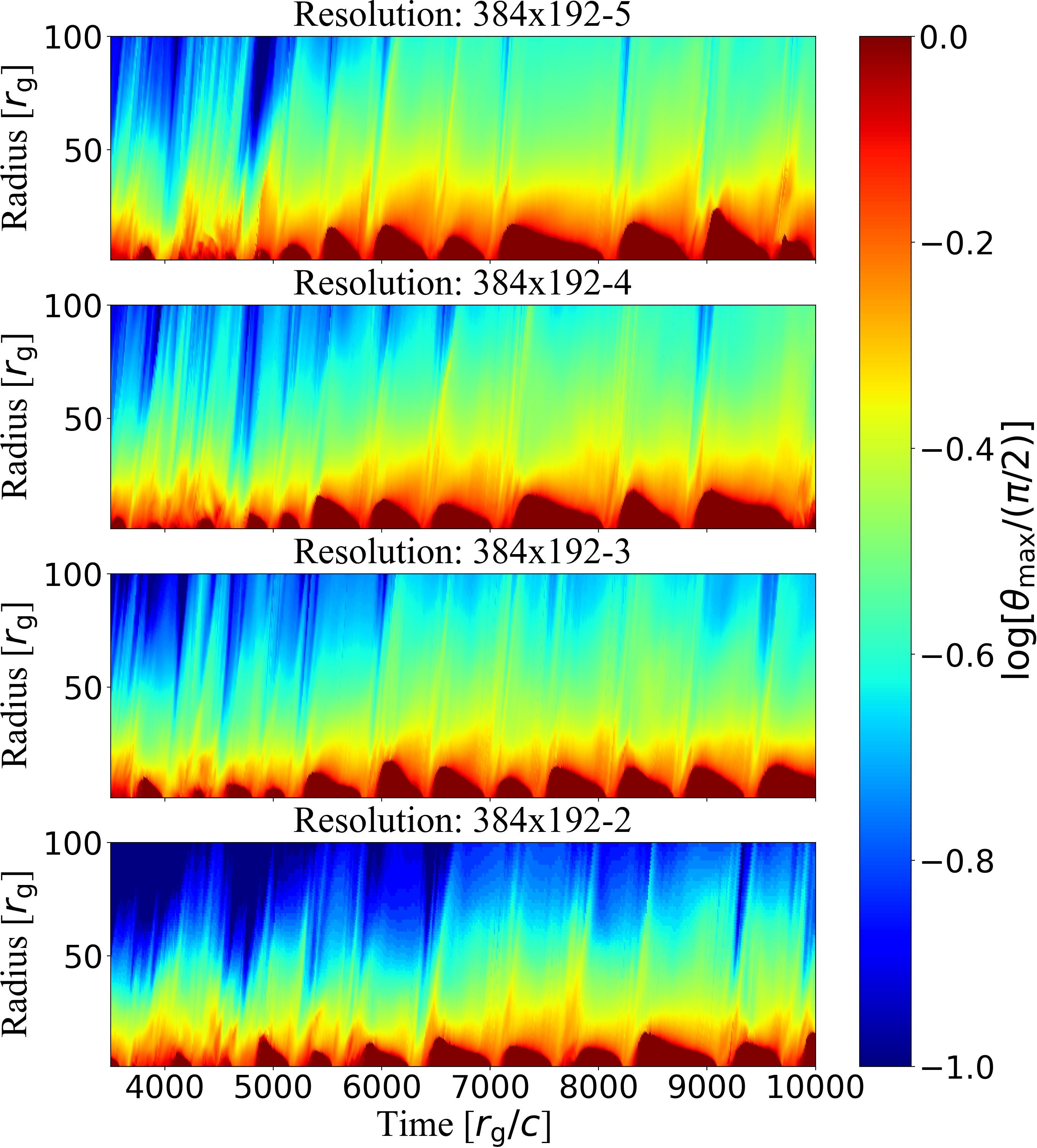}
    \caption{Time–radius diagrams of the maximum polar angle (in radians) in the upper hemisphere $(\theta \leq \pi/2)$ at which the tracer magnetization satisfies $\sigmatr > \sigma_{\rm GJ}=10^{10}$. The horizontal axis shows time, and the 
    vertical axis corresponds to radial distance from the black hole, spanning $r \in [0,100]\rg$. All simulations use the same base grid of $384 \times 192$ in $(\log r,\theta)$-coordinate. The four panels correspond to simulations with different adaptive-mesh-refinement (AMR) levels: from top to bottom, refinement levels 5, 4, 3, and 2, with higher refinement levels providing higher spatial resolution.}
    \label{fig:resol_theta_max_vs_time_resol}
\end{figure}

To assess the numerical convergence of the jet–funnel morphology and baryon loading in our simulations, we performed a resolution study of the prograde case using four grids of increasing resolution: \texttt{384$\times$192-2}, \texttt{384$\times$192-3}, \texttt{384$\times$192-4}, and \texttt{384$\times$192-5}. The final digit $n$ denotes the AMR level, where $n=1$ corresponds to the base grid. Each additional level doubles the resolution in each dimension, yielding effective resolutions of $768 \times 384 \,(n=2)$, $1536\times768 \,(n=3)$, $3072\times1536 \,(n=4)$, and $6144\times3072 \,(n=5)$. All simulations were initialized with identical physical and numerical parameters and differ only in the number of AMR levels.
The jet funnel is characterized by the maximum polar angle (in radians) in the upper hemisphere $(\theta \leq \pi/2)$ for which the tracer magnetization satisfies $\sigmatr > \sigma_{\rm GJ}=10^{10}$, as appropriate for M87*. 
Fig.~\ref{fig:resol_theta_max_vs_time_resol} shows two-dimensional maps of $\log[\thetamax/(\pi/2)]$ as a function of radius and time for the four resolutions. The  three highest resolutions (\texttt{384$\times$192-5}, \texttt{384$\times$192-4}, and \texttt{384$\times$192-3}) exhibit very similar morphologies: the angular extent of the baryon-poor funnel and the time variability of the funnel wall are statistically consistent. \acc{This visual agreement indicates that the large-scale geometry of the charge-starved polar region is numerically converged. It also suggests that the flux-eruption dynamics has reached the asymptotic plasmoid-dominated regime of reconnection, in which the reconnection rate is largely independent of numerical resolution \citep{Ripperda_2020,Grehan_2025}.} 
%In contrast, the \texttt{384$\times$192-2} run shows a noticeable dominance of blue colours at larger radii ($r \gtrsim 70\rg$), indicating systematically lower $\sigmamax$ values and suggesting that the simulation has not yet converged at this resolution.

\begin{comment}
In contrast, the lower-resolution runs \texttt{384$\times$192-2}) show noticeable deviations. Both exhibit intermittent white patches, which correspond to radii and times where no solution for $\thetamax$ is found---i.e., where $\sigma_{\rm tr}<10^{10}$ is not satisfied at any angle, thus suggesting efficient baryon penetration/laoding into the jet. These gaps indicate episodes in which the funnel fails to form or becomes numerically unstable, particularly at larger radii. Moreover, the lower-resolution models show pronounced fragmentation and irregularity in the red, high-$\thetamax$ regions (near the equator), indicating that baryon depletion in the midplane is less reliably captured.
\end{comment}

\begin{figure}
	\centering
	\includegraphics[width=0.48\textwidth]{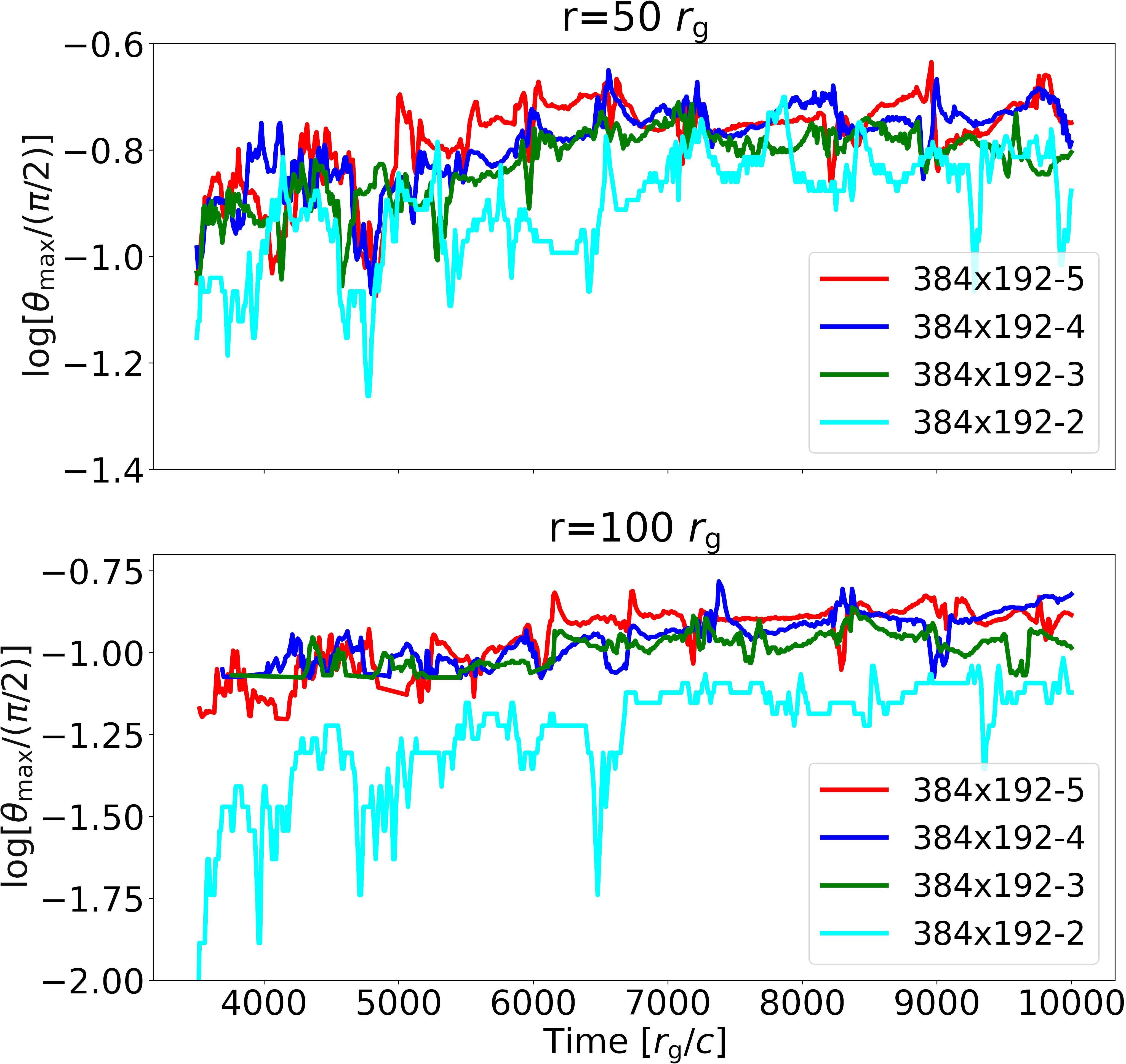}
    \caption{Time evolution of the maximum polar angle $\thetamax$ at a fixed radius of $r=50\rg$ and $r=100\rg$, from the four simulations shown in Fig. \ref{fig:resol_theta_max_vs_time_resol}. The curves show $\log[\thetamax/(\pi/2)]$ for simulations with base grid resolution $384\times 192$ in $(\log r, \theta)$-coordinates but with different AMR refinement levels: 5,4,3 and 2. The horizontal axis shows time in units of $\rg/c.$}
    \label{fig:resol_converge}
\end{figure}

To quantify convergence, Fig.~\ref{fig:resol_converge} shows $\thetamax$ at fixed radii of $r=50\rg$ and $r=100\rg$. Consistent with Fig.~\ref{fig:resol_theta_max_vs_time_resol}, the curves for the three highest resolutions—\texttt{384$\times$192-5} (red), \texttt{384$\times$192-4} (blue), and \texttt{384$\times$192-3} (green)—closely track each other throughout the simulation, with only minor deviations. In contrast, the lowest-resolution run, \texttt{384$\times$192-2} (cyan), systematically yields smaller values of $\thetamax$ than the higher-resolution cases.
These results demonstrate that the \texttt{384$\times$192-3}, \texttt{384$\times$192-4}, and \texttt{384$\times$192-5} grids—the latter adopted in the main body of the paper—have reached convergence for the properties most relevant to this study: (i) the presence and stability of a persistent baryon-depleted jet funnel, (ii) the radial extent of the charge-starved magnetosphere during eruptions, and (iii) the temporal variability of the funnel opening angle.
\acc{The lowest-resolution run systematically overestimates the baryon loading of the jet funnel, likely due either to enhanced numerical diffusion at the jet–disk boundary or to an artificially elevated reconnection rate when current sheets have not yet reached the asymptotic plasmoid-dominated regime. In the latter case, the higher reconnection rate would promote additional mixing and mass loading, potentially suggesting that the faster reconnection rate ($\sim 0.1\,c$) expected in the kinetic regime could further enhance baryon entrainment.}
Given the agreement between the two highest-resolution cases, we conclude that our fiducial highest-resolution simulation (\texttt{384$\times$192-5}) provides a numerically converged description of the baryon content in the jet and black-hole magnetosphere.
%-depleted jet–funnel boundary, and therefore yields tracer-density $\rhotr$ dynamics that are numerically robust.

\section{The Goldreich-Julian \accc{Magnetization}} \label{appendix:GJ_magnetization}
%In the main body of the paper, we have adopted a representative tracer magnetization threshold of $\sigmatr=10^{10}$. This value is  motivated by the characteristic GJ magnetization expected in the plasma surrounding a rapidly spinning supermassive black hole such as M$87^*$. This appendix provides the relevant derivations for the GJ number density and the associated magnetization.
\accc{In the main body of the paper, we adopted a representative tracer magnetization threshold of $\sigmagj=10^{10}$. This value is motivated by the characteristic Goldreich–Julian (GJ) magnetization expected in the plasma surrounding the rapidly spinning supermassive black hole M87*. The GJ magnetization depends on the magnetic-field strength and black-hole mass, however, and therefore varies between sources. In this appendix, we provide the relevant scalings and compare the values expected for M87* and Sgr A*.}

The GJ number density is given by
\begin{align}
    n_{\rm GJ} = \frac{\Omega B}{2\pi e c},
\end{align}
where $e$ is the elementary charge, $c$ is the speed of light, \accc{$B$ is the characteristic magnetic-field strength,} and $\Omega$ is the angular velocity of the black hole magnetosphere. For a rapidly spinning black hole, we approximate the latter as $\Omega \approx c/(2\rg)$. The corresponding GJ magnetization is
\begin{align}
    \sigma_{\rm GJ}=\frac{B^2}{4\pi m_p c^2 n_{\rm GJ}} = \frac{eB}{2m_p c \Omega} = \frac{\omega_B}{2\Omega}.
\end{align}
Here $m_p$ is the proton mass and  $\omega_B=eB/(m_p c)$ is the proton cyclotron frequency.
\accc{Using $\Omega \approx c/(2r_{\rm g})$, the GJ magnetization can be written as
\begin{align}
\sigma_{\rm GJ}
\simeq 3\times10^{10}
\left(\frac{B}{100\,{\rm G}}\right)
\left(\frac{\Mbh}{6.5\times10^9\,M_\odot}\right).
\label{eq:sigma_gj}
\end{align}
Thus, the GJ magnetization scales linearly with both the characteristic magnetic-field strength and the black-hole mass.}

\accc{For M87*, adopting $\Mbh \simeq6.5\times10^9\,M_\odot$ and a characteristic near-horizon magnetic-field strength of $B\sim100$ G gives $\sigmagj\sim3\times10^{10}$. This motivates, to order of magnitude, the fiducial value $\sigmagj =10^{10}$ adopted in the main text.}
\accc{For comparison, Sgr A* has a substantially smaller black-hole mass, $\Mbh \simeq4\times10^6\,M_\odot$. Its magnetic-field strength in quiescence is estimated to be $B\simeq10$--$50$ G in the inner $10 \rg$ \citep{Ripperda_2022}. Substituting these values into Equation~(\ref{eq:sigma_gj}) gives
$
\sigma_{\rm GJ} \simeq 2\times10^6-9\times10^6.
$
The comparison between M87* and Sgr A* illustrates that $\sigmagj$ is source dependent and can differ by several orders of magnitude among different low-luminosity accreting supermassive black holes.}

\begin{comment}
Using representative parameters for M$87^*$---gravitational radius $\rg \approx 10^{15}\rm cm$, angular frequency $\Omega \sim 10^{-5}$Hz, and magnetic field strength on the horizon $B=10^2 B_2$ G \citep{Hakobyan_2025}---the estimate for the GJ magnetization is
\begin{align}
    \sigma_{\rm GJ} \approx 3 \times 10^{10} B_2
\end{align}
This shows that magnetization values of order $10^{10}$ naturally arise when the plasma density approaches the Goldreich–Julian density in a highly magnetized black-hole magnetosphere with field strengths $B\sim 10^2$ G.
% Don't change these lines
\end{comment}

\section{Comparison with earlier works} 
\label{sec:wong}
\begin{figure}
	\centering
	\includegraphics[width=0.48\textwidth]{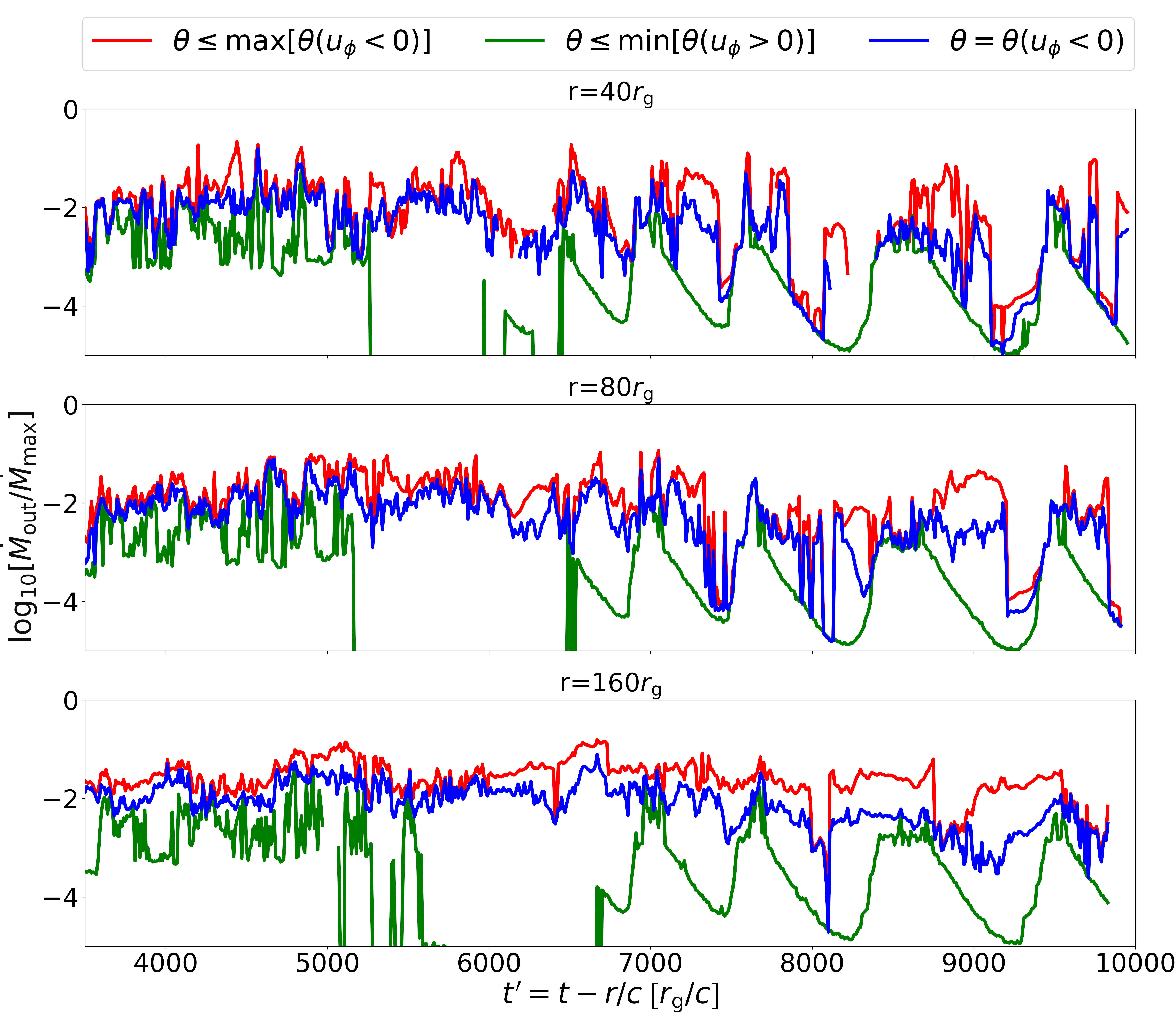}
    \caption{For spin parameter $a=-0.9375$, we compute the mass outflow rate $\dot{M}_{\rm out}$ in the jet at different radii (panels from top to bottom display $r/r_{\rm g}=40$, 80 and 160). The mass outflow rate is calculated using the tracer density, and adopting different definitions of the integration surface (see legend and text), all based on the fact that the jet has negative azimuthal four-velocity. The mass outflow rate is normalized to a representative horizon-scale mass accretion rate during phases of active accretion, which in code units is $\dot{M}_{\rm max}=119.78$ (see Fig.~\ref{fig:mdot_phi}(c)). Time is shifted in a radius-dependent way to account for light-travel time effects.}     \label{fig:wong}
\end{figure}

\acc{In this appendix, we quantify tracer mass entrainment in the jet using a strategy similar to \citet{Wong_2021}, with the goal of comparing to their conclusions. We consider the retrograde case and compute the mass outflow rate $\dot{M}_{\rm out}$ in the jet (upper hemisphere only) at different radii (panels from top to bottom in Fig.~\ref{fig:wong} display $r/r_{\rm g}=40$, 80 and 160). The mass outflow rate is calculated using the tracer density, and adopting different definitions for the integration surface, all based on the fact that the jet has negative azimuthal four-velocity ($u_\phi<0$). In particular, the red line considers all cells up to the maximum polar angle where $u_\phi<0$ (regardless of the fact that some of these cells may have $u_\phi>0$; this should be taken as an upper limit for jet opening angle); the green line takes into account all cells up to the minimum angle at which $u_\phi>0$ (here, all cells have $u_\phi<0$; this is a lower limit for the jet opening angle); and the blue line only considers cells where $u_\phi<0$, regardless of whether they are adjacent, or separated by cells with $u_\phi>0$. Regardless of the specific criterion, the plot confirms that mass entrainment is highly time dependent, and modulated by flux eruption cycles. 
Comparison between the three different criteria suggests that the level of mass entrainment is strongly dependent on how the jet boundary is defined, with variations up to two orders of magnitude. While the red line is roughly consistent with the estimate by \citet{Wong_2021}, the green line is systematically lower by roughly one order of magnitude. Our results then can be reconciled with \citet{Wong_2021} if most of the baryons entering the jet surface (identified by the red line) remain confined near the jet boundary, while the jet spine—carrying most of the electromagnetic flux—remains baryon-poor.}

\section{Passive-Tracer fraction ($\chitr$) Maps} \label{sec:chi_plots}
In the main text, most of our diagnostics are based on $\sigmatr$ (Figs.~\ref{fig:2D_grid_spin+ve}–\ref{fig:2D_grid_spin-ve} and \ref{fig:theta_vs_time}), which depends only on the magnetic field strength and on the physical baryon density supplied by the disk. As such, it is independent of the \textit{ad hoc} numerical mass injection adopted for stability in low-density regions. For completeness, we present here the spatial structure of the tracer fraction $\chitr$.

%only on the unpopulated $\rhotr$ and is independent of any numerical choice of \textit{ad hoc} mass injection. For completeness, we include here the corresponding plots of the standalone 

\begin{figure*}
	\centering
	\includegraphics[width=0.9\textwidth]{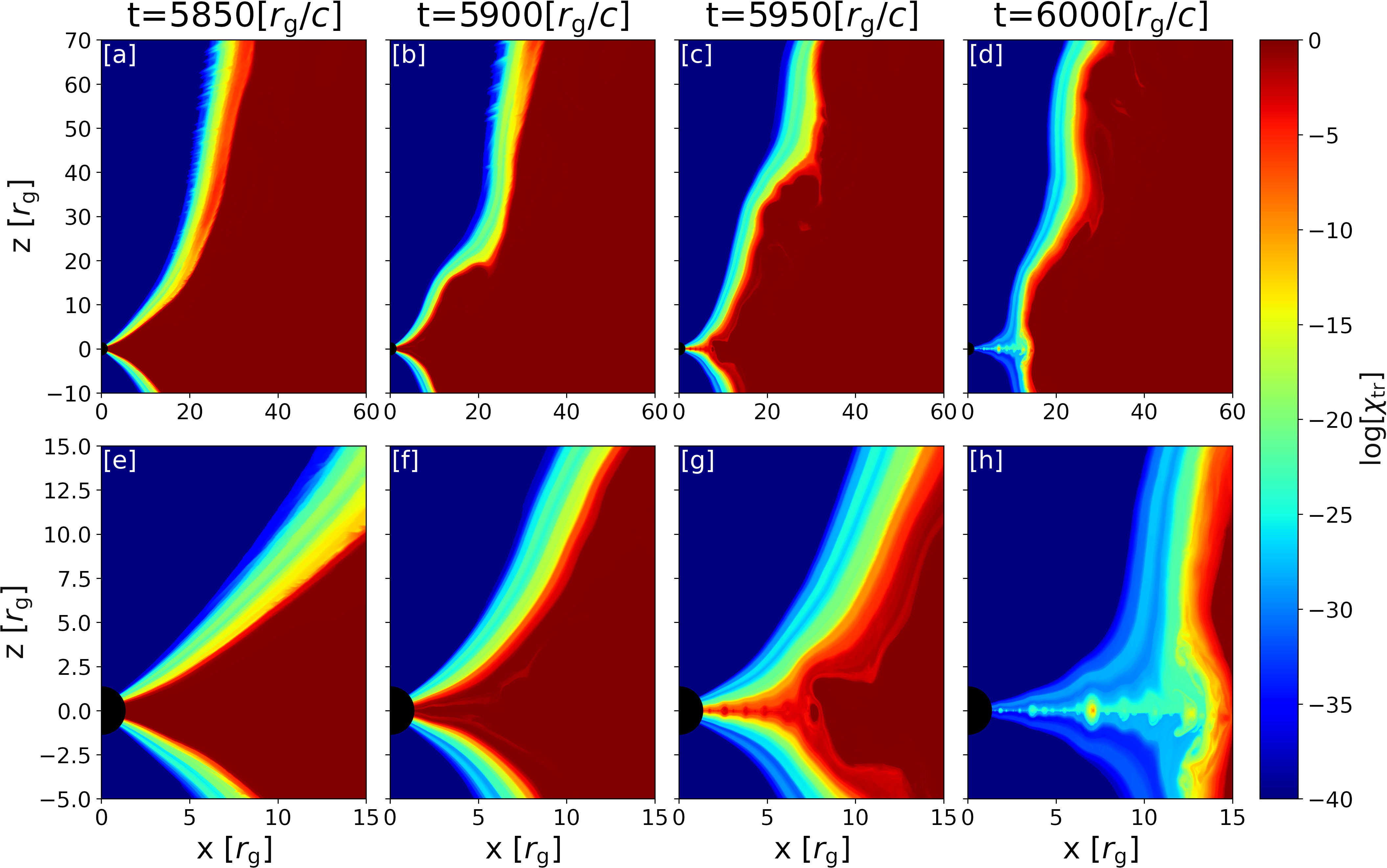}
	\caption{Same as Fig.~\ref{fig:2D_grid_spin+ve} for the prograde case, but plotting the logarithm of the tracer fraction $\chitr$.}
    \label{fig:2D_chi_grid_spin+ve}
\end{figure*}

\begin{figure*}
	\centering
	\includegraphics[width=0.9\textwidth]{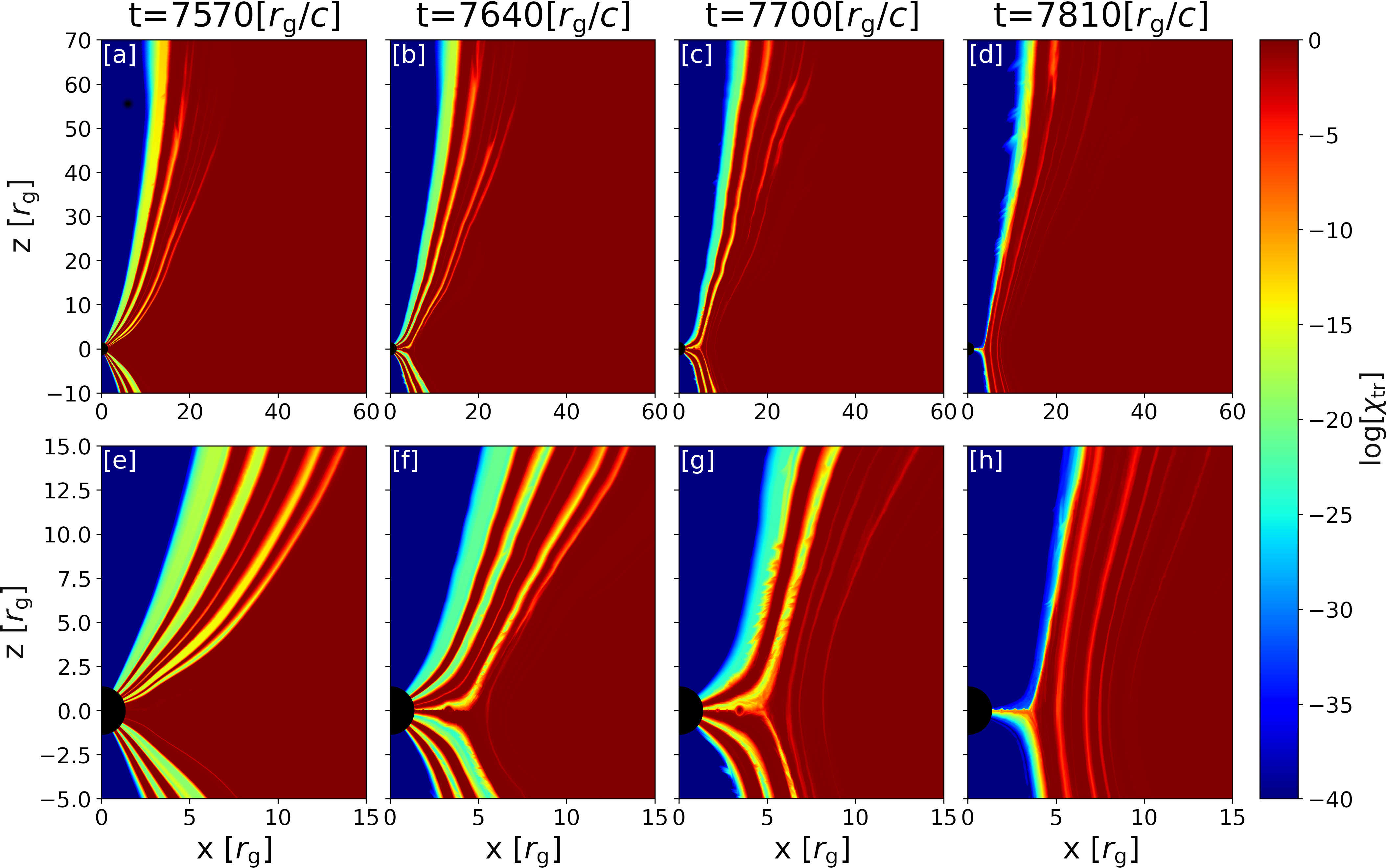}
	\caption{Same as Fig.~\ref{fig:2D_grid_spin0} for the non-rotating case, but plotting the logarithm of the tracer fraction $\chitr$.}
    \label{fig:2D_chi_grid_spin0}
\end{figure*}

\begin{figure*}
	\centering
	\includegraphics[width=0.9\textwidth]{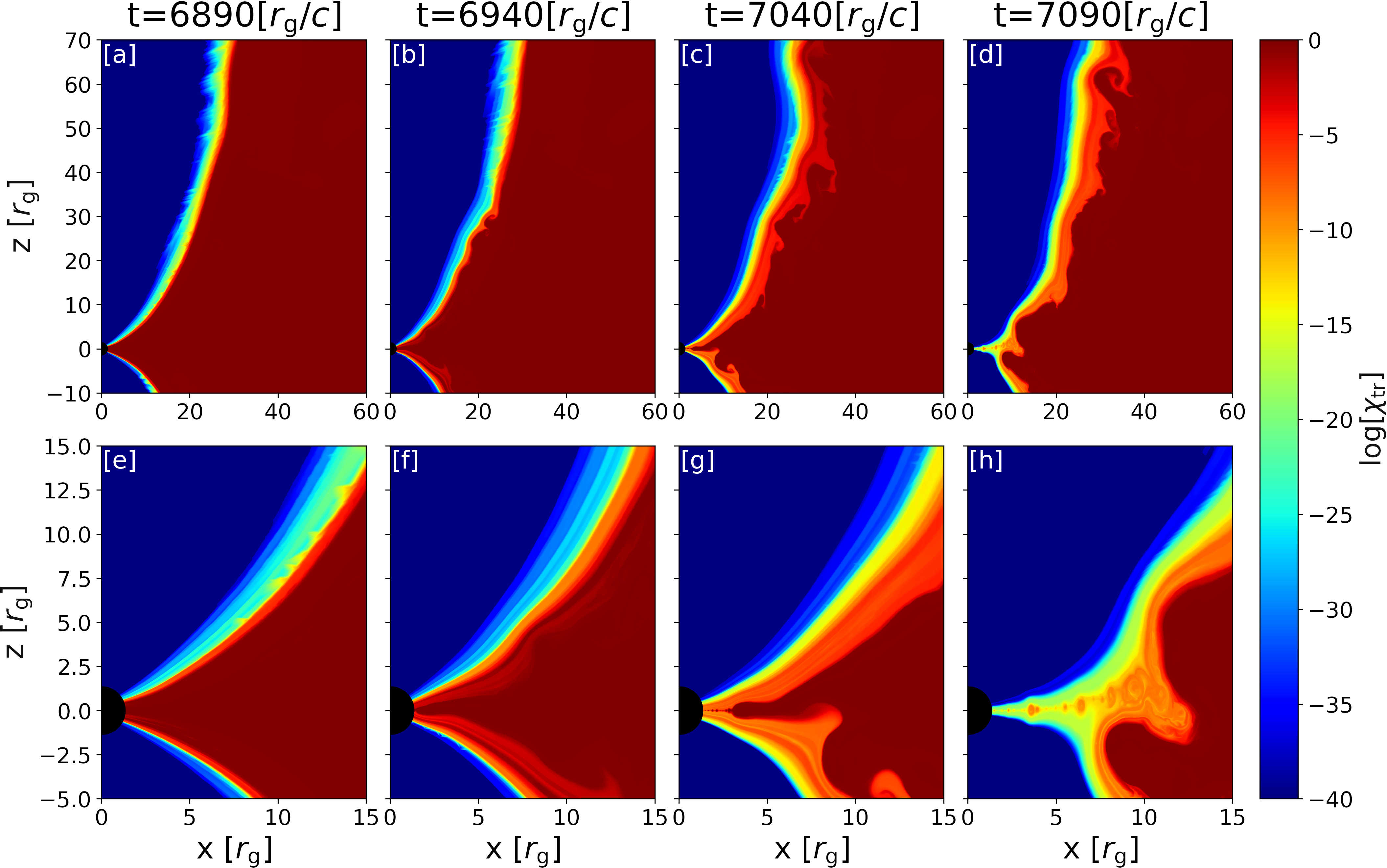}
	\caption{Same as Fig.~\ref{fig:2D_grid_spin-ve} for the retrograde case, but plotting the logarithm of the tracer fraction $\chitr$.}
    \label{fig:2D_chi_grid_spin-ve}
\end{figure*}

\begin{figure}
	\centering
	\includegraphics[width=0.48\textwidth]{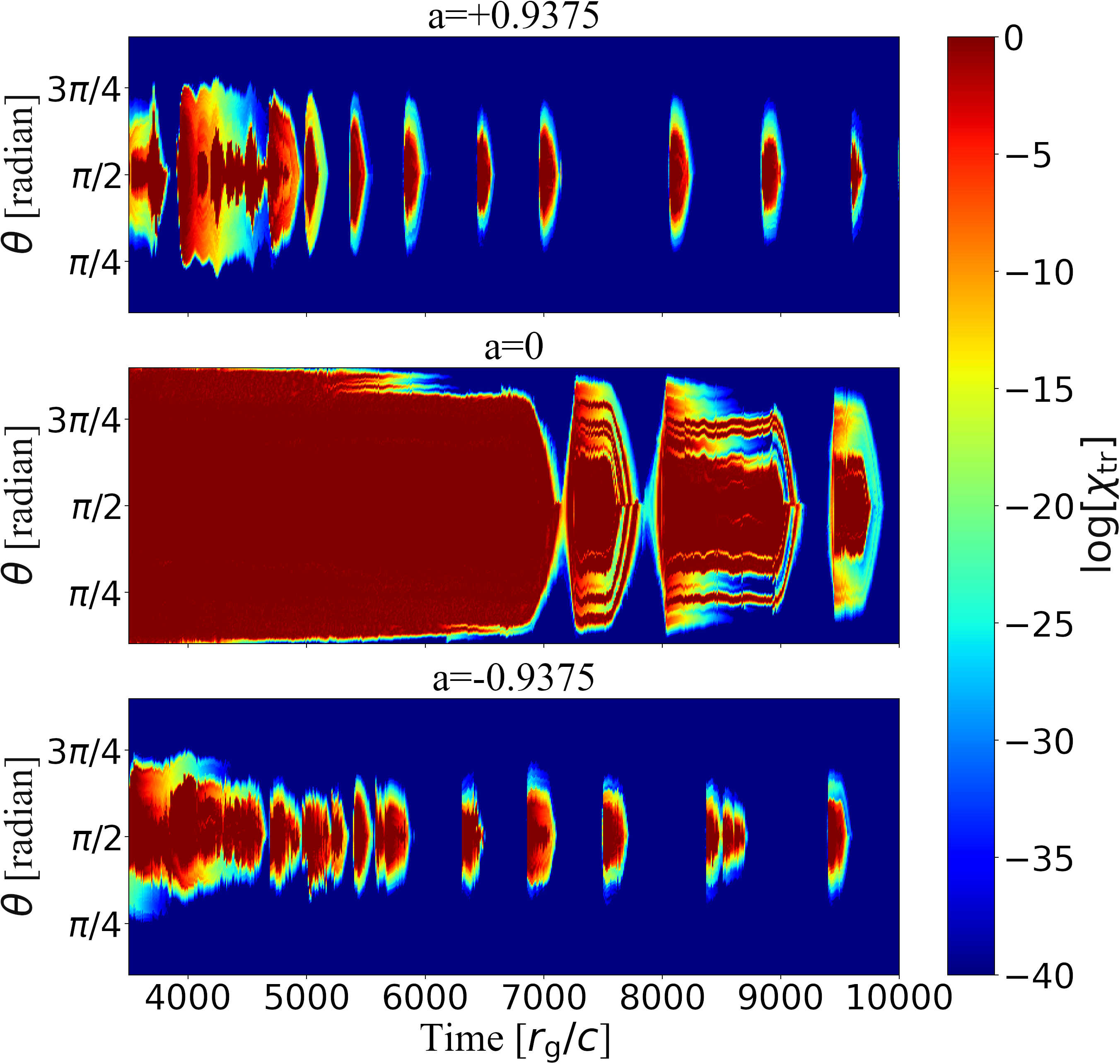}
	\caption{Same as Fig.~\ref{fig:theta_vs_time}, but plotting the logarithm of the tracer fraction $\chitr$.}
    \label{fig:theta_vs_time_tr1}
\end{figure}

Figures~\ref{fig:2D_chi_grid_spin+ve}, \ref{fig:2D_chi_grid_spin0}, and \ref{fig:2D_chi_grid_spin-ve} show the spatial distribution of the tracer fraction $\chitr$ for the snapshots presented in Figs.~\ref{fig:2D_grid_spin+ve}, \ref{fig:2D_grid_spin0}, and \ref{fig:2D_grid_spin-ve}, while Fig.~\ref{fig:theta_vs_time_tr1} presents the $\chitr$ analogue of Fig.~\ref{fig:theta_vs_time}. Because $\chitr$ is restricted to the interval $[0,1]$, the lower bound of $\sim -40$ on the logarithmic colour scale simply corresponds to $\chitr=0$ (i.e., no baryons). In these maps, any colour deviating from dark red (which corresponds to $\chitr=1$) indicates that a fraction of the local mass density originates from numerical density injection.
The $\chitr$ maps therefore provide a  visual measure of how much numerical mass injection has occurred and where the numerically-injected mass has been transported. Regions that remain uniformly dark red consist entirely of disk-originated plasma, whereas blue or cyan patches mark locations affected by numerical mass injection. These diagnostics help interpret the $\sigmatr$ structure discussed in the main text in terms of the underlying tracer-fraction composition.

\bsp	% typesetting comment
\label{lastpage}
\end{document}